\documentclass[11pt]{article} 

\usepackage{fullpage}

\usepackage{graphicx}
\usepackage{upref}
\usepackage{enumerate}
\usepackage{latexsym}
\usepackage{pdfpages}

\usepackage[bookmarks]{hyperref}
\usepackage{color,graphics}
\definecolor{purple}{rgb}{0.5,0,0.5}
\usepackage{graphicx}
\usepackage{comment} 
\usepackage{caption}
\usepackage{subcaption}
\usepackage{wrapfig}
\usepackage{braket}

\usepackage{amssymb}
\usepackage{amsmath}
\usepackage{bbm}

\usepackage{amsthm}
\usepackage{mathrsfs}
\usepackage{mathtools}

\usepackage{epsfig,graphicx}
\usepackage{algorithmic}
\usepackage[ruled,linesnumbered]{algorithm2e}
\usepackage{amsfonts}
\usepackage{tikz} 

\usepackage{varioref}
\usepackage[ansinew]{inputenc}
\usepackage{qcircuit}

\usepackage{authblk}
\usepackage{multirow}
\usepackage{lineno}

\theoremstyle{plain}

\theoremstyle{plain}

\theoremstyle{plain}

\theoremstyle{plain}

\theoremstyle{plain}

\theoremstyle{plain}

\theoremstyle{definition}

\theoremstyle{definition}

\theoremstyle{remark}
\newtheorem{remark}{Remark}[section]

\theoremstyle{definition}

\AtBeginDocument{}

\newcommand{\fld}{\mathbb{F}}

\newcommand{\cliff}{\mathcal{C}}

\newcommand{\X}{\text{X}}

\newcommand{\had}{\text{H}}
\newcommand{\T}{\text{T}}
\newcommand{\CNOT}{\text{CNOT}}
\newcommand{\phase}{\text{S}}

\newcommand{\tof}{\text{TOF}}

\newcommand{\cswap}{\text{CSWAP}}

\newcommand{\conj}[1]{\overline{#1}}

\newcommand{\cns}{\text{COMP-N-SUB}}
\newcommand{\tcount}{\mathcal{T}}
\newcommand{\n}{\mathcal{N}^{cnot}}
\newcommand{\q}{\mathcal{Q}}
\newcommand{\sub}{\text{SUB}}
\newcommand{\cAddSub}{\text{ctrl-AddSub}}
\newcommand{\cAdd}{\text{c-ADD}}

\graphicspath{{./intDiv_fig/}}

\begin{document}

\title{Improved quantum circuits for division}

\author[1]{Priyanka Mukhopadhyay \thanks{Corresponding author : mukhopadhyay.priyanka@gmail.com, Priyanka.Mukhopadhyay@ibm.com}}
\author[1]{Alexandru Gheorghiu \thanks{agheorghiu@ibm.com} }
\author[1]{Hari Krovi \thanks{Hari.Krovi@ibm.com} }

 \affil[1]{IBM Research, Cambridge MA, USA }

\date{}
 \maketitle
 
 \begin{abstract}

 Arithmetic operations are an important component of many quantum algorithms. Optimizing quantum circuits for these operations therefore leads to more efficient implementations of the corresponding algorithms. In this paper, we develop new fault-tolerant quantum circuits for various integer division algorithms (both reversible and non-reversible). These circuits, when implemented in the Clifford+T gate set, achieve  improvements in the asymptotic T-count and CNOT-count from $O(n^2)$ to $O(nm)$, and in the asymptotic T-depth from $O(n^2)$ to $O(n \log m)$, where $n$ and $m$ are the bit lengths of the dividend and divisor, respectively. The qubit counts are significantly lower than those in previous works.  We achieve this through careful modifications of the division algorithms and by expressing them in terms of a primitive we call COMP-N-SUB that compares two integers and conditionally subtracts them. We show that this primitive can be implemented at a cost, in terms of both Clifford and non-Clifford gates, comparable to that of \emph{one addition}. By contrast, performing comparison and conditional subtraction separately would cost roughly as much as a \emph{controlled addition} plus a regular addition.
 
 \end{abstract}

\section{Introduction}
\label{sec:intro}

Much of quantum computing research has been devoted to designing algorithms that can outperform their classical counterparts and, in some cases, solve problems of practical significance. Examples include factoring and computing discrete logarithms \cite{1999_S}, unstructured database search \cite{1996_G}, Hamiltonian simulation \cite{1982_F, lloyd, Childs, Campbell_2022}, solving algebraic problems \cite{RevModPhys.82.1}, and machine learning \cite{2017_RdSRetal}, among others. Many of these algorithms require the computation of classical functions using \emph{algebraic} and \emph{arithmetic} operations \cite{1999_S, 2000_vDH, 2006_vDHI, 2008_vDS, 2010_TGA, 2007_H, 2021_RA, 2020_LFXetal}. As such, designing efficient quantum circuits and procedures for implementing these arithmetic operations is an important and active area of research.
Division, in particular, is a fundamental operation that appears in quantum algorithms for computing shifted quadratic characters \cite{2000_vDH}, solving the principal ideal problem \cite{2007_H}, solving the hidden shift problem \cite{2006_vDHI}, jet clustering \cite{2020_WNHT}, differential evolution \cite{2021_HMEetal}, the finite element method \cite{2025_AKWM}, and objective-function maximization \cite{2019_GI}. Division circuits can also be used to implement functions such as greatest common divisors, square roots, reciprocals, fractional powers, natural logarithms, and bilinear interpolation \cite{2003_BLC, 2007_H, 2016_BHPP, 2017_SRWdM, 2017_ZHFI}, which are used in a number of quantum algorithms \cite{2014_SSX, 2009_HHL, 2007_H, 2003_BLC}.

When considering \emph{fault-tolerant} implementations of division circuits, the choice of gate set is important because the error-correcting code may make some operations, such as \emph{Clifford} gates, easier to perform than others, such as non-Clifford gates. While Clifford+T is the most popular gate set, other universal fault-tolerant gate sets include Clifford+Toffoli and Clifford+CS. Clifford gates are easier to implement fault-tolerantly in most codes because they can be performed \emph{transversally}. By contrast, non-Clifford gates such as T and Toffoli require large ancilla factories and additional operations such as \emph{gate teleportation} and \emph{magic state distillation} \cite{1997_G, 1998_CPMetal, 2005_BK, 2009_FSG, 2013_J, 2013_J2, 2017_PPND, 2019_L, 2018_HH, 2020_BCHK, 2024_BCGetal}. These requirements not only increase the overhead of fault-tolerant implementations but can also make the operations less accurate and more time-consuming. Hence, much of the research on quantum-circuit optimization is devoted to designing fault-tolerant circuits that \emph{minimize the count and depth of non-Clifford gates}; this is also our goal when constructing division circuits.

The arithmetic circuits developed in the literature can be broadly divided into two groups. The first group is inspired by classical reversible circuits. Examples include the quantum ripple-carry adders in \cite{2004_CDKM, 2008_TK, 2010_TTK, 2018_G}, the quantum carry-lookahead adder in \cite{2006_DKRS}, the multipliers in \cite{2018_MT}, and the dividers in \cite{2011_KAF, 2013_JB, 2016_DBJ, 2019_TMVH, 2022_YGWetal, 2024_OPF}. The second group performs phase arithmetic using the quantum Fourier transform (QFT). Examples include the adder and multiplier circuits in \cite{2000_D, 2003_BBF, 2014_MP, 2014_PG, 2017_RG, 2020_S, 2023_RNFetal, 2023_WHH, 2024_KY}. Circuits in the first group are dominated by Toffoli and CNOT gates, whereas those in the second group use many controlled-rotation gates. Toffoli gates can be implemented exactly using the fault-tolerant gate sets mentioned above \cite{2010_NC, 2013_J, 2024_Mcs}, but rotation gates cannot \cite{2015_KMM, 2016_RS, 2022_GMM2, 2024_HL, 2024_Mtof}. In most error-correction schemes, rotation gates must therefore be decomposed further into gates from a universal gate set. The lengths and non-Clifford gate counts of these decompositions increase as the synthesis precision increases, and the decompositions may also require additional ancillas \cite{2015_BRS, 2015_BRS2}. This makes fault-tolerant implementations of rotation gates very costly in most schemes. For NISQ-type implementations, however, QFT-based arithmetic circuits can be more efficient because they have lower gate counts and use fewer ancillas. Their practicality is nevertheless limited to smaller problem sizes by the noise sensitivity of NISQ devices.

 The circuits developed in this paper belong to the first group and avoid expensive rotation gates, making them more amenable to fault-tolerant implementation, the focus of this paper. To the best of our knowledge, no QFT-based divider circuit has previously been constructed.

\subsection{Our results}

 We consider three division algorithms that have been studied extensively in prior work: \emph{long division}, \emph{restoring division} (as given in \cite{2019_TMVH}), and \emph{non-restoring division}. These algorithms divide an integer \emph{dividend} by an integer \emph{divisor} and return an integer \emph{quotient} and an integer \emph{remainder}. Long division is a specific type of restoring algorithm.  Suppose there is a superposition of divisors. If all divisors have the same length (the distance from the most significant 1 to the least significant bit), then long division can be used. Otherwise, the other two division algorithms are useful.

 Integer division can also be used as a subroutine in some floating-point division algorithms \cite{2021_GKDetal, 2022_GKD}. In \cite{2022_YGWetal}, quantum circuits for integer division are used to divide a fraction by an integer. Hence, quantum circuits developed for the operations above are useful for both integer and floating-point arithmetic.

We achieve asymptotic reductions in gate complexity for our long-division (LONG-DIV), restoring-division (RES-DIV), and non-restoring-division (NON-RES-DIV) algorithms compared with the previous algorithms in \cite{2019_TMVH, 2022_YGWetal, 2024_OPF}. Two major factors play a crucial role in realizing these improvements. The first is a careful modification of existing algorithms, and the second is the introduction of a new arithmetic primitive that we call COMP-N-SUB. We briefly summarize our first, algorithmic contribution below.
\begin{enumerate}
    \item \textbf{In-place computation of the quotient:} Unlike previous algorithms, we do not allocate extra space for the quotient. Rather, the qubits that initially store the dividend are used to store the quotient and remainder at the end of the computation. This is possible because, as the number of iterations increases, the more significant bits of the dividend become zero during the residue calculation. This reduces qubit usage.

    \item \textbf{Operations involving fewer qubits:} In RES-DIV and NON-RES-DIV, we do not prepend zeros to the divisor to make its length equal to that of the dividend. Consequently, each intermediate arithmetic operation, such as addition or subtraction, involves fewer qubits. This also leads to an asymptotic improvement in gate complexity.

    \item \textbf{Reducing the number of iterations:} In RES-DIV and NON-RES-DIV, we introduce flexibility in the extent of overlap between the dividend and divisor in the first iteration. This reduces the number of iterations in many scenarios.
\end{enumerate}

 Central to our divider-circuit designs is the development of new quantum circuits for a unitary that performs the following operation. This is our second major contribution.
\begin{eqnarray}
    \mathcal{A}: \quad \text{Given two integers $a$ and $b$, compute $b-a$ if and only if $b\geq a$.}
    \label{eqn:task}
\end{eqnarray}
 In bra--ket notation, the associated unitary acts as follows.
\begin{eqnarray}
    \ket{a}\ket{b}\ket{0}\mapsto\begin{cases}
    \ket{a}\ket{b-a}\ket{0}, & \text{if  } b\geq a \\
    \ket{a}\ket{b}\ket{1}, &\text{otherwise,}
    \end{cases} \nonumber
\end{eqnarray}
 where the last qubit is a ``flag'' indicating whether $b\geq a$. We refer to this unitary as COMP-N-SUB (compare and subtract). Similar logical operations appear in many quantum algorithms, so more efficient quantum circuits for COMP-N-SUB directly yield optimizations for those algorithms as well.
Thus, the COMP-N-SUB unitary can serve as a fundamental arithmetic primitive for implementing many algorithms. In Section \ref{sec:conclude}, we illustrate how COMP-N-SUB can be used to implement a quantum circuit for Stein's GCD algorithm and discuss the resulting advantage.

 We show that COMP-N-SUB can be implemented at roughly the same cost (in terms of T-count, Toffoli-count, CNOT-count, or depth) as one addition of two integers. Previous works have achieved this functionality by first performing a comparison and then a conditional subtraction. The cost is dominated by the conditional subtraction, which is effectively a controlled addition. This operation has much greater overhead than a regular addition because it turns every gate in an addition circuit into a controlled version of itself. Compiling the resulting circuit into a standard gate set then produces a large increase in the number of T or Toffoli gates. Thus, performing comparison and conditional subtraction at a cost comparable to that of a regular addition yields division circuits with fewer non-Clifford gates, fewer qubits, and, in some instances, smaller circuit depth than previous approaches.\footnote{The depth reduction can, in principle, be achieved without COMP-N-SUB; however, this comes at the expense of a significantly larger number of non-Clifford gates. We explain this in more detail at the end of Section~\ref{subsec:cns-II}.}

 Keeping in mind the various trade-offs for an efficient implementation, we have designed multiple circuits for COMP-N-SUB, and these can be broadly classified into three types.
\begin{itemize}
    \item \textbf{Type-I}  circuits are fully reversible and optimize the Toffoli or T count, CNOT count, and qubit count.
    
    \item \textbf{Type-II}  circuits are also fully reversible and optimize the Toffoli or T depth and overall circuit depth at the cost of increased qubit, CNOT, and Toffoli or T counts. Type-II circuits are further divided into two subtypes: Type-IIb, with the smallest known Toffoli or T depth, and Type-IIa, whose metrics are intermediate between those of Type-I and Type-IIb. Specifically, Type-IIa circuits have lower T, CNOT, and qubit counts than Type-IIb circuits but greater T-depth than Type-IIb circuits.
    
    \item \textbf{Type-III}  circuits are not fully reversible and optimize the T and CNOT counts using the logical-AND gadget from \cite{2013_J, 2018_G}. These circuits also have a higher qubit count than Types I and II.
\end{itemize}

Our LONG-DIV and RES-DIV circuits are constructed with mostly COMP-N-SUB unitaries. Hence, depending upon the type of COMP-N-SUB circuit used, we have named our LONG-DIV and RES-DIV circuits as type I, IIa, IIb and III, respectively. Needless to say, that they inherit the trade-offs of the corresponding COMP-N-SUB circuits. 

On the contrary, our NON-RES-DIV circuits are constructed with mostly adders and one COMP-N-SUB unitary. So we name them according to the type of adder used. Again, to reflect the various trade-offs we have considered the following three adders.
\begin{itemize}
    \item The quantum ripple-carry (QRC) adder in \cite{2010_TTK}, that optimize the Toffoli or T-count, CNOT-count and qubit-count. The NON-RES-DIV circuit using this adder is referred to as type-1.

    \item The quantum carry-lookahead adder (QCLA) adder in \cite{2006_DKRS}, that optimizes Toffoli or T-depth, at the cost of an increased Toffoli or T-count, CNOT-count and qubit-count. The NON-RES-DIV circuit using this  adder is referred to as type-2.

    \item The QRC adder in \cite{2018_G}, that used temoporary logical AND gadget, hence reducing T-count, but comes at the cost of an increase in the number of qubits, and extra measurement operations. The NON-RES-DIV circuit using this adder is referred to as type-3.
\end{itemize}

\begin{table}[h]
    \scriptsize
    \centering
    \begin{tabular}{|p{3.5cm}|p{3cm}|p{2.5cm}|p{3.5cm}|p{1.4cm}| }
    \hline
      {\bf Algorithm} & {\bf Circuit} & {\bf T-count} & {\bf T-depth} & {\bf Fully reversible}    \\
      \hline\hline
       \multirow{6}{*}{\bf Long division} & Yuan et al. \cite{2022_YGWetal} & $126m(n-m+1)$ & $54m(n-m+1)$ & No \\
       \cline{2-5}
        & Orts et al. \cite{2024_OPF} & $46m(n-m+1)$ & $20m(n-m+1)$ & No \\
        \cline{2-5}
        &  $LD_I$ ({\bf This paper}) & $18m(n-m+1)$ & $9m(n-m+1)$ & Yes \\
        \cline{2-5}
        & $LD_{IIa}$  ({\bf This paper}) & $76m(n-m+1)$ & $3(m+4\log_2 m)(n-m+1)$ & Yes   \\
        \cline{2-5}
        & $LD_{IIb}$  ({\bf This paper}) & $77m(n-m+1)$ & $12(n-m+1)\log_2m  $ & Yes   \\
        \cline{2-5}
        &  $LD_{III}$  ({\bf This paper}) & $11m(n-m+1)$ & $5m(n-m+1)$ & No   \\
       \hline\hline
       \multirow{5}{*}{\bf Restoring division} & Thapliyal et al. \cite{2019_TMVH} & $35n^2+42n$ & $15n^2+18n$ & Yes  \\
       \cline{2-5}
        & $Res_I$ ({\bf This paper}) & $18m(n-m'+1)$  & $9m(n-m'+1)$  & Yes  \\
        \cline{2-5}
        & $Res_{IIa}$ ({\bf This paper}) & $76m(n-m'+1)$  &  $3(m+4\log_2m)(n-m'+1)$  & Yes    \\
        \cline{2-5}
        & $Res_{IIb}$  ({\bf This paper}) &  $77m(n-m'+1)$  & $12(n-m'+1)\log_2m$  & Yes \\
        \cline{2-5}
        & $Res_{III}$  ({\bf This paper}) & $11m(n-m'+1)$ & $5m(n-m'+1)$ & No \\
       \hline\hline
       \multirow{3}{*}{\bf Non-restoring division} & Thapliyal et al. \cite{2019_TMVH}  & $14n^2+49n $ & $6n^2+21n$ & Yes   \\
       \cline{2-5}
       & $NRes_1$  ({\bf This paper}) & $14m(n-m'+1)$ & $6m(n-m'+1)$ & Yes  \\
       \cline{2-5}
       & $NRes_2$  ({\bf This paper}) & $70m(n-m'+1)$ & $12(n-m'+1)\log_2m$ & Yes  \\
       \cline{2-5}
       & $NRes_3$  ({\bf This paper}) & $4m(n-m'+1)$  & $2m(n-m'+1)$ & No  \\
       \hline\hline
    \end{tabular}
    \caption{ Comparison of the T-count, T-depth, and full reversibility of different quantum division circuits. $LD_x$, $Res_x$, and $NRes_x$ denote LONG-DIV circuit $x$, RES-DIV circuit $x$, and NON-RES-DIV circuit $x$, respectively. Only leading-order terms are shown. Here, $n$ and $m$ are the numbers of bits in the binary representations of the dividend and divisor, respectively.  We assume $n\geq m$, and $m'$ is an integer such that $1\leq m'\leq m$. }
    \label{tab:compare_T}
\end{table}

\begin{table}[h!]
    \scriptsize
    \centering
    \begin{tabular}{|p{3.5cm}|p{3cm}|p{2cm}|p{2.5cm}|p{2.5cm}| }
    \hline
      {\bf Algorithm} & {\bf Circuit} & {\bf Qubit-count} & \multicolumn{2}{|c|}{\bf CNOT-count}    \\
      \cline{4-5}
      & & & {\bf Clifford+Toffoli} & {\bf Clifford+T}   \\
      \hline\hline
       \multirow{6}{*}{\bf Long division} & Yuan et al. \cite{2022_YGWetal} & $4n-m$ & - & -  \\
       \cline{2-5}
        & Orts et al. \cite{2024_OPF} & $3n$ & - & $60m(n-m+1)$  \\
        \cline{2-5}
        & $LD_I$  ({\bf This paper}) & $n+m$  & $4m(n-m+1)$ & $25m(n-m+1)$ \\
        \cline{2-5}
        & $LD_{IIa}$  ({\bf This paper}) & $n+3m$ & $6m(n-m+1)$ & $83m(n-m+1)$  \\
        \cline{2-5}
        & $LD_{IIb}$  ({\bf This paper}) & $n+4m$ & $8m(n-m+1)$ & $85m(n-m+1)$  \\
        \cline{2-5}
        & $LD_{III}$  ({\bf This paper}) & $n+2m$ & - & $19m(n-m+1)$  \\
       \hline\hline
       \multirow{5}{*}{\bf Restoring division} & Thapliyal et al. \cite{2019_TMVH} & $3n$ & $9n^2+10n$ & $44n^2+87n$ \\
       \cline{2-5}
        &  $Res_I$  ({\bf This paper}) &  $n+2m-m'$ & $4m(n-m'+1)$ & $25m(n-m'+1)$ \\
        \cline{2-5}
        &  $Res_{IIa}$  ({\bf This paper}) & $n+4m-m'$ & $6m(n-m'+1)$ & $83m(n-m'+1)$   \\
        \cline{2-5}
        &  $Res_{IIb}$ ({\bf This paper}) & $n+5m-m'$ & $8m(n-m'+1)$ & $85m(n-m'+1)$  \\
        \cline{2-5}
        & $Res_{III}$  ({\bf This paper}) & $n+3m-m'$ & - & $19m(n-m'+1)$   \\
       \hline\hline
       \multirow{3}{*}{\bf Non-restoring division} & Thapliyal et al. \cite{2019_TMVH}  & $3n$ & $5n^2+11n$ & $19n^2+53n$    \\
       \cline{2-5}
       &  $NRes_1$ ({\bf This paper}) &  $n+2m-m'$ & $5m(n-m'+1)$ & $19m(n-m'+1)$ \\
       \cline{2-5}
       &  $NRes_2$  ({\bf This paper}) & $n+5m-m'$  & $4m(n-m'+1)$ & $74m(n-m'+1)$ \\
       \cline{2-5}
       & $NRes_3$  ({\bf This paper}) &  $n+3m-m'$ & - & $16m(n-m'+1)$  \\
       \hline\hline
    \end{tabular}
    \caption{ Comparison of the total qubit and CNOT counts of different quantum division circuits.  $LD_x$, $Res_x$, and $NRes_x$ denote LONG-DIV circuit $x$, RES-DIV circuit $x$, and NON-RES-DIV circuit $x$, respectively. Only leading-order terms are shown. Here, $n$ and $m$ are the numbers of bits in the binary representations of the dividend and divisor, respectively.  We assume $n\geq m$, and $m'$ is an integer such that $1\leq m'\leq m$.  }
    \label{tab:compare_q}
\end{table}

    Tables \ref{tab:compare_T} and \ref{tab:compare_q} summarize and compare important cost metrics for the various circuits developed in this paper.
    Compared with previous constructions in \cite{2019_TMVH}, our restoring (RES-DIV) and non-restoring (NON-RES-DIV) division algorithms achieve asymptotic reductions in T-count, CNOT-count, and T-depth. Suppose $n$ and $m$ are the bit lengths of the dividend and divisor, respectively. The T and CNOT counts of previous circuits scale as $O(n^2)$, whereas those of our circuits scale as $O(nm)$. For $m\in O(n^{1-\delta})$, where $\delta\rightarrow 0$, this yields an asymptotic improvement. The T-depth scaling is reduced from $O(n^2)$ to $O(n\log m)$ in RES-DIV circuit-IIb and NON-RES-DIV circuit-2.

    Our long-division (LONG-DIV) quantum circuits achieve reductions of up to $76.08\%$ and $68.35\%$\footnote{Here, reduction means $\frac{\text{cost of old circuit}-\text{cost of new circuit}}{\text{cost of old circuit}}$.} in T-count and CNOT-count, respectively. We also achieve an asymptotic reduction in T-depth for LONG-DIV circuit-IIb, from $O(nm)$ to $O(n\log m)$.
   
    The qubit count of our LONG-DIV circuit-I is the smallest among all existing divider circuits. Because $n+m$ qubits are required to store the input dividend and divisor, our qubit count of $n+m+3$ may be optimal. In all previous algorithms, the qubit count scales as $kn$ for some constant $k\geq 2$. For each type of algorithm, we reduce the qubit count relative to the previous best estimate.
   
    In Table \ref{tab:ft} (Section \ref{sec:compare}), we compare fault-tolerant resource estimates for implementing the various division algorithms with the surface code, following the procedure in \cite{2012_FMMC}. For $n=512$ and $m=256$, LONG-DIV circuit-I requires the fewest physical qubits, about $6.36\times 10^4$, closely followed by NON-RES-DIV circuit-1 with $m'=m$ ($\approx 6.57\times 10^4$). This represents approximately a twofold reduction relative to previous circuits. NON-RES-DIV circuit-2 has the shortest execution time, about $0.371\,\mathrm{s}$, closely followed by LONG-DIV circuit-IIb ($\approx 0.379\,\mathrm{s}$). These times are significantly shorter than those of previous circuits, which range from about $16\,\mathrm{s}$ to $47\,\mathrm{s}$.

\subsection{Related work}

Several works have developed quantum divider circuits. The circuits in \cite{2009_NHHetal, 2011_DH, 2016_BH, 2016_BM, 2018_GTO} require reversible registers or latches with feedback and are therefore challenging to implement on quantum hardware. Other reversible circuit designs have been proposed in \cite{2011_KAF, 2013_JB, 2016_DBJ}. Fault-tolerant implementations of long division have been given in \cite{2022_YGWetal, 2024_OPF}, while \cite{2019_TMVH} develops fault-tolerant circuits for restoring and non-restoring division. Other fault-tolerant quantum divider circuits have been proposed in \cite{2021_GKDetal, 2022_GKD, 2022_LFXL, 2024_FL}. As mentioned above, we compare our results with the best prior implementations.



\section{Results}

\subsection{Background}
\label{sec:prelim}

\paragraph{Quantum gates and gadgets:}   The $n$-qubit Clifford group, denoted $\cliff_n$, is generated by the one-qubit Hadamard ($\had$) and phase ($\phase$) gates and the two-qubit CNOT gate. The $n$-qubit Clifford+T group is generated by $\cliff_n$ and the one-qubit $\T$ gate. The $n$-qubit Clifford+Toffoli group is generated by $\cliff_n$ and the three-qubit Toffoli gate. These groups contain the unitaries that are \emph{exactly} implementable by the Clifford+T and Clifford+Toffoli gate sets, respectively. Equation \ref{eqn:gates} gives the unitary matrices of the different Clifford+T gates.
\begin{eqnarray}
    \X = \begin{bmatrix}
        0 & 1  \\
        1 & 0
    \end{bmatrix};\qquad
    \had = \frac{1}{\sqrt{2}}\begin{bmatrix}
        1 & 1 \\
        1 & -1
    \end{bmatrix};\qquad
    \T = \begin{bmatrix}
        1 & 0 \\
        0 & e^{i\frac{\pi}{4}}
    \end{bmatrix};\qquad
    \CNOT = \begin{bmatrix}
        1 & 0 & 0 & 0 \\
        0 & 1 & 0 & 0 \\
        0 & 0 & 0 & 1 \\
        0 & 0 & 1 & 0
    \end{bmatrix}
    \label{eqn:gates}
\end{eqnarray}

 The Toffoli unitary can be decomposed into a Clifford+T circuit using either seven T gates (Figure~\ref{fig:tof}) \cite{2010_NC} or four T gates \cite{2013_J}. The reduction in T-count comes at the cost of additional ancillas and measurements. This construction may not be suitable for some applications because the use of measurements makes the resulting circuit non-reversible.

\begin{figure}[h!]
    \centering
    \begin{subfigure}[b]{0.2\textwidth}
        \hspace*{\fill}
        \Qcircuit @C=1.4em @R=1.4em{
        \lstick{q_1} &\ctrl{1} &\qw   \\
        \lstick{q_2} &\ctrl{1} &\qw   \\
        \lstick{q_3} &\targ &\qw
        }
        \hspace*{\fill}        
        \caption{}
        \label{fig:toffoli_symbol}
    \end{subfigure}
    \hspace{0.15in}
    \begin{subfigure}[b]{0.7\textwidth}
        \centering
        \Qcircuit @C=1.4em @R=1.2em{
        \lstick{q_1} &\qw &\gate{T} &\targ &\qw &\ctrl{2} &\qw &\ctrl{1} &\gate{T^{\dagger}} &\qw &\ctrl{2} &\targ &\qw  \\
        \lstick{q_2} &\qw &\gate{T} &\ctrl{-1} &\targ &\qw &\gate{T^{\dagger}} &\targ &\gate{T^{\dagger}} &\targ &\qw &\ctrl{-1} &\qw  \\
        \lstick{q_3} &\gate{H} &\gate{T} &\qw &\ctrl{-1} &\targ &\qw &\qw &\gate{T} &\ctrl{-1} &\targ &\gate{H} &\qw
        }
        \caption{}
        \label{fig:tof_ckt}
    \end{subfigure}
    \caption{(a) Circuit representation of a Toffoli gate. (b) A Clifford+T circuit for a Toffoli gate.}
    \label{fig:tof}
\end{figure}

 The ancilla-and-fix-up construction from \cite{2013_J} was recast as the \emph{temporary logical-AND gadget} in \cite{2018_G}. This construction uses four T gates to compute the logical AND of two qubits and place the result in an ancilla register. Figure \ref{fig:logical_and} illustrates this operation symbolically, while Figure \ref{fig:logical_and_ckt} shows the explicit circuit. There, $\ket{q_1}$ and $\ket{q_2}$ denote the inputs to the logical AND, while the third qubit is an ancilla initialized to $\ket{0}$ that stores the result $\ket{q_1q_2}$ at the end of the computation. That register can subsequently be ``erased'' through measurement at \emph{no additional T-gate cost}. This uncomputation is shown symbolically in Figure \ref{fig:logical_and_uncompute}, and its circuit is given in Figure \ref{fig:measure_fixup}.

\begin{figure}
    \centering
    
    \begin{subfigure}[c]{0.2\textwidth}
        \hspace*{\fill}
        \Qcircuit @C=1.4em @R=1.4em{
        \lstick{\ket{q_1}} &\ctrl{1} &\qw &\rstick{\ket{q_1}}  \\
        \lstick{\ket{q_2}} &\ctrl{1} &\qw &\rstick{\ket{q_2}}  \\
        \lstick{} & &\qw &\rstick{\ket{q_1q_2}} 
        }
        \hspace*{\fill}
        \caption{}
        \label{fig:logical_and}
    \end{subfigure}
    \hfill
    \begin{subfigure}[c]{0.75\textwidth}
        \hspace*{\fill}
        \Qcircuit @C=1.4em @R=1.4em{
        \lstick{\ket{q_1}} &\ctrl{2} &\qw &\targ &\gate{T^{\dagger}} &\targ &\qw &\qw &\qw &\rstick{\ket{q_1}}  \\
        \lstick{\ket{q_2}} &\qw &\ctrl{1} &\targ &\gate{T^{\dagger}} &\targ &\qw &\qw &\qw &\rstick{\ket{q_2}}  \\
        \lstick{\ket{T}} &\targ &\targ &\ctrl{-2} &\gate{T} &\ctrl{-2} &\gate{H} &\gate{S} &\qw &\rstick{\ket{q_1q_2}}
        }
        \hspace*{\fill}
        \caption{}
        \label{fig:logical_and_ckt}
    \end{subfigure}

    \vspace{1em}

    \begin{subfigure}[c]{0.2\textwidth}
        \hspace*{\fill}
        \Qcircuit @C=1.4em @R=1.4em{
        \lstick{\ket{q_1}} &\ctrl{1} &\qw &\rstick{\ket{q_1}}  \\
        \lstick{\ket{q_2}} &\ctrl{1} &\qw &\rstick{\ket{q_2}}  \\
        \lstick{\ket{q_1q_2}} &\qw & &\rstick{} 
        }
        \hspace*{\fill}
        \caption{}
        \label{fig:logical_and_uncompute}
    \end{subfigure}
    \hspace{0.8in}
    \begin{subfigure}[c]{0.35\textwidth}
        \hspace*{\fill}
        \Qcircuit @C=1.4em @R=1.4em{
        \lstick{\ket{q_1}} &\qw &\qw &\ctrl{1} &\qw &\rstick{\ket{q_1}}  \\
        \lstick{\ket{q_2}} &\qw &\qw &\gate{Z} &\qw &\rstick{\ket{q_2}}  \\
        \lstick{\ket{q_1q_2}} &\gate{H} &\meter &\control \cw \cwx
        }
        \hspace*{\fill}
        \caption{}
        \label{fig:measure_fixup}
    \end{subfigure}

    \caption{ (a) Circuit representation of a temporary logical-AND gate. (b) Clifford+T circuit implementing a temporary logical AND. (c) Circuit representation of logical-AND uncomputation. (d) Circuit implementing logical-AND uncomputation.}
    \label{fig:logicalAND}
\end{figure}

\paragraph{Notation:}  We use parenthesized subscripts to indicate the qubits on which a gate acts. For example, $\X_{(q)}$ denotes an X gate applied to qubit $q$. We write $\CNOT_{(q_1;q_2)}$ for a CNOT gate with control qubit $q_1$ and target qubit $q_2$. Similarly, $\tof_{(q_1,q_2;q_3)}$ denotes a Toffoli gate with control qubits $q_1$ and $q_2$ and target qubit $q_3$. This unitary flips the state of $q_3$ if and only if $\ket{q_1} = \ket{q_2} = \ket{1}$. Our designs also use Toffoli gates with zero-controls. A Toffoli gate with a zero-control on qubit $q_1$, a one-control on qubit $q_2$, and target qubit $q_3$ flips the state of $q_3$ if and only if $\ket{q_1} = \ket{0}$ and $\ket{q_2} = \ket{1}$. We denote this gate by $\tof_{(\conj{q_1},q_2;q_3)}$. It can be implemented by conjugating $\tof_{(q_1,q_2;q_3)}$ with $\X_{(q_1)}$, that is,
\begin{eqnarray}
    \tof_{(\conj{q_1},q_2;q_3)} = \X_{(q_1)}\cdot\tof_{(q_1,q_2;q_3)}\cdot\X_{(q_1)}.   
    \label{eqn:tof_0_ctrl}
\end{eqnarray}

\paragraph{Cost metrics:}  We list the metrics used to assess and compare the quantum division circuits in our work and previous work and briefly explain why we consider them.

\begin{itemize}
    \item \textbf{T-count:} This is the number of T gates in a quantum circuit. The cost of fault-tolerantly implementing a T gate is higher than that of implementing a Clifford gate in most quantum error-correction schemes. The T-count of a unitary or circuit is denoted by $\tcount(.)$.

    \item \textbf{Toffoli-count:} This is the number of Toffoli gates in a quantum circuit. A Toffoli gate can be implemented with a Clifford+T circuit (Figure \ref{fig:tof_ckt}) or by magic-state injection. In either case, a fault-tolerant Toffoli gate is costly to implement. Reducing the Toffoli count also reduces the T-count.

    \item \textbf{CNOT-count:} This is the number of CNOT gates in a quantum circuit. Implementing multiqubit entangling operations is typically error-prone, especially on NISQ or other non-fault-tolerant devices. Recent progress on the surface code \cite{2024_GSJ} suggests that, at least in some error regimes, the cost of a fault-tolerant implementation of the T gate may be comparable to that of the CNOT gate, which is a Clifford gate. It is therefore undesirable to reduce the T-count at the expense of a massive increase in the CNOT count. The CNOT counts of Clifford+Toffoli and Clifford+T circuits are denoted by $\n_{Tof}(.)$ and $\n_T(.)$, respectively.

    \item \textbf{Qubit-count:} This is the total number of qubits required to implement a given algorithm. It is an important metric because existing and near-term quantum computers have limited numbers of qubits. We classify qubits as follows.

    \begin{itemize}
        \item \textbf{Input qubits:} These qubits store the input values. We refer to a set of input qubits as an \emph{input register}. For example, in the mapping $U_f \ket{x}\ket{y} = \ket{x}\ket{f(x) \oplus y}$, where the unitary $U_f$ computes a classical function $f$, the first register is the input register.
        \item \textbf{Output qubits:} These are qubits that store information that can be considered as output or useful for further computation. A set of output qubits constitutes an \emph{output register}. In the above example with $U_f$, the second register is the output register.
        \item \textbf{Garbage qubits:} After an operation, these qubits are not restored to their initial states, i.e., their states at the beginning of the operation. They also do not store any useful information at the end of the operation.
        \item \textbf{Ancilla qubits:} These are auxiliary qubits that can help perform certain operations, typically using fewer gates than would be required without ancillas. At the end of the computation, these qubits are restored to their initial states.
    \end{itemize}

    \item \textbf{T-depth and Toffoli-depth:} The T-depth of a circuit is the number of circuit layers (i.e., sets of gates acting in parallel) in which the $\T/\T^{\dagger}$ gate is the only non-Clifford gate. It indicates the maximum number of T gates that the circuit must execute sequentially on any qubit and hence affects the running time of a quantum circuit. The \textbf{Toffoli-depth} is defined analogously. The T-depth of a circuit or unitary is denoted by $\tcount_d(.)$.
\end{itemize}

  In all our resource estimates, we use the Toffoli construction with a T-count of seven and a T-depth of three (Figure \ref{fig:tof}(b)). Although other constructions, such as the ancilla-and-fix-up construction in \cite{2013_J}, use fewer T gates at the cost of additional ancillas and measurements, we choose to perform the computation coherently and do not consider constructions with intermediate measurements.

\subsection{COMP-N-SUB circuits}
\label{sec:cns}

 In this section, we construct efficient quantum circuits for the COMP-N-SUB unitary that implement task $\mathcal{A}$ (Equation \ref{eqn:task}). We assume that the integers are represented using $k$ bits;\footnote{We use $k$ instead of $n$ to avoid confusion when discussing the division algorithms. In those algorithms, which use COMP-N-SUB, $m$ and $n$ denote the bit lengths of the divisor and dividend, respectively.} if either integer is shorter, we prepend it with zeros. The integers $a = (a_{k-1},\ldots,a_0)$ and $b = (b_{k-1},\ldots, b_0)$ are encoded as the quantum states $\ket{a} = \bigotimes_{i=0}^{k-1}\ket{a_i}$ and $\ket{b} = \bigotimes_{i=0}^{k-1}\ket{b_i}$, respectively. These states are stored in input registers $A$ and $B$, respectively. We label the qubits, or memory locations, in $A$ and $B$ by $A_0,\ldots,A_{k-1}$ and $B_0,\ldots,B_{k-1}$, respectively. The initial states of qubits $A_i$ and $B_i$ are $\ket{a_i}$ and $\ket{b_i}$, respectively.

COMP-N-SUB performs the following operation.
\begin{eqnarray}
    \ket{a}\ket{b}\ket{0}\mapsto\begin{cases}
    \ket{a}\ket{b-a}\ket{0}, & \text{if  } b\geq a \\
    \ket{a}\ket{b}\ket{1}, &\text{otherwise}
    \end{cases} \nonumber
\end{eqnarray}
In the difference $b-a$, $b$ and $a$ are called the \emph{minuend} and \emph{subtrahend}, respectively.

\paragraph{Binary subtraction using 1's complement:}  This is one of the most popular methods for subtracting two binary numbers because it expresses subtraction as addition. The \emph{1's complement} of a binary number is obtained by flipping each of its bits. Denote the 1's complement of a number $b$ by $\conj{b}$. To compute the difference $d = b-a$, we proceed as follows.
\begin{enumerate}
    \item Compute the 1's complement of the minuend; that is, compute $\conj{b}$.

    \item Add $a$ to $\conj{b}$; that is, compute $s = (s_k,\ldots,s_0) = \conj{b}+a = (\conj{b_{k-1}},\ldots,\conj{b_0})+(a_{k-1},\ldots,a_0)$.

    \item \begin{enumerate}
        \item If there is no carry-over, that is, $s_k = 0$, then the difference is nonnegative and is obtained by taking the 1's complement of $s$ while omitting the carry-over bit. That is, $d = \conj{s}$; equivalently, $(d_{k-1},\ldots,d_0) = (\conj{s_{k-1}},\ldots,\conj{s_0})$.

        \item If there is a carry-over, that is, $s_k = 1$, then the difference is negative. Its magnitude is obtained by adding 1 to $s$ while omitting the carry-over bit. That is, $|d| = (d_{k-1},\ldots,d_0) = (s_{k-1},\ldots,s_0)+(0,\ldots,1)$.
    \end{enumerate}
\end{enumerate}
 The carry-over bit, $s_k$, is referred to as the \emph{high bit}. Using the method above, we can perform binary subtraction using an adder. We can also compare two integers using an adder because the high bit indicates whether the difference is positive or negative. We modify existing quantum-adder circuits to combine comparison and conditional subtraction and thereby implement the COMP-N-SUB unitary. As noted above, this gives COMP-N-SUB the same cost, in terms of Clifford and non-Clifford gates, as one addition.

 We consider the following three adders for the reasons explained below.
\begin{enumerate}
    \item \emph{In-place QRC adder of \cite{2010_TTK}}: This is an optimized version of the QRC adder introduced in \cite{2004_CDKM}. It uses only one extra qubit to store the high bit, the most significant bit of the sum. Its gate count, qubit count, T-count, and depth are $O(k)$, where $k$ is the bit length of each integer being added. It is \emph{in-place} because the output sum is stored in one of the input registers at the end of the computation. The other input register is restored to the state encoding the other input integer. This design saves qubits relative to \emph{out-of-place} designs, in which the sum is stored in a separate register. The circuit implementation is fully reversible.

    \item \emph{In-place QCLA adder of \cite{2006_DKRS}}: The main appeal of this adder is that it reduces circuit depth to $O(\log k)$, although the gate, T, and qubit counts each increase by an additive factor of $O(k)$. This quantum circuit is also fully reversible.

    \item \emph{Modified QRC adder of \cite{2018_G}}: This variant of the QRC adder is implemented with logical-AND gadgets that reduce the T-count at the cost of $O(k)$ additional ancillas (compared with \cite{2010_TTK}) and classical measurements, which make the circuit non-reversible.
\end{enumerate}

\paragraph{COMP-N-SUB quantum circuit:}  In existing designs of both QRC- and QCLA-type adders, the carry bits and partial sums are computed first; the complete sum is then computed using these carry bits. Suppose that the sum of $\conj{b}$ and $a$ is a $(k+1)$-bit integer. That is,
\begin{eqnarray}
    s = (s_k,\ldots,s_0) = \conj{b}+a = (\conj{b_{k-1}},\ldots,\conj{b_0})+(a_{k-1},\ldots,a_0).    \nonumber
\end{eqnarray}
 Let $c_{i+1}$ be the carry generated after the $i$th bitwise addition when computing $s_i$. Then $c_0=0$, and for every $i>0$, we have
 \begin{eqnarray}
     s_i &=& a_i\oplus \conj{b_i}\oplus c_i  \nonumber \\
     c_{i+1} &=& a_i\conj{b_i}\oplus \conj{b_i}c_i\oplus c_ia_i
     \nonumber
     \label{eqn:si_ci}
 \end{eqnarray}
 Our COMP-N-SUB circuits first compute the high bit, $s_k$, which encodes the sign of the difference $b-a$ and can therefore be used to compare the two integers. The remainder of the circuit is conditioned on this high bit and either computes the difference or uncomputes the intermediate result. Specifically, $s_k=0$ implies that $b\geq a$, in which case the remaining part of the circuit computes the difference $b-a=\conj{\conj{b}+a}$. If $s_k=1$, then $b<a$, and the remaining part of the circuit uncomputes the intermediate result, restoring the input registers to their initial states $\ket{b}$ and $\ket{a}$. Thus, the second part of the circuit performs a subtraction conditioned on the high bit computed in the first part. In this way, we combine comparison and conditional subtraction.

  In the next three subsections, we describe the construction of our quantum circuits using the adders in \cite{2010_TTK} (Section \ref{subsec:cns-I}), \cite{2006_DKRS} (Section \ref{subsec:cns-II}), and \cite{2018_G} (Section \ref{subsec:cns-III}). Our circuits inherit the distinguishing features of these adders, resulting in tradeoffs among the different cost metrics.

\begin{figure}[h]
    \centering
    \scalebox{0.85}{
    \Qcircuit @C=1em @R=1em{
    \lstick{B_0} &\gate{X} &\qw &\qw &\qw &\qw &\qw &\ctrl{1} &\qw &\qw &\qw &\qw &\qw &\qw &\qw &\qw &\qw &\qw &\qw &\ctrl{1} &\targ &\qw &\qw &\qw &\qw &\gate{X} &\qw \\
    \lstick{A_0} &\qw &\qw &\qw &\qw &\qw &\qw &\ctrl{2} &\qw &\qw &\qw &\qw &\qw &\qw &\qw &\qw &\qw &\qw &\qw &\ctrl{2} & \ctrl{-1} &\qw &\qw &\qw &\qw &\qw &\qw  \\
    \lstick{B_1} &\gate{X} &\targ &\qw &\qw &\qw &\qw &\qw &\ctrl{1} &\qw &\qw &\qw &\qw &\qw &\qw &\qw &\qw &\ctrl{1} &\targ &\qw &\qw &\qw &\qw &\qw &\targ &\gate{X} &\qw  \\
    \lstick{A_1} &\qw &\ctrl{-1} &\qw &\qw &\qw &\ctrl{2} &\targ &\ctrl{2} &\qw &\qw &\qw &\qw &\qw &\qw &\qw &\qw &\ctrl{2} &\ctrl{-1} &\targ &\qw &\ctrl{2} &\qw &\qw &\ctrl{-1} &\qw &\qw  \\
    \lstick{B_2} &\gate{X} &\targ &\qw &\qw &\qw &\qw &\qw &\qw &\ctrl{1} &\qw &\qw &\qw &\qw &\qw &\ctrl{1} &\targ &\qw &\qw &\qw &\qw &\qw &\qw &\qw &\targ &\gate{X} &\qw  \\
    \lstick{A_2} &\qw &\ctrl{-1} &\qw &\qw &\ctrl{2} &\targ &\qw &\targ &\ctrl{2} &\qw &\qw &\qw &\qw &\qw &\ctrl{2} &\ctrl{-1} &\targ &\qw &\qw &\qw &\targ &\ctrl{2} &\qw &\ctrl{-1} &\qw &\qw \\
    \lstick{B_3} &\gate{X} &\targ &\qw &\qw &\qw &\qw &\qw &\qw &\qw &\ctrl{1} &\qw &\qw &\ctrl{1} &\targ &\qw &\qw &\qw &\qw &\qw &\qw &\qw &\qw &\qw &\targ &\gate{X} &\qw  \\
    \lstick{A_3} &\qw &\ctrl{-1} &\qw &\ctrl{2} &\targ &\qw &\qw &\qw &\targ &\ctrl{2} &\qw &\qw &\ctrl{2} &\ctrl{-1} &\targ &\qw &\qw &\qw &\qw &\qw &\qw &\targ &\ctrl{2} &\ctrl{-1} &\qw &\qw  \\
    \lstick{B_4} &\gate{X} &\targ &\qw &\qw &\qw &\qw &\qw &\qw &\qw &\qw &\ctrl{1} &\targ &\qw &\qw &\qw &\qw &\qw &\qw &\qw &\qw &\qw &\qw &\qw &\targ &\gate{X} &\qw  \\
    \lstick{A_4} &\qw &\ctrl{-1} &\ctrl{1} &\targ &\qw &\qw &\qw &\qw &\qw &\targ &\ctrl{1} &\ctrl{-1} &\targ &\qw &\qw &\qw &\qw &\qw &\qw &\qw &\qw &\qw &\targ &\ctrl{-1} &\qw &\qw  \\
    \lstick{Z} &\qw &\qw &\targ &\qw &\qw &\qw &\qw &\qw &\qw &\qw &\targ &\ctrlo{-1} &\qw &\ctrlo{-3} &\qw &\ctrlo{-5} &\qw &\ctrlo{-7} &\qw &\ctrlo{-9} &\qw &\qw &\qw &\qw &\qw &\qw \gategroup{1}{3}{11}{12}{1.4em}{--}
    }}
    \caption{COMP-N-SUB circuit-I for 5-bit integers. Registers $B$ and $A$ are initialized in states $\ket{b}$ and $\ket{a}$, respectively. The first layer of X gates computes $\ket{\conj{b}}$. The boxed part of the circuit computes the high bit in register $Z$. If $Z=\ket{0}$, the remaining part of the circuit computes $b-a$ in register $B$; otherwise, both registers are restored to their initial states.}
    \label{fig:compNsub}
\end{figure}

\subsubsection{COMP-N-SUB circuit I}
\label{subsec:cns-I}

In this section, we describe the construction of a COMP-N-SUB circuit using the QRC adder in \cite{2010_TTK}. We call this COMP-N-SUB circuit-I. Figure \ref{fig:compNsub} shows a circuit for 5-bit integers. In addition to the two input registers, we use a qubit $Z$ initialized to $\ket{0}$. As mentioned earlier, we first compute the high bit in register $Z$ using the circuitry in the dashed box of Figure \ref{fig:compNsub}. The high bit contains enough information to compare the input integers $a$ and $b$. The remaining part of the circuit acts as a conditional subtractor that computes the difference $d=b-a$ in register $B$ if the high bit is $\ket{0}$. Otherwise, input registers $A$ and $B$ are restored to their initial states $\ket{a}$ and $\ket{b}$, respectively. To construct the circuit, we perform the following steps.
\begin{enumerate}
    \item For $i= 0 ,\ldots, k-1$: Apply $\X_{(B_i)}$.

    \item For $i = 1,\ldots, k-1$: Apply $\CNOT_{(A_i;B_i)}$.

    \item \begin{enumerate}
        \item Apply $\CNOT_{(A_{k-1};Z)}$.
        \item For $i = k-2,\ldots,1$: Apply $\CNOT_{(A_i;A_{i+1})}$.
    \end{enumerate} 
    
    \item \begin{enumerate}
        \item For $i = 0,\ldots, k-2$: Apply $\tof_{(B_i,A_i;A_{i+1})}$.

        \item Apply $\tof_{(B_{k-1},A_{k-1};Z)}$.
    \end{enumerate}
    
    \item \begin{enumerate}
        \item  Apply $\tof_{(\conj{Z},A_{k-1};B_{k-1})}$.
        \item For $i = k-2,\ldots,0$: Apply $\tof_{(B_i,A_i;A_{i+1})}\cdot\tof_{(\conj{Z},A_i;B_i)}$.
    \end{enumerate}
    
    \item For $i=1,\ldots,k-2$: Apply $\CNOT_{(A_i;A_{i+1})}$.

    \item  For $i = 1,\ldots, k-1$: Apply $\CNOT_{(A_i;B_i)}$.

    \item  For $i= 0 ,\ldots, k-1$: Apply $\X_{(B_i)}$.
\end{enumerate}

\paragraph{Correctness:} To prove that the circuit described above has the intended functionality, we track the changes in the states of the input qubits after each step. The initial state of registers $B$ and $A$ and qubit $Z$ is
    \begin{eqnarray}
        \left(\bigotimes_{i=0}^{k-1}\ket{b_i}_{B_i}\ket{a_i}_{A_i}\right)\ket{0}_Z. \nonumber
    \end{eqnarray}
Appendix \ref{app:correct-I} shows in detail that the final state after applying the layer of X gates is
    \begin{eqnarray}
      \ket{\conj{c_k}d_0\oplus c_k b_0 }_{B_0}\ket{a_0}_{A_0}\left( \bigotimes_{i=1}^{k-1}\ket{ \conj{c_k}d_i\oplus c_k b_i  }_{B_i} \ket{a_i}_{A_i}  \right)\ket{c_k}_Z \nonumber
        \label{cns_qrc_final_main}
    \end{eqnarray}
If $c_k = 0$, implying $b\geq a$, we obtain $d=b-a$ in register $B$; otherwise, we obtain $b$ in register $B$. Hence, the circuit has the desired COMP-N-SUB functionality.

\paragraph{Resource estimates:} We now provide resource estimates for COMP-N-SUB circuit-I.

\begin{itemize}
    \item \textbf{Toffoli-count:} We require $(k-1)+1=k$ Toffolis in step 4 and $1+2(k-1)=2k-1$ Toffolis in step 5. Thus, the Toffoli count of our circuit is $3k-1$.

    \item \textbf{CNOT-count:} We require $k-1$ CNOTs in steps 2, 3, and 7 and $k-2$ CNOTs in step 6. Thus, the CNOT count of the Clifford+Toffoli circuit is $3(k-1)+k-2 = 4k-5$. We require at most seven CNOT gates for the Clifford+T implementation of each Toffoli gate. Thus, the CNOT count of the Clifford+T implementation is at most $(4k-5)+7(3k-1) = 25k-12$.

    \item \textbf{T-count:} We first estimate the T-count of the ladder of Toffoli pairs in step 5b. Consider the following four-qubit unitary, whose superscript specifies the qubits on which it acts.
    \begin{eqnarray}
        U_1^{q_1,q_2,q_3,q_4} = \tof_{(q_1,q_2;q_3)}\cdot\tof_{(q_2,q_3;q_4)}   \nonumber
    \end{eqnarray}
    The T-count of this unitary is 11, as first shown in \cite{2021_MM} and later in \cite{2025_KJM}. Using Equation~\ref{eqn:tof_0_ctrl} in step 5b, we see that the ladder of Toffolis is equivalent to
    \begin{eqnarray}
      \X_{(Z)}\cdot \left( \prod_{i=k-2}^0U_1^{Z,A_i,B_i,A_{i+1}} \right) \cdot\X_{(Z)}. \nonumber
    \end{eqnarray}
We have $k+1$ additional Toffoli gates. Thus, the T-count is at most $11(k-1)+7(k+1) = 18k-4$.

    \item \textbf{Qubit-count:} We require $2k$ input qubits and one additional qubit that stores the high bit.

    \item \textbf{Toffoli-depth:} The Toffoli-depth of this circuit is at most $3k-1$.

    \item \textbf{T-depth:} Each Toffoli has T-depth 3; hence, the T-depth of the complete circuit is at most $3(3k-1)=9k-3$.
\end{itemize}

\subsubsection{COMP-N-SUB circuit II}
\label{subsec:cns-II}

\begin{figure}
    \hspace{1.3in}
    \begin{subfigure}[b]{0.5\textwidth}
        \hspace*{\fill}
        \scalebox{0.75}{
        \Qcircuit @C=1em @R=1em{
        \lstick{A_0} &\qw &\ctrl{1} &\ctrl{1} &\multigate{16}{U_1} &\qw &\qw &\qw &\qw &\qw &\ctrl{1} &\qw &\qw &\multigate{16}{U_2} &\ctrl{1} &\ctrl{1} &\qw &\qw  \\
        \lstick{B_0} &\gate{X} &\ctrl{1} &\targ &\ghost{U_1} &\qw &\qw &\qw &\qw &\qw &\targ &\qw &\targ &\ghost{U_2} &\targ &\ctrl{1} &\targ &\qw  \\
        \lstick{Z_0} &\qw &\targ &\qw &\ghost{U_1} &\ctrl{1} &\qw &\qw &\qw &\qw &\qw &\ctrl{1} &\qw &\ghost{U_2} &\qw &\targ &\qw &\qw   \\
        \lstick{A_1} &\qw &\ctrl{1} &\ctrl{1} &\ghost{U_1} &\targ &\qw &\qw &\qw &\ctrl{1} &\qw &\targ &\qw &\ghost{U_2} &\ctrl{1} &\ctrl{1} &\qw &\qw   \\
        \lstick{B_1} &\gate{X} &\ctrl{1} &\targ &\ghost{U_1} &\qw &\qw &\qw &\qw &\targ &\qw &\qw &\targ &\ghost{U_2} &\targ &\ctrl{1} &\targ &\qw  \\
        \lstick{Z_1} &\qw &\targ &\qw &\ghost{U_1} &\ctrl{1} &\qw &\qw &\qw &\qw &\qw &\ctrl{1} &\qw &\ghost{U_2} &\qw &\targ &\qw &\qw   \\
        \lstick{A_2} &\qw &\ctrl{1} &\ctrl{1} &\ghost{U_1} &\targ &\qw &\qw &\ctrl{1} &\qw &\qw &\targ &\qw &\ghost{U_2} &\ctrl{1} &\ctrl{1} &\qw &\qw   \\
        \lstick{B_2} &\gate{X} &\ctrl{2} &\targ &\ghost{U_1} &\qw &\qw &\qw &\targ &\qw &\qw &\qw &\targ &\ghost{U_2} &\targ &\ctrl{2} &\targ &\qw  \\
        \lstick{W_2} &\qw &\qw &\qw &\ghost{U_1} &\qw &\qw &\qw &\qw &\qw &\qw &\qw &\qw &\ghost{U_2} &\qw &\qw &\qw &\qw   \\
        \lstick{Z_2} &\qw &\targ &\qw &\ghost{U_1} &\ctrl{1} &\qw &\qw &\qw &\qw &\qw &\ctrl{1} &\qw &\ghost{U_2} &\qw &\targ &\qw &\qw   \\
        \lstick{A_3} &\qw &\ctrl{1} &\ctrl{1} &\ghost{U_1} &\targ &\qw &\ctrl{1} &\qw &\qw &\qw &\targ &\qw &\ghost{U_2} &\ctrl{1} &\ctrl{1} &\qw &\qw   \\
        \lstick{B_3} &\gate{X} &\ctrl{1} &\targ &\ghost{U_1} &\qw &\qw &\targ &\qw &\qw &\qw &\qw &\targ &\ghost{U_2} &\targ &\ctrl{1} &\targ &\qw  \\
        \lstick{Z_3} &\qw &\targ &\qw &\ghost{U_1} &\ctrl{1} &\qw &\qw &\qw &\qw &\qw &\ctrl{1} &\qw &\ghost{U_2} &\qw &\targ &\qw &\qw   \\
        \lstick{A_4} &\qw &\ctrl{1} &\ctrl{1} &\ghost{U_1} &\targ &\ctrl{1} &\qw &\qw &\qw &\qw &\targ &\qw &\ghost{U_2} &\ctrl{1} &\qw &\qw &\qw   \\
        \lstick{B_4} &\gate{X} &\ctrl{2} &\targ &\ghost{U_1} &\qw &\targ &\qw &\qw &\qw &\qw &\qw &\targ &\ghost{U_2} &\targ &\qw &\targ &\qw  \\
        \lstick{W_4} &\qw &\qw &\qw &\ghost{U_1} &\qw &\qw &\qw &\qw &\qw &\qw &\qw &\qw &\ghost{U_2} &\qw &\qw &\qw &\qw   \\
        \lstick{Z_4} &\qw &\targ &\qw &\ghost{U_1} &\qw &\ctrlo{-2} &\ctrlo{-5} &\ctrlo{-9} &\ctrlo{-12} &\ctrlo{-15} &\qw &\ctrlo{-15} &\ghost{U_2} &\qw &\qw &\ctrl{-15} &\qw \gategroup{1}{7}{17}{11}{.7em}{--}
        }
        }
        \hspace*{\fill}
        \caption{}
        \label{fig:compnsub_qcla}
    \end{subfigure}

    \vspace{1em}

    \begin{subfigure}[b]{0.48\textwidth}
        \hspace*{\fill}
        \scalebox{0.9}{
        \Qcircuit @C=1em @R=1em{
        \lstick{A_0 (\ket{a_0})} &\qw &\qw &\qw &\qw &\qw &\qw &\rstick{\ket{a_0}}   \\
        \lstick{B_0 (\ket{a_0\oplus b_0})} &\qw &\qw &\qw &\qw &\qw &\qw &\rstick{\ket{a_0\oplus b_0}}   \\
        \lstick{Z_0 (\ket{a_0b_0})} &\qw &\ctrl{2} &\qw &\qw &\qw &\qw &\rstick{\ket{c_1}}   \\
        \lstick{A_1 (\ket{a_1})} &\qw &\qw &\qw &\qw &\qw &\qw &\rstick{\ket{a_1}}   \\
        \lstick{B_1 (\ket{a_1\oplus b_1})} &\qw &\ctrl{1} &\qw &\qw &\qw &\qw &\rstick{\ket{a_1\oplus b_1}}   \\
        \lstick{Z_1 (\ket{a_1b_1})} &\qw &\targ &\ctrl{3} &\ctrl{2} &\qw &\qw &\rstick{\ket{c_2}}   \\
        \lstick{A_2 (\ket{a_2})} &\qw &\qw &\qw &\qw &\qw &\qw &\rstick{\ket{a_2}}   \\
        \lstick{B_2 (\ket{a_2\oplus b_2})} &\ctrl{1} &\qw &\qw &\ctrl{2} &\ctrl{1} &\qw &\rstick{\ket{a_2\oplus b_2}}   \\
        \lstick{W_2 (\ket{0})} &\targ &\qw &\ctrl{4} &\qw &\targ &\qw &\rstick{\ket{0}}   \\
        \lstick{Z_2 (\ket{a_2b_2})} &\qw &\ctrl{2} &\qw &\targ &\qw &\qw &\rstick{\ket{c_3}}   \\
        \lstick{A_3 (\ket{a_3})} &\qw &\qw &\qw &\qw &\qw &\qw &\rstick{\ket{a_3}}   \\
        \lstick{B_3 (\ket{a_3\oplus b_3})} &\ctrl{-3} &\ctrl{1} &\qw &\qw &\ctrl{-3} &\qw &\rstick{\ket{a_3\oplus b_3}}   \\
        \lstick{Z_3 (\ket{a_3b_3})} &\qw &\targ &\targ &\ctrl{2} &\qw &\qw &\rstick{\ket{c_4}}   \\
        \lstick{A_4 (\ket{a_4})} &\qw &\qw &\qw &\qw &\qw &\qw &\rstick{\ket{a_4}}   \\
        \lstick{B_4 (\ket{a_4\oplus b_4})} &\qw &\qw &\qw &\ctrl{2} &\qw &\qw &\rstick{\ket{a_4\oplus b_4}}   \\
        \lstick{W_4 (\ket{0})} &\qw &\qw &\qw &\qw &\qw &\qw &\rstick{\ket{0}}   \\
        \lstick{Z_4 (\ket{a_4b_4})} &\qw &\qw &\qw &\targ &\qw &\qw &\rstick{\ket{c_5}}
        }
        }
        \hspace*{\fill}
        \caption{}
        \label{fig:maj_cla}
    \end{subfigure}
    \hfill
    \begin{subfigure}[b]{0.48\textwidth}
        \hspace*{\fill}
        \scalebox{0.95}{
        \Qcircuit @C=1em @R=1em{
        \lstick{A_0 (\ket{a_0})} &\qw &\qw &\qw &\qw &\qw &\qw &\rstick{\ket{a_0}}   \\
        \lstick{B_0 (\ket{a_0\oplus \conj{s_0}})} &\qw &\qw &\qw &\qw &\qw &\qw &\rstick{\ket{a_0\oplus \conj{s_0}}}   \\
        \lstick{Z_0 (\ket{c_1})} &\qw &\qw &\qw &\ctrl{2} &\qw &\qw &\rstick{\ket{a_0\conj{s_0}}}   \\
        \lstick{A_1 (\ket{a_1})} &\qw &\qw &\qw &\qw &\qw &\qw &\rstick{\ket{a_1}}   \\
        \lstick{B_1 (\ket{a_1\oplus \conj{s_1}})} &\qw &\qw &\qw &\ctrl{1} &\qw &\qw &\rstick{\ket{a_1\oplus \conj{s_1}}}   \\
        \lstick{Z_1 (\ket{c_2})} &\qw &\ctrl{2} &\ctrl{3} &\targ &\qw &\qw &\rstick{\ket{a_1\conj{s_1}}}   \\
        \lstick{A_2 (\ket{a_2})} &\qw &\qw &\qw &\qw &\qw &\qw &\rstick{\ket{a_2}}   \\
        \lstick{B_2 (\ket{a_2\oplus \conj{s_2}})} &\ctrl{1} &\ctrl{2} &\qw &\qw &\ctrl{1} &\qw &\rstick{\ket{a_2\oplus \conj{s_2}}}   \\
        \lstick{W_2 (\ket{0})} &\targ &\qw &\ctrl{4} &\qw &\targ &\qw &\rstick{\ket{0}}   \\
        \lstick{Z_2 (\ket{c_3})} &\qw &\targ &\qw &\ctrl{2} &\qw &\qw &\rstick{\ket{a_2\conj{s_2}}}   \\
        \lstick{A_3 (\ket{a_3})} &\qw &\qw &\qw &\qw &\qw &\qw &\rstick{\ket{a_3}}   \\
        \lstick{B_3 (\ket{a_3\oplus \conj{s_3}})} &\ctrl{-3} &\qw &\qw &\ctrl{1} &\ctrl{-3} &\qw &\rstick{\ket{a_3\oplus \conj{s_3}}}   \\
        \lstick{Z_3 (\ket{c_4})} &\qw &\qw &\targ &\targ &\qw &\qw &\rstick{\ket{a_3\conj{s_3}}}   \\
        \lstick{A_4 (\ket{a_4})} &\qw &\qw &\qw &\qw &\qw &\qw &\rstick{\ket{a_4}}   \\
        \lstick{B_4 (\ket{a_4\oplus \conj{s_4}})} &\qw &\qw &\qw &\qw &\qw &\qw &\rstick{\ket{a_4\oplus \conj{s_4}}}   \\
        \lstick{W_4 (\ket{0})} &\qw &\qw &\qw &\qw &\qw &\qw &\rstick{\ket{0}}   \\
        \lstick{Z_4 (\ket{c_5})} &\qw &\qw &\qw &\qw &\qw &\qw &\rstick{\ket{c_5}}
        }
        }
        \hspace*{\fill}
        \caption{}
        \label{fig:umaj_cla}
    \end{subfigure}

    \caption{(a) COMP-N-SUB circuit-IIa for 5-bit integers. The first part of the circuit, up to and including unitary $U_1$, computes the high bit in register $Z_4$. The remaining circuit computes $b-a$ in register $B$ if $Z_4 = \ket{0}$; otherwise, registers $B$ and $A$ are restored to their initial states $\ket{b}$ and $\ket{a}$, respectively. The Toffoli gates in the boxed part can be parallelized to Toffoli depth 1 by adding extra ancilla qubits, yielding circuit-IIb. (b) Quantum circuit for unitary $U_1$. (c) Quantum circuit for unitary $U_2$.}
    \label{fig:comPsub_qcla}
\end{figure}

In this section, we describe the construction of a COMP-N-SUB circuit using the QCLA adder from \cite{2006_DKRS}. We refer to this construction as COMP-N-SUB circuit-II. A $k$-bit QCLA adder has depth $O(\log k)$, compared with the $O(k)$ depth of a QRC adder. Thus, with this second type of COMP-N-SUB circuit, we aim to reduce the circuit depth.

In addition to the two input registers, we have $k$ auxiliary qubits, indexed by $Z_i$, that are initialized to $\ket{0}$. During the computation, qubit $Z_i$ stores carry bit $c_{i+1}$. At the end of the computation, $Z_{k-1}$ stores the high bit, while the other $k-1$ qubits are restored to $\ket{0}$. We also require $k-\omega(k)-\lfloor\log_2k\rfloor$ additional ancilla qubits initialized to $\ket{0}$. Here, $\omega(k)$ denotes the number of ones in the binary expansion of $k$.
We use the following two unitaries for the reversible computation of the carry bits. The construction of these unitaries is explained in Section 3 of \cite{2006_DKRS}. Here, we briefly describe the main features needed to compute the resource estimates.
\begin{itemize}
    \item $\mathbf{U_1}$: This unitary computes the carries. Specifically, the input state consists of $\ket{a_i}$ in qubit $A_i$, $\ket{a_i\oplus \conj{b_i}}$ in qubit $B_i$, and $\ket{a_i\conj{b_i}}$ in qubit $Z_i$. After the unitary acts, the state of qubit $Z_i$ is $\ket{c_{i+1}}$. The states of the remaining qubits are unchanged. A quantum circuit for $U_1$ acting on 5-bit integers is shown in Figure \ref{fig:maj_cla}. The circuit has $4k-3\omega(k)-3\lfloor\log_2 k\rfloor-1$ Toffoli gates. The Toffoli depth, which is also the circuit depth, is $\lfloor\log_2k\rfloor+\lfloor\log_2\frac{k}{3}\rfloor+3$.

    \item $\mathbf{U_2}$: This unitary uncomputes the carries except for the high bit $c_k$ stored in qubit $Z_{k-1}$. The circuit is obtained by reversing the circuit of $U_1$, excluding the Toffoli gates that act on qubit $Z_{k-1}$. A quantum circuit for $U_2$ acting on 5-bit integers is shown in Figure \ref{fig:umaj_cla}. The input state consists of $\ket{a_i}$ in qubit $A_i$, $\ket{a_i\oplus \conj{s_i}}$ in qubit $B_i$, and $\ket{c_{i+1}}$ in qubit $Z_i$. After $U_2$ acts, the state of qubit $Z_i$ is $\ket{a_i\conj{s_i}}$, except for qubit $Z_{k-1}$, which remains in state $\ket{c_k}$. The states of the remaining qubits are unchanged. The resource estimates for $U_2$ are nearly the same as those for $U_1$.

\end{itemize}
Figure \ref{fig:comPsub_qcla} shows a COMP-N-SUB circuit-II for 5-bit integers. First, the high bit is computed in qubit $Z_{k-1}$. If $Z_{k-1}=\ket{0}$, the remaining part of the circuit computes the difference in register $B$; otherwise, the input registers are restored to their initial states. We describe two variants of this circuit: IIa and IIb. The latter is obtained by parallelizing the Toffoli gates that share the same control (the boxed part of the circuit in Figure \ref{fig:compnsub_qcla}) using extra CNOT gates and ancilla qubits (Figure \ref{fig:tof_par} in Appendix \ref{app:correct-II}). We now describe the construction of a quantum circuit for COMP-N-SUB circuit-IIa.
\begin{enumerate}
    \item For $i = 0,\ldots,k-1$: Apply $\X_{(B_i)}$.
    
    \item For $i = 0,\ldots,k-1$: Apply $\tof_{(A_i,B_i;Z_i)}$.

    \item For $i = 0,\ldots,k-1$: Apply $\CNOT_{(A_i;B_i)}$.

    \item Apply $U_1$.

    \item For $i = 1,\ldots, k-1$: Apply $\CNOT_{(Z_{i-1};A_i )}$.

    \item For $i = k-1,\ldots,0$: Apply $\tof_{(\conj{Z_{k-1}},A_i ;B_i )}$.

    \item For $i = 1,\ldots, k-1$: Apply $\CNOT_{(Z_{i-1};A_i )}$.

    \item For $i = 0,\ldots, k-1$: Apply $\CNOT_{( \conj{Z_{k-1}};B_i )}$.

    \item Apply $U_2$, excluding the Toffolis that uncompute the high bit.

    \item For $i = 0,\ldots,k-1$: Apply $\CNOT_{(A_i;B_i)}$.

    \item For $i = 0,\ldots,k-2$: Apply $\tof_{(A_i,B_i;Z_i)}$.

    \item For $i = 0,\ldots,k-1$: Apply $\CNOT_{(Z_{k-1};B_i)}$.
\end{enumerate}

\paragraph{Correctness:} To prove the correctness of our quantum circuit, we track the qubit states at each step described above. We observe that only the states of the input qubits and the ancilla qubits labeled $Z_i$ change before and after the application of unitaries $U_1$ and $U_2$. Hence, we write only the states of qubits $A_i$, $B_i$, and $Z_i$. The initial state of these qubits is
\begin{eqnarray}
    \bigotimes_{i=0}^{k-1}\ket{a_i}_{A_i}\ket{b_i}_{B_i}\ket{0}_{Z_i}.    \nonumber
\end{eqnarray}
Appendix \ref{app:correct-II} proves in detail that, after the last multi-target CNOT, the state is
    
    \begin{eqnarray}
     \left( \bigotimes_{i=0}^{k-2}\ket{a_i}_{A_i}\ket{ \conj{c_k} d_i \oplus c_k b_i }_{B_i} \ket{0}_{Z_i}\right)\ket{a_{k-1}}_{A_{k-1}}\ket{ \conj{c_k} d_{k-1} \oplus c_k b_{k-1} }_{B_{k-1}} \ket{c_k}_{Z_{k-1}}.    \nonumber
    \end{eqnarray}
Hence, if $c_k=0$, implying $b\geq a$, we obtain $d=b-a$ in register $B$; otherwise, we obtain $b$. Thus, the circuit performs as desired.

\paragraph{Variant:} We can parallelize the Toffoli gates in step 6 by using $k-1$ pairs of CNOTs (Figure \ref{fig:tof_par} in Appendix \ref{app:correct-II}). Using a multi-target CNOT with $k-1$ targets, we make $k-1$ copies of the high bit (the state of qubit $Z_{k-1}$) in $k-1$ additional ancilla qubits initialized to $\ket{0}$. We can also use these ancilla qubits to parallelize the CNOTs in steps 8 and 12. At the end, we uncompute the ancilla qubits. We refer to this variant as COMP-N-SUB circuit-IIb. In a surface-code implementation, a multi-target CNOT has the same cost and execution time as a single-target CNOT. Thus, a multi-target CNOT can be considered one logical CNOT \cite{2012_FMMC}. A multi-target CNOT is also called a fan-out operator \cite{2025_RV}; it can be implemented using only CNOT gates, without ancilla qubits, in depth $O(\log k)$ and size $O(k)$, where $k$ is the number of targets.

Hereafter, COMP-N-SUB circuit-II refers to either variant. We explicitly identify the variant only when necessary.

\paragraph{Resource estimates:} We now estimate the cost metrics of interest.

\begin{itemize}
    \item \textbf{Toffoli-count:} We require $k$ Toffoli gates in each of steps 2 and 6 and $k-1$ Toffoli gates in step 11. Additionally, $U_1$ requires $4k-3\omega(k)-3\lfloor\log_2k\rfloor-1$ Toffoli gates, and $U_2$ requires one fewer. Thus, the Toffoli count of our circuit is at most $11k - 6\omega(k)-6\lfloor\log_2k\rfloor-4$.

    \item \textbf{CNOT-count:} We require $k$ CNOT gates in steps 3, 8, 10, and 12 and $k-1$ CNOT gates in steps 5 and 7. Thus, the total number of CNOT gates in a Clifford+Toffoli implementation of circuit-IIa is at most $6k-2$. Further parallelization increases the CNOT count of COMP-N-SUB circuit-IIb to at most $8k-4$.

    Since we require at most seven CNOT gates to implement each Toffoli, the CNOT count of a Clifford+T implementation of circuit-IIa is at most $83k-42\omega(k)-42\lfloor\log_2k\rfloor-30$, while that of circuit-IIb is at most $85k-42\omega(k)-42\lfloor\log_2k\rfloor-32$.

    \item \textbf{T-count:} Each Toffoli in steps 2 and 11 requires seven T gates. Thus, we require at most $7(2k-1)$ T gates for steps 2 and 11. The number of T gates required to implement $U_1$ and $U_2$ is at most $7(8k-6\omega(k)-6\lfloor\log_2k\rfloor-2)$. In circuit-IIa, the $k$ Toffoli gates in step 6 share a control (taking into account Equation~\ref{eqn:tof_0_ctrl}). Appendix \ref{app:tofPair} shows that the T-count of a unitary comprising a pair of Toffoli gates that share a control is at most 12. Thus, the number of T gates required in step 6 is at most $6k$ if $k$ is even and $6(k-1)+7 = 6k+1$ if $k$ is odd. Hence, the T-count of circuit-IIa is at most $76k-42\omega(k)-42\lfloor\log_2k\rfloor-21$.
    
    In circuit-IIb, the Toffoli gates in step 6 do not share a control and therefore require at most $7k$ T gates. Hence, the total T-count of circuit-IIb is at most $77k-42\omega(k)-42\lfloor\log_2k\rfloor-21$.

    Several Toffoli gates in the implementations of $U_1$ and $U_2$ share a control or target. Hence, the T-count estimates above may be improved.

    \item \textbf{Qubit-count:} We require $3k+k-\omega(k)-\lfloor\log_2k\rfloor = 4k-\omega(k)-\lfloor\log_2k\rfloor$ qubits for circuit-IIa. Of these, $2k$ are input qubits, and one qubit stores the high bit after computation. The remaining $2k-\omega(k)-\lfloor\log_2k\rfloor-1$ ancilla qubits are restored to their initial $\ket{0}$ states after computation. To parallelize the Toffoli gates in step 6 of circuit-IIb, we require $k-1$ more ancilla qubits, bringing the total qubit count to $5k-\omega(k)-\lfloor\log_2k\rfloor-1$. The extra ancilla qubits are restored to their initial states.

    \item \textbf{Toffoli-depth:} The Toffoli depth in steps 2 and 11 is 1. The Toffoli depth of each of $U_1$ and $U_2$ is $\lfloor\log_2k\rfloor+\lfloor\log_2\frac{k}{3}\rfloor+3$. The Toffoli depths in step 6 for circuits-IIa and IIb are $k$ and 1, respectively. Thus, the Toffoli depths are at most $k+2\lfloor\log_2k\rfloor+2\lfloor\log_2\frac{k}{3}\rfloor+8$ and $2\lfloor\log_2k\rfloor+2\lfloor\log_2\frac{k}{3}\rfloor+9$ for circuits-IIa and IIb, respectively.

    \item \textbf{T-depth:} Each Toffoli has T-depth 3. Hence, the T-depths of circuits-IIa and IIb are at most $3k+6\lfloor\log_2k\rfloor+6\lfloor\log_2\frac{k}{3}\rfloor+24$ and $6\lfloor\log_2k\rfloor+6\lfloor\log_2\frac{k}{3}\rfloor+27$, respectively.
\end{itemize}

The QCLA-based COMP-N-SUB circuit presented here will be used to construct division algorithms with depth $O(n \log n)$, where $n$ is the bit length of the dividend. This contrasts with previous work in which division has depth $O(n^2)$. However, it is important to note that this depth reduction does not stem specifically from using COMP-N-SUB. In fact, the standard approach of implementing division as a sequence of $O(n)$ comparisons and conditional subtractions can also achieve depth $O(n \log n)$. This can be accomplished by performing the comparisons in $O(\log n)$ depth and using controlled QCLA adders for the conditional subtractions, which also have $O(\log n)$ depth. The problem with this approach is that a controlled QCLA adder has many more non-Clifford gates than a regular QCLA adder. Performing this controlled adder $O(n)$ times leads to a massive blowup in the number of non-Clifford gates in the division circuit. This may explain why division circuits with depth $O(n \log n)$ have not been discussed in previous work.
We avoid this issue by using COMP-N-SUB, thereby keeping the non-Clifford gate count comparable to that of performing $O(n)$ regular QCLA additions.

\subsubsection{COMP-N-SUB circuit III}
\label{subsec:cns-III}

\begin{figure}[h]
    \centering
    \scalebox{0.9}{
    \Qcircuit @C=0.6em @R=0.8em{
    \lstick{\conj{B_0}} &\gate{X} &\ctrl{1} &\qw &\qw &\qw &\qw &\qw &\qw &\qw &\qw &\qw &\qw &\qw &\qw &\qw &\qw &\qw &\qw &\qw &\qw &\qw &\qw &\qw &\qw &\qw &\qw &\qw &\qw &\qw &\qw &\qw &\qw &\ctrl{1} &\targ &\gate{X} &\qw  \\
    \lstick{A_0} &\qw &\ctrl{1} &\qw &\qw &\qw &\qw &\qw &\qw &\qw &\qw &\qw &\qw &\qw &\qw &\qw &\qw &\qw &\qw &\qw &\qw &\qw &\qw &\qw &\qw &\qw &\qw &\qw &\qw &\qw &\qw &\qw &\qw &\ctrl{1} &\ctrl{-1} &\qw &\qw \\
    \lstick{Z_0} & & &\ctrl{2} &\qw &\ctrl{3} &\qw &\qw &\qw &\qw &\qw &\qw &\qw &\qw &\qw &\qw &\qw &\qw &\qw &\qw &\qw &\qw &\qw &\qw &\qw &\qw &\qw &\qw &\ctrl{3} &\qw &\ctrl{1} &\qw &\ctrl{2} &\qw & &  \\
    \lstick{\conj{B_1}} &\gate{X} &\qw &\targ &\ctrl{1} &\qw &\qw &\qw &\qw &\qw &\qw &\qw &\qw &\qw &\qw &\qw &\qw &\qw &\qw &\qw &\qw &\qw &\qw &\qw &\qw &\qw &\qw &\qw &\qw &\ctrl{1} &\targ &\targ &\qw &\qw &\qw &\gate{X} &\qw  \\
    \lstick{A_1} &\qw &\qw &\targ &\ctrl{1} &\qw &\qw &\qw &\qw &\qw &\qw &\qw &\qw &\qw &\qw &\qw &\qw &\qw &\qw &\qw &\qw &\qw &\qw &\qw &\qw &\qw &\qw &\qw &\qw &\ctrl{1} &\qw &\ctrl{-1} &\targ &\qw &\qw &\qw &\qw \\
    \lstick{Z_1} & & & & &\targ &\ctrl{2} &\qw &\ctrl{3} &\qw &\qw &\qw &\qw &\qw &\qw &\qw &\qw &\qw &\qw &\qw &\qw &\qw &\qw &\ctrl{3} &\qw &\ctrl{1} &\qw &\ctrl{2} &\targ &\qw & & & & & & \\
    \lstick{\conj{B_2}} &\gate{X} &\qw &\qw &\qw &\qw &\targ &\ctrl{1} &\qw &\qw &\qw &\qw &\qw &\qw &\qw &\qw &\qw &\qw &\qw &\qw &\qw &\qw &\qw &\qw &\ctrl{1} &\targ &\targ &\qw &\qw &\qw &\qw &\qw &\qw &\qw &\qw &\gate{X} &\qw   \\
    \lstick{A_2} &\qw &\qw &\qw &\qw &\qw &\targ &\ctrl{1} &\qw &\qw &\qw &\qw &\qw &\qw &\qw &\qw &\qw &\qw &\qw &\qw &\qw &\qw &\qw &\qw &\ctrl{1} &\qw &\ctrl{-1} &\targ &\qw &\qw &\qw &\qw &\qw &\qw &\qw &\qw &\qw \\
    \lstick{Z_2} & & & & & & & &\targ &\ctrl{2} &\qw &\ctrl{3} &\qw &\qw &\qw &\qw &\qw &\qw &\ctrl{3} &\qw &\ctrl{1} &\qw &\ctrl{2} &\targ &\qw & & & & & & & & & & &  \\
    \lstick{\conj{B_3}} &\gate{X} &\qw &\qw &\qw &\qw &\qw &\qw &\qw &\targ &\ctrl{1} &\qw &\qw &\qw &\qw &\qw &\qw &\qw &\qw &\ctrl{1} &\targ &\targ &\qw &\qw &\qw &\qw &\qw &\qw &\qw &\qw &\qw &\qw &\qw &\qw &\qw &\gate{X} &\qw   \\
    \lstick{A_3} &\qw &\qw &\qw &\qw &\qw &\qw &\qw &\qw &\targ &\ctrl{1} &\qw &\qw &\qw &\qw &\qw &\qw &\qw &\qw &\ctrl{1} &\qw &\ctrl{-1} &\targ &\qw &\qw &\qw &\qw &\qw &\qw &\qw &\qw &\qw &\qw &\qw &\qw &\qw &\qw  \\
    \lstick{Z_3} & & & & & & & & & & &\targ &\ctrl{2} &\qw &\ctrl{3} &\ctrl{1} &\qw &\ctrl{2} &\targ &\qw & & & & & & & & & & & & & & & &  \\
    \lstick{\conj{B_4}} &\gate{X} &\qw &\qw &\qw &\qw &\qw &\qw &\qw &\qw &\qw &\qw &\targ &\ctrl{1} &\qw &\targ &\targ &\qw &\qw &\qw &\qw &\qw &\qw &\qw &\qw &\qw &\qw &\qw &\qw &\qw &\qw &\qw &\qw &\qw &\qw &\gate{X} &\qw \\
    \lstick{A_4} &\qw &\qw &\qw &\qw &\qw &\qw &\qw &\qw &\qw &\qw &\qw &\targ &\ctrl{1} &\qw &\qw &\ctrl{-1} &\targ &\qw &\qw &\qw &\qw &\qw &\qw &\qw &\qw &\qw &\qw &\qw &\qw &\qw &\qw &\qw &\qw &\qw &\qw &\qw  \\
    \lstick{Z_4} & & & & & & & & & & & & & &\targ &\qw &\ctrlo{-1} &\qw &\qw &\qw &\qw &\ctrlo{-4} &\qw &\qw &\qw &\qw &\ctrlo{-7} &\qw &\qw &\qw &\qw &\ctrlo{-10} &\qw &\qw &\ctrlo{-14} &\qw &\qw   \gategroup{1}{2}{15}{15}{0.8em}{--}
    }}
    \caption{COMP-N-SUB circuit-III for 5-bit integers. The boxed part of the circuit computes the high bit in qubit $Z_4$. If $Z_4 = \ket{0}$, the remaining part of the circuit computes the difference $b-a$ in register $B$; otherwise, registers $B$ and $A$ are restored to their initial states $\ket{b}$ and $\ket{a}$, respectively.}
    \label{fig:cns_and}
\end{figure}

In this section, we describe the construction of a third type of COMP-N-SUB circuit using the modified QRC adder in \cite{2018_G}, which reduces the T-count through the use of logical-AND gadgets. In addition to the two input registers, COMP-N-SUB circuit-III uses $k$ auxiliary qubits initialized to $\ket{0}$ and labeled $Z_i$. The carry bits are computed in these qubits. Figure \ref{fig:cns_and} illustrates COMP-N-SUB circuit-III for 5-bit integers. The boxed part of the circuit acts as a partial comparator, computing the high bit in qubit $Z_{k-1}$. The unboxed part acts as a conditional subtractor, computing the difference in register $B$ if $Z_{k-1}=\ket{0}$; otherwise, the input registers are restored to their initial states.

We now describe how to construct a quantum circuit for this unitary.
We denote the three-qubit unitary that computes the logical AND of the states of qubits $q_1$ and $q_2$ by $U_{AND(q_1,q_2;q_3)}$, where $q_3$ stores the AND result (Figures \ref{fig:logical_and} and \ref{fig:logical_and_ckt}). We denote the circuit that uncomputes the logical AND by $\mathcal{C}_{(q_1,q_2;q_3)}$ (Figures \ref{fig:logical_and_uncompute} and \ref{fig:measure_fixup}).
\begin{enumerate}
    \item For $i=0,\ldots,k-1$: Apply $\X_{(B_i)}$.

    \item Apply $U_{AND(B_0,A_0;Z_0)}$.

    \item For $i=1,\ldots,k-1$:
    \begin{enumerate}
        \item Apply $\CNOT_{(Z_{i-1};A_i)}\cdot\CNOT_{(Z_{i-1};B_i)} $.

        \item Apply $U_{AND(B_i,A_i;Z_i)}$.

        \item Apply $\CNOT_{(Z_{i-1};Z_i)}$.
    \end{enumerate}
    \item Apply $\CNOT_{(Z_{k-2};B_{k-1})}\cdot\tof_{(\conj{Z_{k-1}},A_{k-1};B_{k-1} )}\cdot \CNOT_{(Z_{k-2};A_{k-1})}$.

    \item For $i=k-2,\ldots,1$:
    \begin{enumerate}
        \item Apply $\CNOT_{(Z_{i-1};Z_i)}$.
        
        \item Apply $\mathcal{C}_{(B_i,A_i;Z_i)}$.

        \item Apply $\CNOT_{(Z_{i-1};B_i)}\cdot\tof_{(\conj{Z_{k-1}},A_i;B_i )}\cdot\CNOT_{(Z_{i-1};A_i)}$.
    \end{enumerate}

    \item \begin{enumerate}
        \item Apply $\mathcal{C}_{(B_0,A_0;Z_0)}$.

        \item Apply $\tof_{(\conj{Z_{k-1}},A_0;B_0)}$.
    \end{enumerate}

    \item For $i = 0,\ldots,k-1$: Apply $\X_{(B_i)}$.
\end{enumerate}

\paragraph{Correctness:} The initial state of the qubits is as follows.
\begin{eqnarray}
    \bigotimes_{i=0}^{k-1}\ket{b_i}_{B_i}\ket{a_i}_{A_i}\ket{0}_{Z_i}.  \nonumber
\end{eqnarray}
Appendix \ref{app:correct-III} shows in detail that the final state after applying the X gates is
    
    \begin{eqnarray}
        \ket{\conj{c_k}\conj{s_0}\oplus c_kb_0}_{B_0}\ket{a_0}_{A_0}\ket{0}_{Z_0}\left(\bigotimes_{i=1}^{k-1}\ket{\conj{c_k}\conj{s_i}\oplus c_kb_i}_{B_i}\ket{a_i}_{A_i}\ket{0}_{Z_i}\right).  \nonumber
    \end{eqnarray}
Hence, if $b\geq a$, that is, $c_k=0$, we obtain $d=b-a$ in register $B$; otherwise, we obtain $b$, as desired.

\paragraph{Resource estimates:} The resource requirements for this circuit are as follows.
\begin{itemize}
    \item \textbf{Toffoli-count:} We require $k$ Toffoli gates in steps 4, 5, and 6.

    \item \textbf{CNOT-count:} We require $3(k-1)$, $2$, and $3(k-2)$ CNOT gates in steps 3, 4, and 5, respectively. Each $U_{AND}$ and each Toffoli require six and seven CNOT gates, respectively. We require a total of $k$ applications of $U_{AND}$ in steps 2 and 3 and a total of $k$ Toffoli gates in steps 4, 5, and 6. Hence, the total number of CNOT gates required is at most $(3k-3)+2+(3k-6)+6k+7k = 19k-7$.

    \item \textbf{T-count:} The T-counts of each $U_{AND}$ and each Toffoli are 4 and 7, respectively. Thus, we require at most $4k+7k = 11k$ T gates.

    \item \textbf{Qubit-count:} We require $2k$ input qubits, one qubit that stores the high bit, and $k-1$ ancilla qubits that are restored to their initial states. Hence, the total qubit count is $3k$.

    \item \textbf{Toffoli-depth:} The Toffoli depth is at most $k$.

    \item \textbf{T-depth:} Each $U_{AND}$ has T-depth 2, while each Toffoli has T-depth 3. Hence, the T-depth of our circuit is at most $2k+3k = 5k$.

    \item \textbf{Measurements:} We require $k$ measurements to uncompute the logical-AND gadgets, unlike COMP-N-SUB circuits I and II, which require no measurements.
\end{itemize}

\begin{table}[h]
\scriptsize
    \centering
    \begin{tabular}{|c|c|c|c|c|c|}
    \hline
    \multicolumn{2}{|c|}{} & \multicolumn{4}{|c|}{\bf COMP-N-SUB circuit}  \\
    \cline{3-6}
       \multicolumn{2}{|c|}{}  &  {\bf I} & {\bf IIa} & {\bf IIb} & {\bf III}\\ 
    \hline\hline
      \multicolumn{2}{|c|}{\bf Toffoli-count}   & $3k-1$ & $11k-6\omega(k)-6\lfloor\log_2k\rfloor-4$ & $11k-6\omega(k)-6\lfloor\log_2k\rfloor-4$ & $k$  \\
      \hline
      \multicolumn{2}{|c|}{\bf Toffoli-depth} & $3k-1$ & $k+2\lfloor\log_2k\rfloor+2\lfloor\log_2\frac{k}{3}\rfloor+8$ & $2\lfloor\log_2k\rfloor+2\lfloor\log_2\frac{k}{3}\rfloor+9$ & $k$  \\
      \hline
      \multicolumn{2}{|c|}{\bf T-count} & $18k-4$ & $76k-42\omega(k)-42\lfloor\log_2k\rfloor-21$ & $77k-42\omega(k)-42\lfloor\log_2k\rfloor-21$ & $11k$  \\
      \hline
      \multicolumn{2}{|c|}{\bf T-depth} & $9k-3$ & $3k+6\lfloor\log_2k\rfloor+6\lfloor\log_2\frac{k}{3}\rfloor+24$ & $6\lfloor\log_2k\rfloor+6\lfloor\log_2\frac{k}{3}\rfloor+27$ & $5k$  \\
      \hline\hline
      \multirow{3}{*}{\bf Qubit-count} & Total & $2k+1$ & $4k-\omega(k)-\lfloor\log_2k\rfloor$ & $5k-\omega(k)-\lfloor\log_2k\rfloor-1$ & $3k$  \\
      \cline{2-6}
       & Ancilla & $0$ & $2k-\omega(k)-\lfloor\log_2k\rfloor-1$ & $3k-\omega(k)-\lfloor\log_2k\rfloor-2$ & $k-1$  \\
      \cline{2-6}
      & Garbage & $1$ & $1$ & $1$ & $1$   \\
      \hline\hline
      \multirow{2}{*}{\bf CNOT-count} & Clifford+Toffoli & $4k-5$ & $6k-2$ & $8k-4$ & - \\
      \cline{2-6}
      & Clifford+T & $25k-12$ & $83k-42\omega(k)-42\lfloor\log_2k\rfloor-30$ & $85k-42\omega(k)-42\lfloor\log_2k\rfloor-32$ & $19k-7$  \\
      \hline\hline
      \multicolumn{2}{|c|}{\bf Fully reversible} & Yes & Yes & Yes & No  \\
      \hline
    \end{tabular}
    \caption{Resource requirements of different quantum circuits for the COMP-N-SUB unitary, which compares and conditionally subtracts $k$-bit integers.}
    \label{tab:compNsub}
\end{table}

\paragraph{Summary:} Table \ref{tab:compNsub} summarizes the resource requirements of the COMP-N-SUB circuits derived above. Circuit-I uses no extra ancilla qubits and has a T-count about 76\% lower than that of circuit-II. The T-count of circuit-III is 38\% lower than that of circuit-I. However, circuit-III uses $k-1$ extra ancilla qubits and is not fully reversible, which may limit its applicability. All the other circuits are fully reversible.

The T-depth of circuit-IIb is exponentially lower than those of the other circuits. We observe that all the circuits are dominated by Toffoli gates and that their overall circuit depths are similar to their T-depths. Thus, the overall circuit depth of circuit-IIb is also exponentially lower than those of the other circuits. However, its CNOT count is about 3.4 and 4.5 times those of circuits-I and III, respectively, and its T-count is about 4.3 and 7 times those of circuits-I and III, respectively. It also uses about three times as many ancilla qubits as circuit-III. Nevertheless, for applications in which the circuit depth or T-depth must be limited, circuit-IIb may be the best option. Circuit-IIa generally offers an intermediate trade-off between circuits I and III, which favor lower counts, and circuit-IIb, which favors lower depth.

\begin{remark}
     
In this paper, we have considered three popular quantum adders to construct quantum circuits for the COMP-N-SUB unitary. Our aim has been to establish a methodology for designing such circuits. Other adder constructions may be considered to develop new and/or more efficient quantum circuits \cite{2020_OOCG, 2020_SZAZ, 2024_R, 2025_R2, 2025_RV}.
    
\end{remark}

\begin{remark}
  Section \ref{sec:prelim} describes two popular implementations of the Toffoli gate \cite{2010_NC,2013_J} and briefly highlights their advantages and disadvantages. In all our resource estimates, we have exclusively used the circuit in Figure \ref{fig:tof}(b). However, depending on the design criteria and application, one may instead consider the Toffoli implementation in \cite{2013_J}. This reduces the T-count and T-depth. For example, for COMP-N-SUB circuit-III (Table \ref{tab:compNsub}), the T-count decreases from $11k$ to $8k$, and the T-depth decreases from $5k$ to $4k$. However, the ancilla count increases from $k-1$ to $2k-2$. The number of measurements also increases, which may slow the computation. These tradeoffs should be considered when selecting the best circuit implementation.
\end{remark}

\subsection{Quantum circuit for division}
\label{sec:div}

In this section, we construct quantum circuits for integer division. There are two main groups of integer division algorithms: (1) restoring and (2) non-restoring. Algorithms in both groups consist of a number of iterations. In each iteration, the divisor is subtracted from either part of the dividend or a residue to which a bit of the dividend has been appended. This residue is obtained from the subtraction performed in the previous iteration. In restoring algorithms, if the difference is negative, then the minuend (from which the divisor is subtracted) becomes the residue; this is the restoration step. In non-restoring algorithms, this restoration does not occur. Instead, subsequent operations are determined by the sign of the difference, and a final correction may be applied.

In this paper, we describe two restoring division algorithms, (1) LONG-DIV and (2) RES-DIV, and one non-restoring algorithm, NON-RES-DIV. Compared with previous work on long division \cite{2022_YGWetal, 2024_OPF} and on restoring and non-restoring division \cite{2019_TMVH}, we introduce a number of algorithmic changes, which we discuss in the following sections. For each algorithm, we show how COMP-N-SUB can be used to design new quantum circuits. Both the algorithmic changes and the use of COMP-N-SUB result in significant reductions in gate counts, depth, and qubit count relative to previous constructions.

 Suppose we want to divide an $n$-bit integer $N$ (the dividend) by an $m$-bit integer $D$ (the divisor). We store these integers in two quantum input registers, each containing at most $n$ qubits. Any register that stores the quotient or remainder is an output register. Any other qubit is an ancilla if it is restored to its initial state; otherwise, it is a garbage qubit. This convention differs from that of previous papers such as \cite{2024_OPF}, in which the authors include the qubits storing the quotient among the garbage qubits.

\paragraph{Notation:} We use $A_{[k:0]}$ to denote the sequence of qubits or bits $(A_k,\ldots,A_0)$. For two strings $A = (A_a,\ldots,A_0)$ and $B = (B_b,\ldots,B_0)$, $(A:B)$ denotes the concatenated string $(A_a,\ldots,A_0,B_b,\ldots,B_0)$. $\cns^{(k)}(\mathcal{B},\mathcal{A},\mathcal{K})$ denotes a COMP-N-SUB unitary that compares and subtracts $k$-bit integers stored in the $k$-qubit registers $\mathcal{B}$ and $\mathcal{A}$. The qubit $\mathcal{K}$ stores the high bit. At the end of the computation, either the registers are unchanged and the high bit is $\ket{1}$, or the difference is stored in $\mathcal{B}$, $\mathcal{A}$ is unchanged, and $\mathcal{K}$ is $\ket{0}$.

\subsubsection{LONG-DIV: A long division algorithm}
\label{subsec:longDiv}

\begin{algorithm}[H]
    \scriptsize
    \caption{LONG-DIV}
    \label{alg:longDiv}
    
    \KwIn{(i) An $n$-bit dividend $N = (N_{n-1},\ldots,N_0)$ and (ii) an $m$-bit divisor $D = (D_{m-1},\ldots,D_0)$, with $n\geq m$.}

    \KwOut{(i) Quotient $Q$ and (ii) remainder $R$.}

    $B \leftarrow (0:0:N)$    \;

    $A \leftarrow (0:D)$    \;

    \If{ $B_{[n-1:n-m]} \geq A_{[m-1:0]} $  }
    {
        $B \leftarrow  (0:0:( B_{[n-1:n-m]} - A_{[m-1:0]}):B_{ [n-m-1:0] } )$   \;
        $B_{n+1} \leftarrow 1$    \;
    }

    \For{$i = 1,\ldots, n-m$}
    {
        \If{ $B_{[n-i:n-m-i]} \geq A $  }
        {
            $B \leftarrow  (B_{[n+1:n-i+1]}:( B_{[n-i:n-m-i]} - A):B_{ [n-m-i-1:0] }) $   \;
            $B_{n+1-i} \leftarrow 1$    \;
        }
    }

    \textbf{return} $B_{[n+1:m+1]}$ and $B_{[m-1:0]}$ as the quotient and remainder, respectively. \;
    
\end{algorithm}

Long division is a textbook algorithm with numerous applications. It is a restoring division algorithm consisting of $n-m+1$ iterations. In each iteration, we subtract the divisor from a minuend. If the difference is negative, the minuend is retained as the residue and the corresponding quotient bit is 0; otherwise, the difference becomes the residue and the quotient bit is 1. In the first iteration, the minuend consists of the $m$ most significant bits of the dividend. Beginning with the next iteration, the minuend is formed by appending the next unprocessed bit of the dividend to the residue from the previous iteration.

The computation always starts with the most significant bit of the dividend. In binary long division, the dividend segment must be less than twice the divisor, and the residue at each iteration must be less than the divisor.  Thus, the subtraction in the first iteration is between $m$-bit integers, whereas beginning with the second iteration it is between $(m+1)$-bit integers. We therefore prepend a $0$ to the divisor in order to subtract binary numbers of equal length.

Dividing an $n$-bit dividend $N$ by an $m$-bit divisor $D$ results in an $(n-m+1)$-bit quotient $Q$ and an $m$-bit remainder $R$. In previous papers \cite{2019_TMVH, 2022_YGWetal, 2024_OPF}, separate qubits are allocated to compute the quotient, while the remainder is computed in place of the dividend. In our LONG-DIV algorithm, the quotient is also computed in place. This substantially reduces the qubit requirement, as explained in more detail below.

\paragraph{In-place computation of the quotient and remainder:} We do not allocate any extra space for the quotient. Instead, some of the qubits that store the dividend at the beginning of the computation store part of the quotient at the end; the remaining qubits store the remainder. This is possible because a portion of the dividend changes in each iteration. In the first iteration, we either replace $N_{[n-1:n-m]}$ by $N_{[n-1:n-m]}-D$ or leave it unchanged. Denoting this $m$-bit residue by $d^{(1)}$, the dividend string becomes $N\leftarrow(d_{[m-1:0]}^{(1)}:N_{[n-m-1:0]})$, and we store $Q_{n-m}$ in an extra bit. In the second iteration, we compute $N_{[n-1:n-m-1]}-(0:D)$, replace the first $m+1$ bits of $N$ by the residue $d^{(2)}$, and store $Q_{n-m-1}$ in an extra bit. After this operation, bit $N_{n-1}$ must be $0$ because the residue is less than $(0:D)$. In the third iteration, we can therefore use this bit to store $Q_{n-m-2}$. By similar reasoning, bit $N_{n-2}$ must be $0$ after the third iteration and can store the quotient bit in the next iteration. The same argument applies to every iteration, allowing us to store part of the quotient in place of a portion of the dividend.

Pseudocode for LONG-DIV is shown in Algorithm \ref{alg:longDiv}. Although we assume that $n\geq m$, the procedure also works when $n<m$. In that case, we prepend zeros to the dividend to make its length equal to $m$.
We now describe a quantum circuit for LONG-DIV using COMP-N-SUB unitaries. It has two registers. The $(m+1)$-qubit register $A = (A_{m},\ldots,A_0)$ stores the divisor $D$ prepended with one zero. The $(n+2)$-qubit register $B = (B_{n+1},\ldots,B_0)$ stores the dividend $N$ prepended with two zeros. At the end of the computation, qubits $B_{[n+1:m+1]}$ store the quotient, qubits $B_{[m-1:0]}$ store the remainder, and $B_m=0$. The following steps describe the construction of the quantum circuit based on Algorithm \ref{alg:longDiv}.

\begin{enumerate}
    \item Apply $\cns^{(m)}(B_{[n-1:n-m]},A_{[m-1:0]},B_{n+1})$.

    \item For $i = 1,\ldots, n-m$:
    \begin{enumerate}
        \item Apply $ \cns^{(m+1)} ( B_{[n-i:n-m-i]}, A, B_{n+1-i} )  $.
    \end{enumerate}

    \item For $i = n+1,\ldots,m+1$: Apply $\X_{(B_i)}$.
\end{enumerate}
First, we implement steps 3--6 of LONG-DIV (Algorithm \ref{alg:longDiv}). After applying the $\cns^{(m)}$ unitary, the residue is stored in $B_{[n-1:n-m]}$, and $B_{n+1}$ stores the high bit. Specifically, $B_{n+1} = \ket{1}$ if the dividend segment is less than the divisor; otherwise, it is $\ket{0}$. Register $A$ remains unchanged.

Next, we implement steps 7--12 by applying a COMP-N-SUB unitary to $(m+1)$-qubit integer segments. Specifically, in iteration $i$, it acts on qubits $B_{n-i},\ldots,B_{n-m-i}$ of register $B$ and on $A=\ket{0}\otimes\ket{D}$, which remains unchanged. The qubit $B_{n+1-i}$ is $\ket{1}$ if the dividend segment is less than $A$; otherwise, it is $\ket{0}$.

After all iterations, applying $\X$ gates to qubits $B_{n+1},\ldots,B_{m+1}$ yields the quotient. The remainder is obtained from qubits $(B_{m-1},\ldots,B_0)$. Figure \ref{fig:longDiv} shows a LONG-DIV quantum circuit for dividing a 3-bit dividend by a 2-bit divisor.

\begin{figure}[h!]
    \centering
    \scalebox{0.8}{
    \begin{subfigure}[b]{0.4\textwidth}
    \Qcircuit @C=1em @R=1em{
    \lstick{\ket{0}_{B_4}} &\qw &\multigate{7}{U^{(2)}_{(B_{[2:1]},A_{[1:0]},B_4 )} } &\qw &\multigate{7}{U^{(3)}_{(B_{[2:0]},A_{[2:0]},B_3 )} } &\gate{X} &\rstick{\ket{Q_1}}\qw \\
    \lstick{\ket{0}_{B_3}} &\qw &\ghost{U^{(2)}_{(B_{[2:1]},A_{[1:0]},B_4 )} } &\qw &\ghost{U^{(3)}_{(B_{[2:0]},A_{[2:0]},B_3 )} } &\gate{X} &\rstick{\ket{Q_0}}\qw \\
    \lstick{\ket{N_2}_{B_2}} &\qw &\ghost{U^{(2)}_{(B_{[2:1]},A_{[1:0]},B_4 )} } &\qw &\ghost{U^{(3)}_{(B_{[2:0]},A_{[2:0]},B_3 )} } &\qw &\rstick{\ket{0}}\qw   \\
    \lstick{\ket{N_1}_{B_1}} &\qw &\ghost{U^{(2)}_{(B_{[2:1]},A_{[1:0]},B_4 )} } &\qw &\ghost{U^{(3)}_{(B_{[2:0]},A_{[2:0]},B_3 )} } &\qw &\rstick{\ket{R_1}}\qw   \\
    \lstick{\ket{N_0}_{B_0}} &\qw &\ghost{U^{(2)}_{(B_{[2:1]},A_{[1:0]},B_4 )} } &\qw &\ghost{U^{(3)}_{(B_{[2:0]},A_{[2:0]},B_3 )} } &\qw &\rstick{\ket{R_0}}\qw   \\
    \lstick{\ket{0}_{A_2}} &\qw &\ghost{U^{(2)}_{(B_{[2:1]},A_{[1:0]},B_4 )} } &\qw &\ghost{U^{(3)}_{(B_{[2:0]},A_{[2:0]},B_3 )} } &\qw &\rstick{\ket{0}}\qw \\
    \lstick{\ket{D_1}_{A_1}} &\qw &\ghost{U^{(2)}_{(B_{[2:1]},A_{[1:0]},B_4 )} } &\qw &\ghost{U^{(3)}_{(B_{[2:0]},A_{[2:0]},B_3 )} } &\qw &\rstick{\ket{D_1}}\qw   \\
    \lstick{\ket{D_0}_{A_0}} &\qw &\ghost{U^{(2)}_{(B_{[2:1]},A_{[1:0]},B_4 )} } &\qw &\ghost{U^{(3)}_{(B_{[2:0]},A_{[2:0]},B_3 )} } &\qw &\rstick{\ket{D_0}}\qw
    }
    \caption{}
    \label{fig:longDiv}
    \end{subfigure}
    }
    \hfill
    \scalebox{0.8}{
    \begin{subfigure}[b]{0.4\textwidth}
    \Qcircuit @C=1em @R=1em{
    \lstick{\ket{0}_{B_5}} &\qw &\multigate{8}{U^{(2)}_{(B_{[3:2]},A_{[1:0]},B_5 )} } &\qw &\multigate{8}{U^{(3)}_{(B_{[3:1]},A_{[2:0]},B_4 )} } &\qw &\multigate{8}{U^{(3)}_{(B_{[2:0]},A_{[2:0]},B_3 )} } &\gate{X} &\rstick{\ket{Q_2}}\qw  \\
    \lstick{\ket{0}_{B_4}} &\qw &\ghost{U^{(2)}_{(B_{[3:2]},A_{[1:0]},B_5 )} } &\qw &\ghost{U^{(3)}_{(B_{[3:1]},A_{[2:0]},B_4 )} } &\qw &\ghost{U^{(3)}_{(B_{[2:0]},A_{[2:0]},B_3 )} } &\gate{X} &\rstick{\ket{Q_1}}\qw    \\
    \lstick{\ket{0}_{B_3}} &\qw &\ghost{U^{(2)}_{(B_{[3:2]},A_{[1:0]},B_5 )} } &\qw &\ghost{U^{(3)}_{(B_{[3:1]},A_{[2:0]},B_4 )} } &\qw &\ghost{U^{(3)}_{(B_{[2:0]},A_{[2:0]},B_3 )} } &\gate{X} &\rstick{\ket{Q_0}}\qw   \\
    \lstick{\ket{N_2}_{B_2}} &\qw &\ghost{U^{(2)}_{(B_{[3:2]},A_{[1:0]},B_5 )} } &\qw &\ghost{U^{(3)}_{(B_{[3:1]},A_{[2:0]},B_4 )} } &\qw &\ghost{U^{(3)}_{(B_{[2:0]},A_{[2:0]},B_3 )} } &\qw &\rstick{\ket{0}}\qw  \\
    \lstick{\ket{N_1}_{B_1}} &\qw &\ghost{U^{(2)}_{(B_{[3:2]},A_{[1:0]},B_5 )} } &\qw &\ghost{U^{(3)}_{(B_{[3:1]},A_{[2:0]},B_4 )} } &\qw &\ghost{U^{(3)}_{(B_{[2:0]},A_{[2:0]},B_3 )} } &\qw &\rstick{\ket{R_1}}\qw  \\
    \lstick{\ket{N_0}_{B_0}} &\qw &\ghost{U^{(2)}_{(B_{[3:2]},A_{[1:0]},B_5 )} } &\qw &\ghost{U^{(3)}_{(B_{[3:1]},A_{[2:0]},B_4 )} } &\qw &\ghost{U^{(3)}_{(B_{[2:0]},A_{[2:0]},B_3 )} } &\qw &\rstick{\ket{R_0}}\qw  \\
    \lstick{\ket{0}_{A_2}} &\qw &\ghost{U^{(2)}_{(B_{[3:2]},A_{[1:0]},B_5 )} } &\qw &\ghost{U^{(3)}_{(B_{[3:1]},A_{[2:0]},B_4 )} } &\qw &\ghost{U^{(3)}_{(B_{[2:0]},A_{[2:0]},B_3 )} } &\qw &\rstick{\ket{0}}\qw   \\
    \lstick{\ket{D_1}_{A_1}} &\qw &\ghost{U^{(2)}_{(B_{[3:2]},A_{[1:0]},B_5 )} } &\qw &\ghost{U^{(3)}_{(B_{[3:1]},A_{[2:0]},B_4 )} } &\qw &\ghost{U^{(3)}_{(B_{[2:0]},A_{[2:0]},B_3 )} } &\qw &\rstick{\ket{D_1}}\qw \\
    \lstick{\ket{D_0}_{A_0}} &\qw &\ghost{U^{(2)}_{(B_{[3:2]},A_{[1:0]},B_5 )} } &\qw &\ghost{U^{(3)}_{(B_{[3:1]},A_{[2:0]},B_4 )} } &\qw &\ghost{U^{(3)}_{(B_{[2:0]},A_{[2:0]},B_3 )} } &\qw &\rstick{\ket{D_0}}\qw
    }
        \caption{}
        \label{fig:resDiv}
    \end{subfigure}
    }
    \caption{Quantum circuits for (a) LONG-DIV and (b) RES-DIV. The dividend $N = (N_2,N_1,N_0)$ and the divisor $D = (D_1,D_0)$ are 3-bit and 2-bit integers, respectively. The remainder is $R = (R_1,R_0)$. The quotient is $Q = (Q_1,Q_0)$ for (a) and $Q = (Q_2,Q_1,Q_0) = (0,Q_1,Q_0)$ for (b). For (b), $m'=1$. The unitary $U^{(k)}$ is a COMP-N-SUB unitary that compares and subtracts $k$-bit integers; its subscript identifies the qubits on which it acts.}
    \label{fig:resDivCkt}
\end{figure}

\paragraph{Resource estimates:} A LONG-DIV quantum circuit for dividing an $n$-bit dividend by an $m$-bit divisor consists of $(n-m+1)$ COMP-N-SUB unitaries. The first operates on $m$-bit integers ($\cns^{(m)}$), and the remaining $n-m$ operate on $(m+1)$-bit integers ($\cns^{(m+1)}$). Depending on the type of COMP-N-SUB circuit used, we refer to these as LONG-DIV circuits I, IIa, IIb, and III. The resulting bounds on the different cost metrics are given below. The resource requirements of $\cns^{(k)}$ for arbitrary $k$ can be obtained from Table \ref{tab:compNsub}.

\begin{table}[h]
\scriptsize
    \centering
    \begin{tabular}{|p{1.75cm}|c|p{2cm}|p{3.5cm}|p{2.5cm}|p{2cm}|}
    \hline
    \multicolumn{2}{|c|}{} & \multicolumn{4}{|c|}{\bf RES-DIV circuit }  \\
    \cline{3-6}
       \multicolumn{2}{|c|}{}  &  {\bf I} & {\bf IIa} & {\bf IIb} & {\bf III} \\ 
    \hline\hline
      \multicolumn{2}{|c|}{\bf Toffoli-count}   & $3m(n-m'+1)$ & $11m(n-m'+1)$ & $11m(n-m'+1)$ & $m(n - m'+1)$ \\
      \hline
      \multicolumn{2}{|c|}{\bf Toffoli-depth} & $3m(n-m'+1)$ & $(m+4\log_2m)(n-m'+1)$ & $4(n-m'+1)\log_2m$ & $m(n-m'+1)$ \\
      \hline
      \multicolumn{2}{|c|}{\bf T-count} & $18m(n-m'+1)$ & $76m(n-m'+1)$ & $77m(n-m'+1)$ & $11m(n-m'+1)$  \\
      \hline
      \multicolumn{2}{|c|}{\bf T-depth} & $9m(n-m'+1)$ & $3(m+4\log_2m )(n-m'+1)$ & $12(n-m'+1)\log_2m $ & $5m(n-m'+1)$  \\
      \hline\hline
      \multirow{3}{*}{\bf Qubit-count} & Total & $n+2m-m'$ & $n+4m-m'$ & $n+5m-m'$ &  $n+3m-m'$ \\
      \cline{2-6}
       & Ancilla & $2$ & $2m$ & $3m$ &  $m$ \\
      \cline{2-6}
      & Garbage & $0$ & $0$ & $0$ & $0$  \\
      \hline \hline
      \multirow{2}{*}{\bf CNOT-count} & Clifford+Toffoli & $4m(n-m'+1)$ & $6m(n-m'+1)$ & $8m(n-m'+1)$ & -  \\
      \cline{2-6}
      & Clifford+T & $25m(n-m'+1)$ & $83m(n-m'+1)$ & $85m(n-m'+1)$ & $19m(n-m'+1)$ \\
      \hline\hline
      \multicolumn{2}{|c|}{\bf Fully reversible} & Yes & Yes & Yes & No \\
      \hline
    \end{tabular}
    \caption{Resource requirements of quantum circuits for RES-DIV. Here, $n$ and $m$ are the numbers of bits in the binary representations of the dividend and divisor, respectively. The parameter $m'$ is an integer such that $1\leq m'\leq m$. When $m'=m$, we obtain the resource requirements of LONG-DIV. Only leading-order terms are shown.}
    \label{tab:resDiv}
\end{table}

\begin{itemize}
    \item \textbf{Toffoli-count:} Using COMP-N-SUB circuit I, the Toffoli count for LONG-DIV circuit I is
    \begin{eqnarray}
     (3m-1)+(n-m)(3(m+1)-1) = 3nm-3m^2+2n+m-1.   \nonumber
    \end{eqnarray}
     For each of LONG-DIV circuits IIa and IIb, the Toffoli count is at most
     \begin{eqnarray}
     &&(11m-6\omega(m)-6\lfloor\log_2m\rfloor-4)+(n-m)(11(m+1)-6\omega(m+1)-6\lfloor\log_2(m+1)\rfloor-4) \nonumber \\
     &=& 11nm-11m^2+7n+4m-4-6\big(\omega(m)+(n-m)\omega(m+1)+\lfloor\log_2m\rfloor+(n-m)\lfloor\log_2(m+1)\rfloor\big)\nonumber \\
     &\leq& 11nm-11m^2+7n+4m-4. \nonumber
     \end{eqnarray}
     Using COMP-N-SUB circuit III, the Toffoli count for LONG-DIV circuit III is at most
     \begin{eqnarray}
         &&m+(n-m)(m+1) = nm - m^2+n.   \nonumber
     \end{eqnarray}

    \item \textbf{CNOT-count:} The CNOT counts of Clifford+Toffoli implementations of the LONG-DIV circuits are as follows.
    \begin{eqnarray}
        \text{Circuit-I:}&\quad& \n_{Tof}(LD_I)=(4m-5)+(n-m)(4(m+1)-5) \nonumber \\
        &&= 4nm-4m^2-n+5m-5     \label{eqn:cnot_tof_LDI} \\
        \text{Circuit-IIa:}&\quad& \n_{Tof}(LD_{IIa})=(6m-2)+(n-m)(6(m+1)-2) \nonumber \\
        &&= 6nm-6m^2+4n+2m-2   \label{eqn:cnot_tof_LDIIa} \\
        \text{Circuit-IIb:}&\quad& \n_{Tof}(LD_{IIb})=(8m-4)+(n-m)(8(m+1)-4) \nonumber\\
        &&= 8nm-8m^2+4n+4m-4    \label{eqn:cnot_tof_LDIIb}
    \end{eqnarray}
    The CNOT counts for Clifford+T implementations of the various LONG-DIV circuits are as follows.
    \begin{eqnarray}
        \text{Circuit-I:}&\quad&\n_T(LD_I)=(25m-12)+(n-m)(25(m+1)-12) \nonumber \\
        &&= 25nm-25m^2+13n+12m-12     \label{eqn:cnot_t_LDI} \\
        \text{Circuit-IIa:}&\quad& \n_T(LD_{IIa})=(83m-30)+(n-m)(83(m+1)-30) \nonumber   \\
    &&= 83nm-83m^2+53n+30m-30  \label{eqn:cnot_t_LDIIa} \\
    \text{Circuit-IIb:}&\quad&  \n_T(LD_{IIb})=(85m-32)+(n-m)(85(m+1)-32) \nonumber   \\
    &&= 85nm-85m^2+53n+32m-32    \label{eqn:cnot_t_LDIIb}   \\
    \text{Circuit-III:}&\quad& \n_T(LD_{III})=(19m-7)+(n-m)(19(m+1)-7) \nonumber   \\
    &&= 19nm-19m^2+12n+7m-7
    \label{eqn:cnot_t_LDIII}
    \end{eqnarray}

    \item \textbf{T-count:} The numbers of T gates required to implement the various LONG-DIV circuits are as follows.
    \begin{eqnarray}
        \text{Circuit-I:}&\quad& \tcount(LD_I)=(18m-4)+(n-m)(18(m+1)-4) \nonumber  \\
        &&= 18nm-18m^2+14n+4m-4        \label{eqn:t_LDI}    \\
        \text{Circuit-IIa:}&\quad& \tcount(LD_{IIa})= (76m-21-42\omega(m)-42\lfloor\log_2m\rfloor)    \nonumber \\
     &&+(n-m)(76(m+1)-21-42\omega(m+1)-42\lfloor\log_2(m+1)\rfloor) \nonumber \\
     &&\leq 76nm-76m^2+55n+21m-21   \label{eqn:t_LDIIa} \\
     \text{Circuit-IIb:}&\quad& \tcount(LD_{IIb})= (77m-21-42\omega(m)-42\lfloor\log_2m\rfloor) \nonumber \\
    &&+(n-m)(77(m+1)-42\omega(m+1)-42\lfloor\log_2(m+1)\rfloor-21)     \nonumber \\
    &&\leq 77nm-77m^2+56n+21m-21   \label{eqn:t_LDIIb}  \\
    \text{Circuit-III:}&\quad& \tcount(LD_{III}) = 11m+(n-m)11(m+1) = 11nm-11m^2+11n  \label{eqn:t_LDIII} 
    \end{eqnarray}

    \item \textbf{Qubit-count:} For LONG-DIV circuit I, registers $B$ and $A$ require $n+2$ and $m+1$ qubits, respectively. Of these, $A_m$ remains in the $\ket{0}$ state, and $B_m=\ket{0}$ at the end of the computation. These two qubits can therefore be regarded as ancillas, and the total qubit count is $n+m+3$.

    For LONG-DIV circuits IIa and IIb, registers $A$ and $B$ require $n+m+3$ qubits in total, as before. In addition, circuits IIa and IIb require $2(m+1)-\omega(m+1)-\lfloor\log_2(m+1)\rfloor-1$ and $2(m+1)-\omega(m+1)-\lfloor\log_2(m+1)\rfloor-1+m$ ancilla qubits, respectively. These ancillas are restored to their initial $\ket{0}$ states. Thus, the total qubit count for circuit IIa is at most
    \begin{eqnarray}
        n+3m+4-\omega(m+1)-\lfloor\log_2(m+1)\rfloor \leq n+3m+4.    \nonumber
    \end{eqnarray}
    The total qubit count for circuit IIb is at most
    \begin{eqnarray}
        n+4m+4-\omega(m+1)-\lfloor\log_2(m+1)\rfloor\leq n+4m+4.  \nonumber
    \end{eqnarray}
    For circuit III, in addition to the $n+m+3$ qubits in registers $A$ and $B$, we require at most $m$ ancilla qubits to uncompute the logical-AND gadgets. These ancillas are restored to their initial states after each iteration and can therefore be reused in the next iteration. Thus, the ancilla count of this circuit is at most $m+2$, and its total qubit count is $n+2m+3$.

    \item \textbf{Toffoli-depth:} The Toffoli depths of the various LONG-DIV circuits are as follows.
    \begin{eqnarray}
        \text{Circuit-I:}&\quad& (3m-1)+(n-m)(3(m+1)-1) = 3nm-3m^2+2n+m-1.   \nonumber \\
        \text{Circuit-IIa:}&\quad& (m+2\lfloor\log_2m\rfloor+2\lfloor\log_2\frac{m}{3}\rfloor+8 )    \nonumber   \\
    &&+(n-m)\left(m+1+2\lfloor\log_2(m+1)\rfloor+2\lfloor\log_2\frac{m+1}{3}\rfloor+8\right)  \nonumber   \\
    &&\leq nm-m^2+9n-8m+8+4(n-m+1)\log_2(m+1)  \nonumber   \\
    \text{Circuit-IIb:}&\quad& 2\lfloor\log_2m\rfloor+2\lfloor\log_2\frac{m}{3}\rfloor+9+(n-m)\left(2\lfloor\log_2(m+1)\rfloor+2\lfloor\log_2\frac{m+1}{3}\rfloor+9\right)   \nonumber\\
    &&\leq (n-m+1)\left(9+4\log_2(m+1)\right)  \nonumber    \\
    \text{Circuit-III:}&\quad& m+(n-m)(m+1) = nm - m^2+n   \nonumber
    \end{eqnarray}

    \item \textbf{T-depth:} The T-depths of the various LONG-DIV circuits are as follows.
    \begin{eqnarray}
        \text{Circuit-I:}&\quad& \tcount_d(LD_I)=9nm-9m^2+6n+3m-3   \label{eqn:tdepth_LDI} \\
        \text{Circuit-IIa:}&\quad& \tcount_d(LD_{IIa})=(3m+6\lfloor\log_2m\rfloor+6\lfloor\log_2\frac{m}{3}\rfloor+24 )  \nonumber \\
    &&+(n-m)\left(3(m+1)+6\lfloor\log_2(m+1)\rfloor+6\lfloor\log_2\frac{m+1}{3}\rfloor+24\right)    \nonumber\\
    &&\leq 3nm-3m^2+27n-24m+24+12(n-m+1)\log_2(m+1)    \label{eqn:tdepth_LDIIa}  \\
    \text{Circuit-IIb:}&\quad& \tcount_d(LD_{IIb})=3(2\lfloor\log_2m\rfloor+2\lfloor\log_2\frac{m}{3}\rfloor+9)+3(n-m) \nonumber   \\
    &&\cdot\left(2\lfloor\log_2(m+1)\rfloor+2\lfloor\log_2\frac{m+1}{3}\rfloor+9\right)   \nonumber \\
    &&\leq 3(n-m+1)\left(9+4\log_2(m+1)\right)     \label{eqn:tdepth_LDIIb} \\
    \text{Circuit-III:}&\quad& \tcount_d(LD_{III}) =   5m+(n-m)5(m+1) = 5nm-5m^2+5n   \label{eqn:tdepth_LDIII}
    \end{eqnarray}

    \item \textbf{Measurements:} LONG-DIV circuits I, IIa, and IIb require no measurements. Because COMP-N-SUB circuit III requires measurements, LONG-DIV circuit III does as well. The number of measurements required is
    \begin{eqnarray}
        m + (n-m)(m+1) = nm+n-m^2.  \nonumber
    \end{eqnarray}
    
\end{itemize}

A summary of the resource requirements of the various LONG-DIV circuits can be obtained from Table \ref{tab:resDiv} by setting $m'=m$.

\subsubsection{RES-DIV: A restoring division algorithm}
\label{subsec:restoreDiv-A}

\begin{algorithm}[H]
    \scriptsize
    \caption{ RES-DIV}
    \label{alg:resDiv}
    
    \KwIn{(i) An $n$-bit dividend $N = (N_{n-1},\ldots,N_0)$; (ii) an $m$-bit divisor $D = (D_{m-1},\ldots,D_0)$, with $n\geq m$; and (iii) an integer $m'$ such that $1\leq m'\leq m$.}

    \KwOut{(i) Quotient $Q$ and (ii) remainder $R$.}

    $B\leftarrow ( 0^{m-m'+2} : N   )$    \;

    $A\leftarrow(0:D)$ \;

    \If{ $B_{[n+m-m'-1:n-m']} \geq A_{[m-1:0]} $  }
    {
        $B \leftarrow  (0:0 : ( B_{[n+m-m'-1:n-m']} - A_{[m-1:0]} ):B_{ [n-m'-1:0] } )$   \;
        $B_{n+m-m'+1} \leftarrow 1$    \;
    }

    \For{$i = 1,\ldots, n-m'$}
    {
        \If{ $B_{[n+m-m'-i:n-m'-i]} \geq A $  }
        {
            $B \leftarrow  (B_{[n+m-m'+1:n+m-m'+1-i]}:( B_{[n+m-m'-i:n-m'-i]} - A):B_{ [n-m'-1-i:0] }) $   \;
            $B_{n+m-m'+1-i} \leftarrow 1$    \;
        }
    }

    \textbf{return} $B_{[n+m-m'+1:m+1]}$ and $B_{[m-1:0]}$ as the quotient and remainder, respectively. \;
        
\end{algorithm}

 In this section, we describe a restoring division algorithm that, unlike long division, does not require the most significant bits (MSBs) of the divisor and dividend to align in the first iteration. Instead, we prepend enough zeros to the dividend that its $m'$ most significant bits overlap the $m'$ least significant bits of the divisor, where $m'\leq m$. Equivalently, the MSB of the dividend aligns with the divisor bit in position $m'$, counted from the least significant end. Beginning with the second iteration, we append the next significant bit of the dividend to the residue. The restoration procedure is the same as in long division (LONG-DIV).

Because the MSBs need not align in the first iteration, this restoring algorithm is useful when divisors of varying lengths occur in superposition. Here, the length is the number of bits from the most significant 1 through the least significant bit, inclusive. Thus, the length of the binary integer $0000111000$ is $6$, not $10$. For RES-DIV, $m'$ is the minimum length of a divisor in the superposition.
For a divisor of length $m''\geq m'$, this procedure ensures that its MSB aligns with the MSB of the dividend in iteration $m''-m'+1$. Before this iteration, the minuend remains unchanged as the residue, so no nontrivial computation occurs. Beginning with iteration $m''-m'+1$, the computation proceeds exactly as in LONG-DIV. Thus, LONG-DIV is the special case of RES-DIV in which $m'=m$, meaning that all divisors in the superposition have the same length.

In \cite{2019_TMVH}, the authors describe a restoring division algorithm in which the dividend is in two's-complement form and is therefore prepended with a sign bit (0 for positive integers). The divisor, also in two's-complement form, is prepended with enough zeros that the dividend and divisor have equal lengths. Consequently, all arithmetic operations in that algorithm involve $(n+1)$-bit integers. Below, we explain several important differences between our RES-DIV algorithm and the restoring division algorithm in \cite{2019_TMVH}.
\begin{enumerate}
    \item \textbf{Unequal lengths of the dividend and divisor:} Throughout our algorithm, we prepend at most one zero to the divisor. This ensures that all arithmetic operations involve integers with bit lengths of at most $m+1$, reducing the implementation cost.

    \item \textbf{Register allocation:} We change the registers that store the quotient and remainder. At the end of our computation, the quotient is stored in place of the zeros prepended to the dividend, while the remainder is stored in place of the dividend. In \cite{2019_TMVH}, these register roles are reversed.

    \item \textbf{In-place computation of the quotient:} As in LONG-DIV, we reuse dividend qubits that become zero during the residue calculation in one iteration to store quotient bits in the next iteration. We therefore need to prepend at most $m-m'+2$ zero bits to the dividend. By contrast, the restoring division algorithm in \cite{2019_TMVH} prepends the dividend with $n$ zeros. Our construction therefore requires fewer qubits.

    \item \textbf{Number of iterations:} In \cite{2019_TMVH}, the MSB of the dividend aligns with the LSB of the divisor in the first iteration, so $n$ iterations are required to slide the full length of the divisor across the dividend. Our algorithm instead allows the $m'$ most significant bits of the dividend, where $1\leq m'\leq m$, to overlap the $m'$ least significant bits of the divisor in the first iteration. This is useful when divisors of length at least $m'$ occur in superposition. We therefore require $n-m'+1$ iterations. Because $n-m+1\leq n-m'+1\leq n$, the implementation cost can be lower than that of \cite{2019_TMVH} when $m'>1$.
\end{enumerate}

The pseudocode for our restoring division algorithm, RES-DIV, is shown in Algorithm \ref{alg:resDiv}. Steps 3--6 and 8--11 can be implemented by $\cns^{(m)}$ and $\cns^{(m+1)}$ unitaries, respectively. The qubits $B_{n+m-m'+1}$ and $B_{n+m-m'+1-i}$ in steps 5 and 10, respectively, can store the high bit produced by a COMP-N-SUB unitary. As explained in the previous section, these bits must be flipped because our COMP-N-SUB construction sets the high bit to 1 when the difference is negative and to 0 otherwise. The restoring-division circuit therefore uses an $(m+1)$-qubit register $A = (A_m,\ldots,A_0)$ storing $(0:D)$ and an $(n+m-m'+2)$-qubit register $B = (B_{n+m-m'+1},\ldots,B_0)$ whose last $n$ qubits initially store the dividend. At the end of the computation, the first $n-m'+1$ qubits of $B$ store the quotient, while the last $m$ qubits store the remainder.

\begin{enumerate}
    \item Apply $\cns^{(m)}(B_{[n+m-m'-1:n-m']},A_{[m-1:0]},B_{n+m-m'+1})$.

    \item For $i=1,\ldots,n-m'$:
    \begin{enumerate}
        \item Apply $\cns^{(m+1)}(B_{[n+m-m'-i:n-m'-i]},A,B_{n+m-m'+1-i})$.
    \end{enumerate}

    \item For $i=0,\ldots,n-m'$, apply $\X_{(B_{n+m-m'+1-i})}$.
\end{enumerate}
 Figure \ref{fig:resDiv} illustrates the algorithm with a circuit for dividing a 3-bit dividend by a 2-bit divisor.

\paragraph{Resource estimates:} Depending on the COMP-N-SUB circuit used, we refer to these circuits as RES-DIV circuits I, IIa, IIb, and III. The bounds on the various gate counts and depths can be computed in the same way as for LONG-DIV. Appendix \ref{app:resource_resDiv} provides detailed calculations of these estimates. Table \ref{tab:resDiv} summarizes the resource requirements of the various quantum circuits for the RES-DIV algorithm.

\subsubsection{Non-restoring division}
\label{subsec:nonRestoreDiv}

\begin{algorithm}[H]
    \scriptsize
    \caption{NON-RES-DIV}
    \label{alg:nonResDiv}
            
    \KwIn{(i) An $n$-bit dividend $N = (N_{n-1},\ldots,N_0)$; (ii) an $m$-bit divisor $D = (D_{m-1},\ldots,D_0)$, with $n\geq m$; and (iii) $m'$ such that $1\leq m'\leq m$.}
    \KwOut{(i) Quotient $Q$ and (ii) remainder $R$.}

    $B\leftarrow \left( 0^{m-m'+2} : N\right)$   \;

    $A\leftarrow (0:0:D)$   \;

    $B\leftarrow \left( ( B_{[n+m-m'+1:n-m']} - A ) : B_{[n-m'-1:0]}   \right)$ \;

    \For{$i = 0,\ldots, n-m'-1 $}
    {
        $B_{n+m-m'+1-i}\leftarrow \conj{B_{n+m-m'+1-i}} $   \;

        \eIf{ $B_{n+m-m'+1-i} == 0$ }
        {
            $B\leftarrow \left( B_{[n+m-m'+1:n+m-m'+1-i]} : ( B_{[n+m-m'-i:n-m'-i-1]} + A ) : B_{[n-m'-i-2:0]}  \right) $   \;
        }
        {
            $B\leftarrow \left( B_{[n+m-m'+1:n+m-m'+1-i]} : ( B_{[n+m-m'-i:n-m'-i-1]} - A ) : B_{[n-m'-i-2:0]}  \right) $   \;
        }
    }
    $B_{m+1}\leftarrow \conj{B_{m+1}} $ \;

    \If{ $B_m == 1$ }
    {
        $ B\leftarrow \left( B_{[n+m-m'+1:m]} : ( B_{[m-1:0]} + A_{[m-1:0]  } )  \right)  $   \;
    }

    \textbf{return} $B_{[n+m-m'+1:m+1]}$ and $B_{[m-1:0]}$ as the quotient and remainder, respectively.  \;

\end{algorithm}

\begin{figure}[h]
    \centering
    \scalebox{0.8}{
    \Qcircuit @C=1em @R=1em{
    \lstick{\ket{0}_{B_5}} &\qw &\multigate{9}{V^{(4)}_{(B_{[5:2]},A )} } &\gate{X} &\ctrl{1} &\qw &\qw &\qw &\qw &\qw &\rstick{\ket{Q_2}} \qw \\
    \lstick{\ket{0}_{B_4}} &\qw &\ghost{V^{(4)}_{(B_{[5:2]},A )} } &\qw &\multigate{8}{\hat{V}^{(4)}_{(B_{[4:1]},A )} } &\gate{X} &\ctrl{1} &\qw &\qw &\qw &\rstick{\ket{Q_1}} \qw \\
    \lstick{\ket{0}_{B_3}} &\qw &\ghost{V^{(4)}_{(B_{[5:2]},A )} } &\qw &\ghost{\hat{V}^{(4)}_{(B_{[4:1]},A )} } &\qw &\multigate{7}{\hat{V}^{(4)}_{(B_{[3:0]},A )}} &\gate{X} &\qw &\qw &\rstick{\ket{Q_0}} \qw \\
    \lstick{\ket{N_2}_{B_2}} &\qw &\ghost{V^{(4)}_{(B_{[5:2]},A )} } &\qw &\ghost{\hat{V}^{(4)}_{(B_{[4:1]},A )} } &\qw &\ghost{\hat{V}^{(4)}_{(B_{[3:0]},A )}} &\qw &\ctrl{1} &\qw &\qw \\
    \lstick{\ket{N_1}_{B_1}} &\qw &\ghost{V^{(4)}_{(B_{[5:2]},A )} } &\qw &\ghost{\hat{V}^{(4)}_{(B_{[4:1]},A )} } &\qw &\ghost{\hat{V}^{(4)}_{(B_{[3:0]},A )}} &\qw &\multigate{5}{ \tilde{V}^{(2)}_{(B_{[1:0]},A_{[1:0]})} } &\qw &\rstick{\ket{R_1}}\qw \\
    \lstick{\ket{N_0}_{B_0}} &\qw &\ghost{V^{(4)}_{(B_{[5:2]},A )} } &\qw &\ghost{\hat{V}^{(4)}_{(B_{[4:1]},A )} } &\qw &\ghost{\hat{V}^{(4)}_{(B_{[3:0]},A )}} &\qw &\ghost{ \tilde{V}^{(2)}_{(B_{[1:0]},A_{[1:0]})} } &\qw &\rstick{\ket{R_0}}\qw  \\
    \lstick{\ket{0}_{A_3}} &\qw &\ghost{V^{(4)}_{(B_{[5:2]},A )} } &\qw &\ghost{\hat{V}^{(4)}_{(B_{[4:1]},A )} } &\qw &\ghost{\hat{V}^{(4)}_{(B_{[3:0]},A )}} &\qw &\ghost{ \tilde{V}^{(2)}_{(B_{[1:0]},A_{[1:0]})} } &\qw &\rstick{\ket{0}}\qw \\
    \lstick{\ket{0}_{A_2}} &\qw &\ghost{V^{(4)}_{(B_{[5:2]},A )} } &\qw &\ghost{\hat{V}^{(4)}_{(B_{[4:1]},A )} } &\qw &\ghost{\hat{V}^{(4)}_{(B_{[3:0]},A )}} &\qw &\ghost{ \tilde{V}^{(2)}_{(B_{[1:0]},A_{[1:0]})} } &\qw &\rstick{\ket{0}}\qw \\
    \lstick{\ket{D_1}_{A_1}} &\qw &\ghost{V^{(4)}_{(B_{[5:2]},A )} } &\qw &\ghost{\hat{V}^{(4)}_{(B_{[4:1]},A )} } &\qw &\ghost{\hat{V}^{(4)}_{(B_{[3:0]},A )}} &\qw &\ghost{ \tilde{V}^{(2)}_{(B_{[1:0]},A_{[1:0]})} } &\qw &\rstick{\ket{D_1}}\qw \\
    \lstick{\ket{D_0}_{A_0}} &\qw &\ghost{V^{(4)}_{(B_{[5:2]},A )} } &\qw &\ghost{\hat{V}^{(4)}_{(B_{[4:1]},A )} } &\qw &\ghost{\hat{V}^{(4)}_{(B_{[3:0]},A )}} &\qw &\ghost{ \tilde{V}^{(2)}_{(B_{[1:0]},A_{[1:0]})} } &\qw &\rstick{\ket{D_0}}\qw
    }
     }
    \caption{Quantum circuit for NON-RES-DIV. The dividend $N = (N_2,N_1,N_0)$ and the divisor $D = (D_1,D_0)$ are 3-bit and 2-bit integers, respectively. The remainder and quotient are $R = (R_1,R_0)$ and $Q = (Q_2,Q_1,Q_0)$, respectively, and $m'=1$. The unitaries $V^{(k)}$, $\hat{V}^{(k)}$, and $\tilde{V}^{(k)}$ implement subtraction, Ctrl-AddSub, and c-ADD, respectively, on $k$-bit integers. Their subscripts identify the qubits on which they act.}
    \label{fig:nonResDivCkt}
\end{figure}

\begin{table}[h]
\scriptsize
    \centering
    \begin{tabular}{|c|c|c|c|c|}
    \hline
      \multicolumn{2}{|c|}{}   &  \multicolumn{3}{|c|}{\bf NON-RES-DIV circuit} \\
      \cline{3-5}
       \multicolumn{2}{|c|}{}  & {\bf A} & {\bf B} & {\bf C} \\ 
    \hline\hline
         \multicolumn{2}{|c|}{\bf Toffoli-count}  & $2m(n-m'+1)$ & $10m(n-m'+1)$ & -  \\
      \hline
      \multicolumn{2}{|c|}{\bf Toffoli-depth} & $2m(n-m'+1)$ & $4(n-m'+1)\log_2(m+2)$ & - \\
      \hline
      \multicolumn{2}{|c|}{\bf T-count} & $14m(n-m'+1)$ & $70m(n-m'+1)$ & $4m(n-m'+1)$ \\
      \hline
      \multicolumn{2}{|c|}{\bf T-depth} & $6m(n-m'+1)$ & $12(n-m'+1)\log_2(m+2)$ & $2m(n-m'+1)$   \\
      \hline\hline
      \multirow{3}{*}{\bf Qubit-count} & Total & $n+2m-m'$ & $n+5m-m'$ & $n+3m-m'$  \\
      \cline{2-5}
       & Ancilla & $2$ & $2m$  & $m$  \\
      \cline{2-5}
      & Garbage & $0$ & $0$ & $0$  \\
      \hline\hline
      \multirow{2}{*}{\bf CNOT-count} & Clifford+Toffoli & $5m(n-m'+1)$ & $4m(n-m'+1)$ & -  \\
      \cline{2-5}
      & Clifford+T & $19m(n-m'+1)$ & $74m(n-m'+1)$ & $16m(n-m'+1)$ \\
      \hline\hline
      \multicolumn{2}{|c|}{\bf Fully reversible} & Yes & Yes & No \\
      \hline
    \end{tabular}
    \caption{Resource requirements of quantum circuits for the NON-RES-DIV algorithm. Only leading-order terms are shown. Here, $n$ and $m$ are the numbers of bits in the binary representations of the dividend and divisor, respectively. The parameter $m'$ is an integer such that $1\leq m'\leq m$.}
    \label{tab:nonRestore}
\end{table}

Non-restoring division algorithms are iterative algorithms in which the divisor is subtracted from or added to a minuend in each iteration. Unlike restoring division algorithms, they have no restoration phase: if the difference is negative, we do not restore the residue to the minuend. Instead, we adjust the negative difference in later iterations until we obtain a positive difference. A positive difference implies that the corresponding quotient bit is 1; otherwise, it is 0.

Algorithm \ref{alg:nonResDiv} gives pseudocode for our non-restoring division algorithm, NON-RES-DIV. As in RES-DIV, the parameter $m'$ determines how many bits of the dividend overlap the divisor in the first iteration and hence determines the number of iterations. When divisors occur in superposition, $m'$ is again their minimum length. Register $B$ contains the dividend prepended with $m-m'+2$ zeros, while register $A$ contains the divisor prepended with two zeros. We begin by subtracting $A$ from the $m+2$ most significant bits of $B$. In subsequent iterations, we slide this window by appending the next significant bit of $B$ to the residue. If the MSB of the residue is 1, the difference is negative and is adjusted by adding the divisor in the next iteration; if the MSB is 0, the reverse operation is performed (steps 6--10). The MSB is also flipped and used to store the corresponding quotient bit (step 5). After all iterations, if bit $B_{m+1}$ is 1, the residue is still negative. Adding the divisor in steps 13--15 corrects it and yields the remainder.

The non-restoring division procedure is similar to the algorithm described in \cite{2019_TMVH}, but we introduce the following important changes to enable a more efficient implementation.

\begin{enumerate}
    \item \textbf{Unequal lengths of the dividend and divisor:} As in RES-DIV, we ensure that operations (subtractions or controlled additions) involve integers with bit lengths of at most $m+2$ by appending at most two zeros to the divisor. By contrast, \cite{2019_TMVH} appends enough zeros to the divisor and dividend to produce operations on $(n+1)$-bit integers, including the sign bit, which increases the gate complexity per operation.

    \item \textbf{Number of iterations:} As in RES-DIV, we allow at most $m'$ bits of the dividend to overlap the divisor in the first iteration, resulting in $n-m'+1$ iterations. In \cite{2019_TMVH}, the MSB of the dividend aligns with the LSB of the divisor in the first iteration, so $n$ iterations are required to slide the full length of the divisor across the dividend.
\end{enumerate}

NON-RES-DIV can be implemented with the following two unitaries, which are also discussed in \cite{2019_TMVH}.
\begin{enumerate}
    \item \textbf{Ctrl-AddSub:} The inputs to this unitary are two integers $a$ and $b$ and a control bit $c$. It subtracts the two integers if the control is $\ket{1}$ and adds them if the control is $\ket{0}$. As explained above, $a-b = \conj{\conj{a}+b}$. Thus, this unitary can be implemented with an adder conjugated by CNOT gates whose control is $c$ and whose targets are the qubits of $a$. Further details about this unitary and its implementation appear in \cite{2019_TMVH}.

    \item \textbf{Ctrl-AddNOP or c-ADD:} Ctrl-AddNOP, as described in \cite{2019_TMVH}, is a controlled adder (c-ADD). Any of the COMP-N-SUB circuits can be modified to implement a controlled adder. Suppose the addition is conditioned on the state of a qubit $c$. Because our construction produces no high bit, we need not allocate a qubit for it. From the COMP-N-SUB circuit, we remove all gates that change the high bit. We retain gates controlled on the high bit but instead control them on qubit $c$. We also remove all X gates at the beginning and end of the circuit.
\end{enumerate}
A quantum circuit implementing NON-RES-DIV (Algorithm \ref{alg:nonResDiv}) can be constructed as follows.
\begin{enumerate}
    \item Apply $\sub^{(m+2)}\left( B_{[n+m-m'+1:n-m']} , A \right)$.

    \item For $i = 0,\ldots, n-m'-1$ :
    \begin{enumerate}
        \item Apply $\X_{ (B_{n+m-m'+1-i}) }$.

        \item Apply $\cAddSub^{(m+2)} \left(  B_{[n+m-m'-i:n-m'-i-1]} , A \right)$, controlled on $B_{n+m-m'+1-i}$.
    \end{enumerate}

    \item Apply $\X_{ (B_{m+1} )}$.

    \item Apply $\cAdd^{(m)}\left( B_{[m-1:0]} , A_{[m-1:0]} \right)$ controlled on $B_m$.
\end{enumerate}
SUB is a subtractor. As noted above, Ctrl-AddSub can be implemented with an adder and CNOT gates. We consider the following types of adders and give their resource estimates for operations on $k$-bit integers.

\begin{enumerate}
    \item The quantum ripple-carry (QRC) adder of \cite{2010_TTK} has a Toffoli count and Toffoli depth of at most $2k-1$. The CNOT counts for Clifford+Toffoli and Clifford+T implementations are at most $5k-5$ and $19k-12$, respectively. In addition to the $2k$ input qubits, it requires one qubit to store the high bit.

    \item The quantum carry-lookahead (QCLA) adder of \cite{2006_DKRS} has a Toffoli count of at most $10k-3\omega(k)-3\omega(k-1)-3\lfloor\log_2k\rfloor-3\lfloor\log_2(k-1)\rfloor-7$ and a Toffoli depth of at most $\lfloor\log_2k\rfloor+\lfloor\log_2(k-1)\rfloor+\lfloor\log_2\frac{k}{3}\rfloor+\lfloor\log_2\frac{k-1}{3}\rfloor+8$. Its CNOT count is $4k-5$ for a Clifford+Toffoli implementation and at most $(4k-5)+7(10k-7) = 74k-54$ for a Clifford+T implementation. It requires $2k$ input qubits, one additional qubit to store the high bit, and $2k-\omega(k)-\lfloor\log_2k\rfloor-1$ ancilla qubits that are restored to their initial $\ket{0}$ states at the end of the computation.

    \item The QRC adder of \cite{2018_G}, which uses logical-AND gadgets, has T-count $4k-4$, T-depth $2k-2$, and CNOT count $16k-15$. In addition, it uses $k-1$ ancilla qubits that are restored to their initial $\ket{0}$ states at the end of the computation.
\end{enumerate}
 We do not use the high bits of any of the controlled adders or subtractors and therefore do not need to allocate those extra qubits.

\paragraph{Resource estimates:} The NON-RES-DIV circuits obtained using the adders in (1), (2), and (3) above are called circuits 1, 2, and 3, respectively. The contributions from Ctrl-AddNOP (c-ADD) or COMP-N-SUB are lower-order terms. So, the variant of the NON-RES-DIV circuits depend mostly on the type of adder used, and hence the nomenclature. We use COMP-N-SUB circuits I, IIb, and III with NON-RES-DIV circuits 1, 2, and 3, respectively. The adder in \cite{2018_G} has only a Clifford+T implementation, so we do not report the Toffoli count or depth of circuit 3. Each circuit requires $n-m'+1$ adders acting on $(m+2)$-bit integers, $2(n-m')(m+2)$ CNOT gates for Ctrl-AddSub, and one $\cns^{(m)}$.

\begin{itemize}
    \item \textbf{Toffoli-count:} The Toffoli counts for NON-RES-DIV circuits 1 and 2 are bounded above by
    \begin{eqnarray}
    (n-m'+1)(2(m+2)-1)+(3m-1) , \nonumber 
    \end{eqnarray}
    and
    \begin{eqnarray}
     (n-m'+1)(10(m+2)-7)+11m-4 ,    \nonumber   
    \end{eqnarray}
    respectively.

    \item \textbf{CNOT-count:} The CNOT counts for Clifford+Toffoli implementations of the various NON-RES-DIV circuits are as follows.
    \begin{eqnarray}
    \text{Circuit 1:}&\quad& \n_{Tof}(NRes_1) = (n-m'+1)(5(m+2)-5)+4m-5    \label{eqn:cnot_tof_NRes_a}  \\
     \text{Circuit 2:}&\quad&\n_{Tof}(NRes_2) =  (n-m'+1)(4(m+2)-5)+8m-4 \label{eqn:cnot_tof_NRes_b}
    \end{eqnarray} 
    For Clifford+T implementations of the various NON-RES-DIV circuits, the CNOT counts are as follows.
    \begin{eqnarray}
        \text{Circuit 1:}&\quad& \n_T(NRes_1) = (n-m'+1)(19(m+2)-12)+25m-12 \label{eqn:cnot_t_NRes_a}   \\
        \text{Circuit 2:}&\quad& \n_{T}(NRes_2) = (n-m'+1)(74(m+2)-54)+85m-32      \label{eqn:cnot_t_NRes_b}    \\
        \text{Circuit 3:}&\quad& \n_T(NRes_3)= (n-m'+1)(16(m+2)-15)+19m-7   \label{eqn:cnot_t_NRes_c}
    \end{eqnarray}

    \item \textbf{T-count:} The T-counts for the various NON-RES-DIV circuits are as follows.
    \begin{eqnarray}
    \text{Circuit 1:}&\quad& \tcount(NRes_1)\leq 7(n-m'+1)(2(m+2)-1) + 18m-4  \label{eqn:t_NRes_a}    \\
        \text{Circuit 2:}&\quad& \tcount(NRes_2) \leq 7(n-m'+1)(10(m+2)-7)+77m-21   \label{eqn:t_NRes_b}    \\
    \text{Circuit 3:}&\quad& \tcount(NRes_3) = (n-m'+1)(4(m+2)-4)+11m
    \label{eqn:t_NRes_c}
    \end{eqnarray}

    \item \textbf{Qubit-count:} Registers $A$ and $B$ consist of $m+2$ and $n+m-m'+2$ qubits, respectively. At the end of the computation, the quotient and remainder are stored in register $B$. The two qubits prepended to the divisor in $A$ are restored to the $\ket{0}$ state and are therefore ancilla qubits. All three types of circuits require these qubits.
    
    NON-RES-DIV circuit-1 uses no other qubits, so its total qubit count is $n+2m-m'+4$, of which two are ancillas.

    The QCLA adders in circuit-2 require $2(m+2)-\omega(m+2)-\lfloor\log_2(m+2)\rfloor-1$ ancilla qubits. Because these qubits are restored to their initial states, they can be shared among the adders and COMP-N-SUB unitaries. COMP-N-SUB circuit IIb uses $m$ additional ancilla qubits. Hence, the total number of qubits required for NON-RES-DIV circuit-2 is at most $n+5m-m'+7-\omega(m+2)-\lfloor\log_2(m+2)\rfloor \leq n+5m-m'+7$.

    Each adder in circuit-3 requires $m+1$ ancilla qubits that are restored to their initial states and can therefore be reused by later adders and by COMP-N-SUB circuit III. Thus, the total number of qubits required for NON-RES-DIV circuit-3 is $n+3m-m'+5$.

    \item \textbf{Toffoli-depth:} The Toffoli depth of circuit 1 is at most
    \begin{eqnarray}
        (n-m'+1)(2(m+2)-1)+(3m-1),
    \end{eqnarray}
    and that of circuit 2 is at most
    \begin{eqnarray}
    &\leq& (n-m'+1)\left(4\lfloor\log_2(m+2)\rfloor+8\right) +4\lfloor\log_2m\rfloor+9. \nonumber \\
    \end{eqnarray}

    \item \textbf{T-depth:} The T-depths of the various NON-RES-DIV circuits are as follows.
    \begin{eqnarray}
    \text{Circuit 1:}&\quad&\tcount_d(NRes_1)\leq 3(n-m'+1)(2(m+2)-1)+9m-3   \label{eqn:tdepth_NRes_a}   \\
        \text{Circuit 2:}&\quad&\tcount_d(NRes_2) \leq 3(n-m'+1)\left(4\lfloor\log_2(m+2)\rfloor+8\right)   +3\left(4\lfloor\log_2m\rfloor+9\right)     \label{eqn:tdepth_NRes_b}   \\
    \text{Circuit 3:}&\quad& \tcount_d(NRes_3) = (n-m'+1)(2(m+2)-2)+5m  \label{eqn:tdepth_NRes_c}
    \end{eqnarray}

    \item \textbf{Measurements:} NON-RES-DIV circuits 1 and 2 require no measurements. For circuit-3, the number of measurements required is at most
    \begin{eqnarray}
        (n-m'+1)(m+1)+m = nm+n+2m-m'-mm'+1. \nonumber
    \end{eqnarray}
\end{itemize}

Table \ref{tab:nonRestore} summarizes the resource requirements of the various quantum circuits for the NON-RES-DIV algorithm.

\subsection{Comparison}
\label{sec:compare}

\begin{figure}
    \centering
    
    \begin{subfigure}[b]{0.4\textwidth}
        \centering
        \scalebox{0.6}{
        \begin{tabular}{|c||c|c|c|c|c|}
            \hline
            \multicolumn{6}{|c|}{\textbf{Long division}} \\
            \hline
            $\mathcal{T}$ & \textbf{OPF24} &  $\mathbf{LD_I}$  &  $\mathbf{LD_{IIa}}$ &  $\mathbf{LD_{IIb}}$  &  $\mathbf{LD_{III}}$  \\
            \hline\hline
            \textbf{OPF24} & 1 & 0.39 & 1.64 & 1.67 & 0.24 \\
             $\mathbf{LD_I}$  & 2.56 & 1 & 4.17 & 4.35 & 0.61 \\
            $\mathbf{LD_{IIa}}$  & 0.61 & 0.24 & 1 & 1.01 & 0.14 \\
            $\mathbf{LD_{IIb}}$ & 0.60 & 0.23 & 0.99 & 1 & 0.14 \\
            $\mathbf{LD_{III}}$ & 4.18 & 1.64 & 6.91 & 7 & 1 \\
            \hline
        \end{tabular}
        }
        \subcaption{}
        \label{tab:long_div_T}
    \end{subfigure}
    \hfill
    \begin{subfigure}[b]{0.4\textwidth}
        \centering
        \scalebox{0.6}{
        \begin{tabular}{|c||c|c|c|c|c|}
            \hline
            \multicolumn{6}{|c|}{\textbf{Long division}} \\
            \hline
            $\mathcal{N}_T^{cnot}$ & \textbf{OPF24} & $\mathbf{LD_I}$ & $\mathbf{LD_{IIa}}$ & $\mathbf{LD_{IIb}}$ & $\mathbf{LD_{III}}$ \\
            \hline
            \hline
            \textbf{OPF24} & 1 & 0.35 & 1.39 & 1.41 & 0.32 \\
             $\mathbf{LD_I}$  & 2.86 & 1 & 4 & 4 & 0.90 \\
            $\mathbf{LD_{IIa}}$ & 0.72 & 0.25 & 1 & 1.02 & 0.23 \\
            $\mathbf{LD_{IIb}}$ & 0.71 & 0.25 & 0.98 & 1 & 0.22 \\
            $\mathbf{LD_{III}}$ & 3.16 & 1.11 & 4.37 & 4.47 & 1 \\
            \hline
        \end{tabular}
        }
        \subcaption{}
        \label{tab:long_div_N}
    \end{subfigure}
    \hfill
    \begin{subfigure}[b]{0.4\textwidth}
        \centering
        \scalebox{0.6}{
        \begin{tabular}{|c||c|c|c|c|c|}
            \hline
            \multicolumn{6}{|c|}{\textbf{Restoring division}} \\
            \hline
            $\mathcal{T}$ & \textbf{TMCVH19} & $\mathbf{Res_I}$ & $\mathbf{Res_{IIa}}$ & $\mathbf{Res_{IIb}}$ & $\mathbf{Res_{III}}$ \\
            \hline\hline
            \textbf{TMCVH19} & 1 & $\frac{0.52m}{n}$ & $\frac{2.17m}{n}$ & $\frac{2.22m}{n}$ & $\frac{0.31m}{n}$ \\
            $\mathbf{Res_I}$ & $\frac{1.94n}{m}$ & 1 & 4.17 & 4.35 & 0.61 \\
            $\mathbf{Res_{IIa}}$ & $\frac{0.46n}{m}$ & 0.24 & 1 & 1.01 & 0.14 \\
            $\mathbf{Res_{IIb}}$ & $\frac{0.45n}{m}$ & 0.23 & 0.99 & 1 & 0.14 \\
            $\mathbf{Res_{III}}$ & $\frac{3.18n}{m}$ & 1.64 & 6.91 & 7 & 1 \\
            \hline
        \end{tabular}
        }
        \subcaption{}
        \label{tab:restore_div_T}
    \end{subfigure}

    \vspace{1em}

    \begin{subfigure}[b]{0.4\textwidth}
        \centering
        \scalebox{0.6}{
        \begin{tabular}{|c||c|c|c|c|c|}
            \hline
            \multicolumn{6}{|c|}{\textbf{Restoring division}} \\
            \hline
            $\mathcal{N}_T^{cnot}$ & \textbf{TMCVH19} & $\mathbf{Res_I }$ & $\mathbf{Res_{IIa}}$  & $\mathbf{Res_{IIb}}$ & $\mathbf{Res_{III}}$  \\
            \hline\hline
            \textbf{TMCVH19} & 1 & $\frac{0.57m}{n}$ & $\frac{1.89m}{n}$ & $\frac{1.92m}{n}$ & $\frac{0.43m}{n}$ \\
            $\mathbf{Res_I }$  & $\frac{1.76n}{m}$ & 1 & 3.33 & 3.45 & 0.76 \\
            $\mathbf{Res_{IIa}}$ & $\frac{0.53n}{m}$ & 0.30 & 1 & 1.02 & 0.23 \\
            $\mathbf{Res_{IIb}}$ & $\frac{0.52n}{m}$ & 0.29 & 0.98 & 1 & 0.22 \\
            $\mathbf{Res_{III}}$ & $\frac{2.32n}{m}$ & 1.32 & 4.37 & 4.47 & 1 \\
            \hline
        \end{tabular}
        }
        \subcaption{}
        \label{tab:restore_div_N}
    \end{subfigure}
    \hspace{0.3in}
    \begin{subfigure}[b]{0.32\textwidth}
        \centering
        \scalebox{0.6}{
        \begin{tabular}{|c||c|c|c|c|}
            \hline
            \multicolumn{5}{|c|}{\textbf{Non-restoring division}} \\
            \hline
            $\mathcal{T}$ & \textbf{TMCVH19} & $\mathbf{NRes_1}$ & $\mathbf{NRes_2}$ & $\mathbf{NRes_3}$  \\
            \hline\hline
            \textbf{TMCVH19} & 1 & $\frac{m}{n}$ & $\frac{5m}{n}$ & $\frac{0.29m}{n}$ \\
            $\mathbf{NRes_1}$  & $\frac{n}{m}$ & 1 & 5 & 0.29 \\
            $\mathbf{NRes_2}$ & $\frac{0.2n}{m}$ & 0.2 & 1 & 0.06 \\
            $\mathbf{NRes_3}$ & $\frac{3.5n}{m}$ & 3.5 & 17.5 & 1 \\
            \hline
        \end{tabular}
        }
        \subcaption{}
        \label{tab:nonrestore_div_T}
    \end{subfigure}
    \hfill
    \begin{subfigure}[b]{0.32\textwidth}
        \centering
        \scalebox{0.6}{
        \begin{tabular}{|c||c|c|c|c|}
            \hline
            \multicolumn{5}{|c|}{\textbf{Non-restoring division}} \\
            \hline
            $\mathcal{N}_T^{cnot}$ & \textbf{TMCVH19} & $\mathbf{NRes_1}$ & $\mathbf{NRes_2}$ & $\mathbf{NRes_3}$ \\
            \hline\hline
            \textbf{TMCVH19} & 1 & $\frac{m}{n}$ & $\frac{3.85m}{n}$ & $\frac{0.84m}{n}$ \\
            $\mathbf{NRes_1}$  & $\frac{n}{m}$ & 1 & 3.85 & 0.84 \\
            $\mathbf{NRes_2}$  & $\frac{0.26n}{m}$ & 0.26 & 1 & 0.22 \\
            $\mathbf{NRes_3}$ & $\frac{1.19n}{m}$ & 1.19 & 4.63 & 1 \\
            \hline
        \end{tabular}
        }
        \subcaption{}
        \label{tab:nonrestore_div_N}
    \end{subfigure}
    \caption{Ratios of the T-counts (a, c, e) and CNOT-counts (b, d, f) of quantum circuits for different division algorithms. Each row except the first and each column is labeled by a circuit. OPF24 and TMCVH19 denote the circuits in \cite{2024_OPF} and \cite{2019_TMVH}, respectively. $LD_x$, $Res_x$, and $NRes_x$ denote LONG-DIV circuit $x$, RES-DIV circuit $x$, and NON-RES-DIV circuit $x$, respectively. The metric, either T-count ($\tcount$) or CNOT-count ($\n_T$), is specified in the leftmost cell of the second row. The cell at the intersection of column $i$ and row $j$ stores $\tcount(i)/\tcount(j)$ for a T-count table or $\n_T(i)/\n_T(j)$ for a CNOT-count table.}
    \label{fig:compare}
\end{figure}

In this section, we compare our results with previous work using the cost metrics for fault-tolerant implementations described earlier. Nearly all papers focused on fault-tolerant implementations use Clifford+T gates and derive bounds on T-count, T-depth, and qubit count. We consider papers that provide the best known bounds, comparing our circuits with the long-division circuits in \cite{2024_OPF} and the restoring and non-restoring division circuits in \cite{2019_TMVH}. We first summarize the relative applicability of the division algorithms to determine suitable parameters for comparison.

\textbf{Applicability:} The long-division algorithm (\cite{2022_YGWetal, 2024_OPF}, LONG-DIV), which is a specific type of restoring algorithm, can be applied when there is a single divisor or when all divisors in a superposition have the same length. It requires the MSBs of the dividend and divisor to align in the first iteration. This occurs, for example, in algorithms that perform repeated division by a fixed integer $p$ in a finite field $\fld_p$. When divisors of varying lengths occur in superposition, non-restoring algorithms and restoring algorithms other than long division can be used. The algorithms in \cite{2019_TMVH} ensure that only the MSB of the dividend overlaps the LSB of the divisor in the first iteration. Our RES-DIV and NON-RES-DIV algorithms introduce the parameter $m'$, which provides greater flexibility in the amount of overlap.

 For a meaningful comparison with any long-division algorithm, we therefore set $m'=m$. Below, we give the T-count, T-depth, and CNOT-count of the circuits in \cite{2019_TMVH} and \cite{2024_OPF}. The algorithms in \cite{2019_TMVH} operate on integers in two's-complement form. For positive integers, their binary representations are therefore prepended with a 0. Thus, for $n$-bit integers, the effective input length is $n+1$ bits. The unitaries consequently operate on $(n+1)$-bit integers, and the expressions have been scaled accordingly. The T-counts of those circuits are as follows.
\begin{eqnarray}
    \text{Long division \cite{2024_OPF}:}&\quad& \tcount(LD_{OPF24}) = 46mn-46m^2+48m-2n-2  \nonumber \\
    \text{Restoring division \cite{2019_TMVH}:}&\quad& \tcount(Res_{TMCVH19}) = 35n^2+42n+7    \nonumber    \\
    \text{Non-restoring division \cite{2019_TMVH}:}&\quad& \tcount(NRes_{TMCVH19}) = 14n^2+49n+7  \nonumber
\end{eqnarray}
The T-depths of those circuits are as follows.
\begin{eqnarray}
    \text{Long division \cite{2024_OPF}:}&\quad& \tcount_d(LD_{OPF24}) = 20mn-20m^2+20m  \nonumber \\
    \text{Restoring division \cite{2019_TMVH}:}&\quad& \tcount_d(Res_{TMCVH19}) = 15n^2+18n+3    \nonumber    \\
    \text{Non-restoring division \cite{2019_TMVH}:}&\quad& \tcount_d(NRes_{TMCVH19}) = 6n^2+21n+11  \nonumber
\end{eqnarray}
The CNOT-counts are as follows.
\begin{eqnarray}
  \text{Long division \cite{2024_OPF}} &:& \n_T(LD_{OPF24}) = 60nm-60m^2-13n+73m-13   \nonumber \\
  \text{Restoring division \cite{2019_TMVH}} &:& \n_{Tof}(Res_{TMCVH19}) = 9n^2+10n+1 \nonumber \\
  &:&  \n_T(Res_{TMCVH19}) = 44n^2+87n+43  \nonumber \\
  \text{Non-restoring division \cite{2019_TMVH}} &:& \n_{Tof}(NRes_{TMCVH19}) = 5n^2+11n    \nonumber \\
  &:& \n_T(NRes_{TMCVH19}) = 19n^2+53n+42  \nonumber
\end{eqnarray}
Tables \ref{tab:compare_T} and \ref{tab:compare_q} compare the resource requirements of circuits for the different division algorithms. We first discuss the full reversibility of each circuit, because many quantum applications require this feature.

\textbf{Full reversibility:} LONG-DIV circuit III, RES-DIV circuit III, and NON-RES-DIV circuit 3 are not fully reversible (Table \ref{tab:compare_T}). The quantum circuit in \cite{2024_OPF} is not fully reversible, whereas those in \cite{2019_TMVH} are.

Next, we identify the circuits with the best metrics for each algorithm using the tables in Figure \ref{fig:compare}. Each table corresponds to a particular division algorithm and metric, either T-count ($\tcount$) or CNOT count ($\n_T$). In every table, each row except the first and each column is labeled by a circuit. OPF24 and TMCVH19 denote the circuits in \cite{2024_OPF} and \cite{2019_TMVH}, respectively. Each cell gives the ratio of the metric for the circuit labeling its column to that for the circuit labeling its row. For example, in a T-count table, the cell at column $i$ and row $j$ contains $\tcount(i)/\tcount(j)$. These ratios are computed from the bounds derived in Section \ref{sec:div} and summarized in Tables \ref{tab:compare_T}, \ref{tab:compare_q}, \ref{tab:resDiv}, and \ref{tab:nonRestore}. We assume that $n-m+1>0$ and $n-m'+1>0$ and take the limit $n\rightarrow\infty$.

\textbf{Long-division algorithm:} Figure \ref{tab:long_div_T} shows that LONG-DIV circuit III has the lowest T-count, approximately 76.08\% lower than that of \cite{2024_OPF}, the previous best long-division circuit. Among the reversible circuits, LONG-DIV circuit I has the lowest T-count, a 60.94\% reduction relative to \cite{2024_OPF}. Both circuits also significantly reduce the CNOT count relative to \cite{2024_OPF} (Figure \ref{tab:long_div_N}), by approximately 68.35\% and 65.03\% for circuits III and I, respectively. Circuit IIb has the smallest T-depth (Table \ref{tab:compare_T}), improving on \cite{2024_OPF} by a factor of $O\left(m/\log m\right)$. Circuit I has the smallest qubit count (Table \ref{tab:compare_q}) because it uses no extra ancillas. We are not aware of any work that establishes the optimal qubit count of a divider circuit. Because $n+m$ qubits are required to store the dividend and divisor, the qubit count of LONG-DIV circuit I may be optimal.

\textbf{Restoring division algorithm:} From the tables in Figures \ref{tab:restore_div_T} and \ref{tab:restore_div_N} and Table \ref{tab:compare_T}, we observe that, for $m\in O(n^{1-\delta})$, where $\delta\rightarrow0$, all our RES-DIV circuits achieve asymptotic reductions in T-count, T-depth, and CNOT count relative to the previous best restoring-division circuit from \cite{2019_TMVH}. Circuit I has the smallest qubit count (Table \ref{tab:compare_q}), circuit III has the smallest T-count, and circuit IIb has a significantly smaller T-depth (Table \ref{tab:compare_T}) than the other circuits.

\textbf{Non-restoring division algorithm:} From the tables in Figures \ref{tab:nonrestore_div_T} and \ref{tab:nonrestore_div_N} and Table \ref{tab:compare_q}, we observe that, for $m\in O(n^{1-\delta})$, where $\delta\rightarrow0$, all three NON-RES-DIV circuits achieve asymptotic reductions in the stated cost metrics relative to the previous best circuit in \cite{2019_TMVH}. Circuit-1 has the smallest qubit count, circuit-2 has the smallest T-depth, and circuit-3 has the smallest T-count and CNOT count.

Hence, for all three types of algorithms, the circuits derived in this paper achieve significant reductions in the various cost metrics relative to the previous best circuits in the literature. We now select the best circuit for each cost metric. For this purpose, we use the bounds derived in Section \ref{sec:div} and summarized to leading order in Tables \ref{tab:compare_T} and \ref{tab:compare_q}.

\textbf{T-count and CNOT-count:} NON-RES-DIV circuit-3 has the smallest T-count and CNOT count among all the circuits, while NON-RES-DIV circuit-1 has the smallest T-count and CNOT count among the fully reversible circuits.

\textbf{T-depth:} The T-depths of LONG-DIV circuit IIb, RES-DIV circuit IIb, and NON-RES-DIV circuit-2 are comparable and are the smallest.

\textbf{Qubit-count:} LONG-DIV circuit I, RES-DIV circuit I, and NON-RES-DIV circuit-1 have comparable qubit counts. More specifically, LONG-DIV circuit I uses one fewer qubit than the others and, as noted above, may achieve the optimal qubit count.

\subsubsection*{Fault-tolerant implementation}

\begin{table}[h]
    \scriptsize
    \centering
    \begin{tabular}{|c|c|c|c|c|c|c|}
    \hline
    {\bf Algorithm} & {\bf Circuit} &  & {\bf \# Physical qubits} & {\bf Time} & $\mathbf{d}$ & $\mathbf{d_c}$    \\
      \hline\hline
      \multirow{5}{*}{\bf Long division} & Orts et al. \cite{2024_OPF} & & $1.78\times 10^5$ & $15.79\,\mathrm{s}$ & 6 & 6 \\
        \cline{2-7}
        &  $LD_I$ ({\bf This paper}) & & $6.36\times 10^4$ & $7\,\mathrm{s}$ & 6 & 5 \\
        \cline{2-7}
        & $LD_{IIa}$  ({\bf This paper}) & & $1.84\times 10^5$ & $2.75\,\mathrm{s}$ & 6 & 6   \\
        \cline{2-7}
        & $LD_{IIb}$  ({\bf This paper}) & & $4.63\times 10^5$ & $0.379\,\mathrm{s}$ & 6 & 6   \\
        \cline{2-7}
        & $LD_{III}$  ({\bf This paper}) & & $8.56\times 10^4$ & $3.97\,\mathrm{s}$ & 6 & 5   \\
       \hline\hline
       \multirow{5}{*}{\bf Restoring division} & Thapliyal et al. \cite{2019_TMVH} & & $1.78\times 10^5$ & $47.3\,\mathrm{s}$ & 6 & 6  \\
       \cline{2-7}
        & $Res_I$ ({\bf This paper}) & $m'=1$ & $1.19\times 10^5$ & $14.19\,\mathrm{s}$ & 6 & 6  \\
        \cline{2-7}
        & $Res_{IIa}$ ({\bf This paper}) & $m'=1$ & $2.13\times 10^5$ & $5.47\,\mathrm{s}$ & 6 & 6    \\
        \cline{2-7}
        &  $Res_{IIb}$  ({\bf This paper}) & $m'=1$ & $4.914\times 10^5$ & $0.756\,\mathrm{s}$ & 6 & 6 \\
        \cline{2-7}
        & $Res_{III}$  ({\bf This paper}) & $m'=1$ & $1.495\times 10^5$ & $7.89\,\mathrm{s}$ & 6 & 6 \\
       \hline\hline
       \multirow{3}{*}{\bf Non-restoring division} & Thapliyal et al. \cite{2019_TMVH} & & $1.782\times 10^5$ & $19\,\mathrm{s}$ & 6 & 6   \\
       \cline{2-7}
       & \multirow{1}{*}{$NRes_1$  ({\bf This paper})} & $m'=m$ & $6.57\times 10^4$ & $4.79\,\mathrm{s}$ & 6 & 5  \\
       \cline{3-7}
       & & $m'=1$ & $9.23\times 10^4$ & $9.548\,\mathrm{s}$ & 6 & 6  \\
       \cline{2-7}
       & \multirow{1}{*}{ $NRes_2$ ({\bf This paper})} & $m'=m$ & $4.45\times 10^5$ & $0.371\,\mathrm{s}$ & 6 & 6  \\
       \cline{3-7}
       & & $m'=1$ & $4.45\times 10^5$ & $0.739\,\mathrm{s}$ & 6 & 6  \\
       \cline{2-7}
       & \multirow{1}{*}{ $NRes_3$ ({\bf This paper}) } & $m'=m$ & $ 8.29\times 10^4$ & $1.334\,\mathrm{s}$ & 5 & 5  \\
       \cline{3-7}
       & & $m'=1$ & $ 1.19\times 10^5$ & $3.173\,\mathrm{s}$ & 6 & 6 \\
       \hline
    \end{tabular}
    \caption{Fault-tolerant resource estimates for various division circuits encoded with the surface code. The quantities $d$ and $d_c$ are the code distances used for non-Clifford and Clifford gates, respectively.}
    \label{tab:ft}
\end{table}

Following the procedures in \cite{2012_FMMC}, we estimate the resources required to implement the various division circuits fault-tolerantly with the surface code. We consider dividing a 512-bit dividend by a 256-bit divisor, so $n=512$ and $m=256$. We assume a magic-state injection error rate $p_{in}=10^{-4}$, a per-gate error rate $p_g=10^{-5}$, a surface-code cycle time $t_{sc}=200\,\mathrm{ns}$, and a measurement time $t_M=100\,\mathrm{ns}$.

\paragraph{LONG-DIV-I:} First, we consider LONG-DIV circuit I. From Equations \ref{eqn:t_LDI} and \ref{eqn:tdepth_LDI}, the T-count and T-depth are bounded as follows.
\begin{eqnarray}
    \tcount(LD_I) \leq 1.188\times 10^6;\qquad \tcount_d(LD_I)\leq 5.937\times 10^5.  \nonumber
\end{eqnarray}
Each T gate requires one magic state. The output error rate for state distillation is at most
\begin{eqnarray}
    p_{out} = \frac{1}{\tcount(LD_I)} = 8.419\times 10^{-7}.  \nonumber
\end{eqnarray}
To calculate the surface-code distance, we use Algorithm 4 in \cite{2016_AdMGetal} with parameter $\epsilon=1$. For the stated values of $p_{in}$ and $p_g$, the algorithm suggests one layer of distillation at distance $d=6$. This stage requires 16 logical qubits to generate one magic state. One logical qubit requires $N_p=2.5\times1.25\times d^2$ physical qubits. Thus, the total is approximately $16\times N_p=1800$ physical qubits, and one round of distillation takes $\sigma=10d=60$ surface-code cycles. One magic state is therefore produced every $\sigma t_{sc}=1.2\times10^{-5}\,\mathrm{s}$. We require $\tcount(LD_I)/\tcount_d(LD_I)\approx2$ magic states in parallel per T layer. Consequently, the number of physical qubits required to implement the non-Clifford gates is
\begin{eqnarray}
    2 \times 1800 = 3600, \label{eqn:phys_LDI}
\end{eqnarray}
and the execution time is
\begin{eqnarray}
    \tcount_d(LD_I)\times 1.2\times 10^{-5}\,\mathrm{s} \approx 7\,\mathrm{s}. \label{eqn:time_LDI}
\end{eqnarray}

Next, we consider the Clifford-gate implementation. Because the Clifford gates in our circuits are mostly CNOT gates, we estimate the cost of implementing them. From Equation \ref{eqn:cnot_t_LDI}, the CNOT count and qubit count are bounded as follows.
\begin{eqnarray}
    \n_T(LD_I) \leq 1.648\times 10^6; \qquad \mathcal{Q}(LD_I) = n+m+3 = 771.  \nonumber
\end{eqnarray}
Thus, the overall error rate of the Clifford gates should be less than $\frac{1}{\n_T(LD_I)} = \frac{1}{1.648\times 10^6}\approx 6.07\times 10^{-7}$. The distance is the smallest $d_c$ such that
\begin{eqnarray}
    \left( \frac{p_{in}}{ 0.0125 } \right)^{\frac{d_c+1}{2}} = 0.008^{\frac{d_c+1}{2}} < 6.07\times 10^{-7}, \label{eqn:dc}
\end{eqnarray}
and hence $d_c=5$. Encoding the Clifford operations requires at most $(n+m+3)\times2.5\times1.25\times5^2=60{,}000$ physical qubits. Thus, implementing LONG-DIV-I requires approximately $60{,}000+3{,}600=63{,}600$ physical qubits in total. We expect to perform approximately $\n_T(LD_I)/((n+m+3)\tcount_d(LD_I))\approx0.004$ Clifford operations per qubit per T layer. Each logical CNOT takes two surface-code cycles, whereas each T layer requires $\sigma=60$ surface-code cycles, so the running time is dominated by the non-Clifford T gates.

In summary, implementing LONG-DIV circuit I with $n=512$ and $m=256$ requires approximately $6.36\times10^4$ physical qubits and $7\,\mathrm{s}$.

\paragraph{LONG-DIV-IIb:} A similar analysis for LONG-DIV circuit IIb gives the following bounds on its T-count (Equation \ref{eqn:t_LDIIb}), T-depth (Equation \ref{eqn:tdepth_LDIIb}), output error rate for distillation, CNOT count (Equation \ref{eqn:cnot_t_LDIIb}), and qubit count.
\begin{eqnarray}
    &&\tcount(LD_{IIb}) \leq 5.08\times 10^6 ;\qquad \tcount_d(LD_{IIb})\leq 31611  ;\qquad p_{out} = \frac{1}{\tcount(LD_{IIb})} \approx 1.968\times 10^{-7}   \nonumber   \\
    &&\n_T(LD_{IIb})\leq 5.61\times 10^6 ;\qquad \mathcal{Q}(LD_{IIb})\leq 1541 \nonumber 
\end{eqnarray}
Using Algorithm 4 of \cite{2016_AdMGetal}, we apply one layer of distillation at distance $d=6$. This stage requires $16\times2.5\times1.25\times6^2=1800$ physical qubits, and one round takes $\sigma=10d=60$ surface-code cycles. Thus, one magic state is produced every $\sigma t_{sc}=60\times200\times10^{-9}\,\mathrm{s}=1.2\times10^{-5}\,\mathrm{s}$. We require $\tcount(LD_{IIb})/\tcount_d(LD_{IIb})\approx161$ magic states in parallel per T layer. Implementing the non-Clifford gates therefore requires $161\times1800=289{,}800$ physical qubits and takes $\tcount_d(LD_{IIb})\times1.2\times10^{-5}\,\mathrm{s}=0.379\,\mathrm{s}$.

Because the Clifford gates in this circuit are mostly CNOT gates, their overall error rate should be less than $1/\n_T(LD_{IIb})=1.784\times10^{-7}$. The distance $d_c$ is therefore the smallest integer satisfying the following inequality.
\begin{eqnarray}
    0.008^{\frac{d_c+1}{2}} < 1.784\times 10^{-7}.  \nonumber
\end{eqnarray}
We obtain $d_c=6$. Encoding the Clifford operations requires at most $\mathcal{Q}(LD_{IIb})\times2.5\times1.25\times d_c^2\approx173{,}362$ physical qubits, bringing the total to $289{,}800+173{,}362\approx4.63\times10^5$. We expect to perform approximately $\n_T(LD_{IIb})/(\mathcal{Q}(LD_{IIb})\tcount_d(LD_{IIb}))\approx0.115$ Clifford operations per qubit per T layer. Each logical CNOT takes two surface-code cycles, whereas each T layer requires $\sigma=60$ surface-code cycles, so the running time is dominated by the non-Clifford T gates.

In summary, implementing LONG-DIV circuit IIb with $n=512$ and $m=256$ requires approximately $4.63\times10^5$ physical qubits and $0.379\,\mathrm{s}$.

We apply similar procedures to estimate the physical-resource requirements of the other division circuits, as summarized in Table \ref{tab:ft}. NON-RES-DIV circuit-2, which has the minimum T-depth, requires the least time. LONG-DIV circuit IIb and RES-DIV circuit IIb are also significantly faster than the other circuits. These circuits achieve an asymptotic reduction in T-depth relative to the remaining circuits (Table \ref{tab:compare_T}). LONG-DIV circuit I requires the fewest physical qubits, closely followed by NON-RES-DIV circuit-1. Their qubit requirements are nearly optimal (Table \ref{tab:compare_q}); in fact, both use only $O(1)$ ancillas.

\section{Discussion}
\label{sec:conclude}

 In this paper, we have designed novel quantum circuits for several division algorithms. We have analyzed the resource requirements of each circuit and compared them with those of the best circuits in the literature. We have focused on cost metrics such as T-count, T-depth, CNOT-count, qubit count, and full reversibility because these are important from a fault-tolerant perspective. Each of our circuits outperforms previous implementations, achieving asymptotic reductions in these metrics. Two main factors contribute to these improvements. First, we modify the algorithms to reuse qubits, ensure that intermediate operations act on fewer qubits, and reduce the number of iterations, thereby lowering the qubit requirements and gate complexity. Second, we introduce the COMP-N-SUB primitive, which can be implemented at roughly the same cost as an ordinary addition of two integers. Expressing the division algorithms in terms of this primitive further reduces gate complexity. We have estimated the number of physical qubits required to implement the various circuits using the surface code when the dividend and divisor are 512 and 256 bits long, respectively. We have also estimated the time required to perform division using the encoded circuits. Our circuits outperform previous ones by a significant margin with respect to these metrics as well. 

To the best of our knowledge, no QFT-based quantum divider circuit has been constructed. Although we expect such circuits to use many rotation gates and therefore to be inefficient for fault-tolerant implementation, it would be interesting to develop one and compare its resource requirements with those of existing circuits. In particular, it would be interesting to develop a QFT-based version of COMP-N-SUB, which plays a crucial role in the implementation of our divider circuits. Comparing the noise sensitivity of the various circuits is another interesting direction.

 We also envision using COMP-N-SUB in other algorithms, for instance to compute the greatest common divisor, which is used in quantum algorithms for computing discrete logarithms on elliptic curves~\cite{cryptoeprint:2026/280} and for Jacobi factoring~\cite{2025_KRVK}. A detailed resource analysis of this application appears in Appendix~\ref{app:gcd}. We conclude by illustrating how COMP-N-SUB can be used to implement Stein's GCD algorithm. We do not optimize the complete algorithm or discuss its efficient implementation in detail. Instead, our aim is to highlight the usefulness of COMP-N-SUB, so we focus on the part of the algorithm in which this primitive can be used and discuss its advantages. 

 More broadly, COMP-N-SUB illustrates a design philosophy in which a predicate and the data transformation conditioned on that predicate are treated as a single coherent operation, rather than as separate comparison and controlled-arithmetic modules. Treating such decision-and-update operations as first-class primitives can help recast classical algorithms with data-dependent branches into forms that are more natural for reversible computation. This perspective can guide the design of new quantum algorithms by encouraging the algorithm and its arithmetic primitives to be developed together, with recurring decision-and-update patterns determining the appropriate computational building blocks. 

\subsection{Application of COMP-N-SUB in the binary GCD algorithm}
\label{subsec:gcd}

\begin{algorithm}
    \scriptsize
    \caption{Binary GCD algorithm~\cite{2012_LPDS}.}
      \label{alg:gcd}

    \KwIn{Positive integers $a$ and $b$.}

    \KwOut{Positive integer $g=\gcd(a,b)$.}

    $u\leftarrow a$; $\quad v\leftarrow b$; $\quad g\leftarrow 1$   \;

    \While{$u$ and $v$ are even}
    {
        $u\leftarrow u/2$; $\quad v\leftarrow v/2$; $\quad g\leftarrow 2g$  \;
    }

    \While{$v\neq 0$}
     {
        \While{$u$ is even}
        {
            $u\leftarrow u/2$   \;
        }
        \While{$v$ is even}
        {
            $v\leftarrow v/2$   \;
          }
        \eIf{$u>v$}
        {
            $u\leftarrow (u-v)/2$   \;
        }
        {
            $v\leftarrow (v-u)/2$   \;
        }
    }
     \textbf{return} $g\cdot u$  \;
    
\end{algorithm}

\begin{figure}[h]
    \centering
    \makebox[\textwidth][c]{%
    \Qcircuit @C=0.6em @R=0.8em{
    \lstick{t_{\lceil\log n\rceil-1}} &\qw &\qw &\multigate{8}{U_{(2:4)}} &\qw &\qw &\qw &\qw &\qw &\multigate{5}{c-SHIFT} &\qw &\qw &\rstick{}  \\
    \lstick{\vdots} &\vdots & &\ghost{U_{(2:4)}} &\vdots & & &\vdots & &\ghost{c-SHIFT} &\vdots & &   \\
    \lstick{t_0} &\qw &\qw &\ghost{U_{(2:4)}} &\qw &\qw &\qw &\qw &\qw &\ghost{c-SHIFT} &\qw &\qw &\rstick{}  \\
    \lstick{u_{n-1}} &\qw &\qw &\ghost{U_{(2:4)}} &\qw &\qw &\multigate{5}{U_{(5:17)}} &\qw &\qw &\ghost{c-SHIFT} &\qw &\qw &\rstick{\ket{g_{n-1}}}     \\
    \lstick{\vdots} &\vdots & &\ghost{U_{(2:4)}} &\vdots & &\ghost{U_{(5:17)}} &\vdots & &\ghost{c-SHIFT} &\vdots & &    \\
    \lstick{u_0} &\qw &\qw &\ghost{U_{(2:4)}} &\qw &\qw &\ghost{U_{(5:17)}} &\qw &\qw &\ghost{c-SHIFT} &\qw &\qw &\rstick{\ket{g_0}}     \\
    \lstick{v_{n-1}} &\qw &\qw &\ghost{U_{(2:4)}} &\qw &\qw &\ghost{U_{(5:17)}} &\qw &\qw &\qw &\qw &\qw &\rstick{\ket{0}}  \\
    \lstick{\vdots} &\vdots & &\ghost{U_{(2:4)}} &\vdots & &\ghost{U_{(5:17)}} &\vdots & & &\vdots & &  \\
    \lstick{v_0} &\qw &\qw &\ghost{U_{(2:4)}} &\qw &\qw &\ghost{U_{(5:17)}} &\qw &\qw &\qw &\qw &\qw &\rstick{\ket{0}} 
    }
    }
    \caption{A quantum circuit implementing the binary GCD algorithm specified by the pseudocode in Algorithm~\ref{alg:gcd}. The input integers $u$ and $v$ are stored in the $n$-qubit registers $(u_{n-1},\ldots,u_0)$ and $(v_{n-1},\ldots,v_0)$, respectively. The unitaries $U_{(2:4)}$ and $U_{(5:17)}$ implement the loops in steps 2--4 and 5--17, respectively. The register $(t_{\lceil\log n\rceil-1},\ldots,t_0)$ stores the number $t$ of iterations of the loop in steps 2--4. The unitary c-SHIFT cyclically shifts the contents of register $u$ by $t$ places. The $n$-bit GCD $g$ is output in register $u$. Other qubits are omitted for clarity.}
    \label{fig:gcd}
\end{figure}

\begin{figure}[h]
    \centering
    \makebox[\textwidth][c]{%
    \Qcircuit @C=0.6em @R=0.8em{
    \lstick{u_{n-1}} &\qw &\qw &\multigate{2}{U_{odd}} &\qw &\qw &\multigate{6}{CNS} &\qw &\targ &\qw &\qw &\qw &\qw &\multigate{5}{c-ADD} &\qw &\targ &\qw &\qw &\qw &\qw    \\
    \lstick{\vdots} &\vdots & &\ghost{U_{odd}} &\vdots & &\ghost{CNS} &\vdots & & &\vdots & & &\ghost{c-ADD} &\vdots & & &\vdots & &   \\
    \lstick{u_0} &\qw &\qw &\ghost{U_{odd}} &\qw &\qw &\ghost{CNS} &\qw &\qw &\qw &\qw &\targ &\qw &\ghost{c-ADD} &\qw &\qw &\qw &\qw &\targ &\qw    \\
    \lstick{v_{n-1}} &\qw &\qw &\multigate{2}{U_{odd}} &\qw &\qw &\ghost{CNS} &\qw &\qw &\qw &\qw &\qw &\qw &\ghost{c-ADD} &\qw &\qw &\qw &\qw &\qw &\qw \\
    \lstick{\vdots} &\vdots & &\ghost{U_{odd}} &\vdots & &\ghost{CNS} &\vdots & & &\ddots & & &\ghost{c-ADD} &\vdots & & &\ddots & &  \\
    \lstick{v_0} &\qw &\qw &\ghost{U_{odd}} &\qw &\qw &\ghost{CNS} &\qw &\qw &\qw &\qw &\qw &\qw &\ghost{c-ADD} &\qw &\qw &\qw &\qw &\qw &\qw    \\
    \lstick{\ket{0}_c} &\qw &\qw &\qw &\qw &\qw &\ghost{CNS} &\qw &\ctrl{-6} &\qw &\qw &\ctrl{-4} &\qw &\ctrl{-1} &\qw &\ctrl{-6} &\qw &\qw &\ctrl{-4} &\qw 
    }
    }
    \caption{A quantum circuit implementing one iteration of the loop in steps 5--17 of Algorithm~\ref{alg:gcd}. The unitary $U_{odd}$ removes the trailing zeros from an input integer. The CNS unitary is a COMP-N-SUB unitary that computes the difference in register $v$ if and only if $v\geq u$. The controlled adder c-ADD conditionally computes a sum in register $u$. Other qubits are omitted for clarity.}
    \label{fig:gcd_iter}
\end{figure}

 In this section, we discuss how COMP-N-SUB unitaries can be used to implement the binary GCD algorithm whose pseudocode is given in Algorithm~\ref{alg:gcd}. This popular GCD algorithm is often referred to as Stein's algorithm. We follow the pseudocode given in \cite{2012_LPDS}. The inputs are two $n$-bit integers, $u$ and $v$, and the output is their GCD, another $n$-bit integer denoted by $g$. 

 We emphasize at the outset that several variants of the classical GCD algorithm appear in the literature~\cite{1993_J, 1994_S, 1995_W, 2004_SZ, 2008_M, 2019_BY}. A thorough comparative study of these algorithms and their quantum implementations is beyond the scope of this work. Our goal is to demonstrate the usefulness of the COMP-N-SUB unitary developed here. We do so by considering a widely used algorithm and constructing a quantum circuit for it. We highlight the steps that use COMP-N-SUB and explain its advantage over alternative implementations of those steps. Because the remaining steps are implemented identically, this approach permits a fair comparison of the resource reductions provided by COMP-N-SUB. For completeness, detailed resource estimates for a complete circuit implementation are given in Appendix~\ref{app:gcd}. 

 Figure~\ref{fig:gcd} illustrates a quantum circuit that implements Algorithm~\ref{alg:gcd}. We allocate two $n$-qubit input registers, $(u_{n-1},\ldots,u_0)$ and $(v_{n-1},\ldots,v_0)$, which store the input integers $u$ and $v$, respectively. With a slight abuse of notation, we also denote these registers by $u$ and $v$. The register $(t_{\lceil\log n\rceil-1},\ldots,t_0)$ stores the number $t$ of iterations of the loop in steps 2--4. The unitaries $U_{(2:4)}$ and $U_{(5:17)}$ implement the loops in steps 2--4 and 5--17, respectively. In Algorithm~\ref{alg:gcd}, the integer $g$, initialized to $1$, is updated in every iteration of the first loop. Instead of updating $g$ in each iteration, it is sufficient to store $t$ and later apply the unitary c-SHIFT, which cyclically shifts the bits in register $u$ by $t$ places. This operation implements step 18 and yields the GCD $g=(g_{n-1},\ldots,g_0)$ in register $u$. 

 One iteration of the loop in steps 5--17 can be implemented by the circuit shown in Figure~\ref{fig:gcd_iter}. The unitary $U_{odd}$ removes the trailing zeros from the binary representation of an integer. It is applied to both registers $u$ and $v$ to implement steps 6--8 and 9--11. CNS is a COMP-N-SUB unitary, and c-ADD is a controlled adder. CNS computes $v-u$ in register $v$ if and only if $v\geq u$; otherwise, it leaves the registers unchanged and flips qubit $c$ to $\ket{1}$. Together, the remaining CNOT gates and c-ADD implement controlled subtraction, computing $u-v$ in register $u$ if and only if $u>v$. Thus, CNS, c-ADD, and the CNOT gates implement steps 12--16. If the loop in steps 5--17 has $N_{iter}$ iterations, this circuit must be repeated $N_{iter}$ times. 

 We now describe several unitaries used to implement $U_{(2:4)}$ and $U_{(5:17)}$. 
\begin{itemize}
    \item \textbf{$\mathbf{U_{odd}^{(2)}}$:} Let $u$ and $v$ be two $n$-bit integers such that $u=2^t\cdot u'$ and $v=2^t\cdot v'$ for integers $u'$ and $v'$, at least one of which is odd. Thus, the trailing $t$ bits of both $u$ and $v$ are $0$, and the next bit is $1$ in at least one of them. Therefore, $t$ is the number of iterations of the loop in steps 2--4 of Algorithm~\ref{alg:gcd}. Our goal is to compute $t$, $u'$, and $v'$. Appendix~\ref{app:Upair} describes a quantum circuit implementing a unitary that takes $u$, $v$, and $\ket{0}^{\lceil\log n\rceil}$ as input and outputs $u'$, $v'$, and $t$ in the corresponding registers. We denote this unitary by $U_{odd}^{(2)}$.

    \item \textbf{$\mathbf{U_{odd}}$:} This unitary takes as input an $n$-bit integer $v$ such that $v=2^{t'}\cdot v'$ for some odd integer $v'$ and nonnegative integer $t'$. It outputs $v'$ in the register storing $v$ and $t'$ in a separate register. A quantum circuit implementing $U_{odd}$ is described in Appendix~\ref{app:Uodd}.

    \item \textbf{c-SHIFT:} We described the operation of c-SHIFT above. A quantum circuit implementing c-SHIFT is described in Appendix~\ref{app:Uodd}. Specifically, it is circuit $\mathcal{C}_{shift}$, the boxed portion of Figure~\ref{fig:Uodd}.

    \item \textbf{c-ADD:} This controlled adder has two $n$-qubit input registers that store two $n$-bit integers. Conditioned on another qubit $c$, it adds these integers and stores the sum in one of the input registers. An implementation of c-ADD using COMP-N-SUB is described in Section~\ref{subsec:nonRestoreDiv}.
    
\end{itemize}

 Appendix~\ref{app:gcd} provides a thorough analysis of the resource requirements of the GCD circuit in Figure~\ref{fig:gcd}. We have designed several circuits for the COMP-N-SUB unitary (Section~\ref{sec:cns}) that offer different resource trade-offs. A GCD circuit can be implemented using any of these variants. We refer to the resulting GCD implementations as circuits I, IIa, IIb, and III. 

\subsubsection*{Advantage of COMP-N-SUB}

 The COMP-N-SUB unitary is used to implement steps 12--16 of Algorithm~\ref{alg:gcd}. Specifically, it is used twice in every iteration of the loop in steps 5--17 (Figure~\ref{fig:gcd_iter}). Alternatively, steps 12--16 can be implemented with a comparator and two controlled subtractors, as follows. 
\begin{enumerate}
    \item Apply a comparator to registers $u$ and $v$, using an additional qubit $c$ initialized to $\ket{0}$. After the comparison, $c=\ket{0}$ if $v\geq u$; otherwise, $c=\ket{1}$.

    \item Apply a controlled subtractor to $u$ and $v$, controlled on the $\ket{0}$ state of $c$. The difference $v-u$ is stored in register $v$ if $v\geq u$.

    \item Apply a controlled subtractor to $u$ and $v$, controlled on the $\ket{1}$ state of $c$. The difference $u-v$ is stored in register $u$ if $u>v$.
\end{enumerate}
 As explained earlier, we assume that the cost of a controlled subtractor is approximately equal to that of COMP-N-SUB. The alternative implementation above therefore uses one extra comparator per iteration. Because a comparator costs roughly as much as an adder, our implementation saves approximately the cost of one adder each time steps 12--16 are executed. Specifically, let $\tcount(.)$ denote the T-count of a circuit. Under these approximate cost assumptions, the ratio of the T-count of our implementation to that of the alternative implementation is 
\begin{eqnarray}
   \frac{2\tcount(\cns)}{\tcount(\text{ADDER})+2\tcount(\cns)} \approx \frac{3}{4}.   \nonumber
\end{eqnarray}
A similar conclusion holds for other cost metrics, such as Toffoli-count, CNOT-count, Toffoli-depth, and T-depth.\footnote{For a QRC adder, we compare against COMP-N-SUB circuit I; for a QCLA adder, we compare against COMP-N-SUB circuit IIa or IIb; and for a logical-AND-based adder, we compare against COMP-N-SUB circuit III.} The qubit count, however, remains the same. 

 There can be at most $2n$ executions of steps 12--16 because, in each execution, the bit length of the integer in at least one of the two input registers decreases by at least one. Fewer executions are possible, but this does not affect the ratio above. Further optimizations may be possible when implementing the remaining steps of Algorithm~\ref{alg:gcd}. Because our focus is the usefulness of COMP-N-SUB, we restrict the comparison to the steps in which it is used. We do not quantify the percentage savings for the complete circuit because this figure depends heavily on the implementation of the other steps, which we have not optimized. Such an analysis is beyond the scope of this work. 


\section*{Acknowledgments}

 We thank Robbie King and Matt Pocrnic for helpful suggestions. We received no support from external funding agencies. We also acknowledge the use of ChatGPT 5.6 for fixes in typos, grammatical errors and copyediting.

\section*{Competing Interests}

The authors have no competing interests to declare.

\section*{Code and Data Availability}

All relevant data are included in the manuscript.

\section*{Author Contributions}

 All authors developed the initial ideas, discussed the results, and prepared the manuscript. P.M. designed the circuits and estimated their resource requirements.

\bibliographystyle{unsrt}
\bibliography{arith_ref}

\newpage 

\appendix

\section*{Appendix}

\section{Proofs of correctness}
\label{app:correct}

 In this section, we prove in detail that the COMP-N-SUB circuits described in Section~\ref{sec:cns} are correct; that is, they produce the desired output states from a given initial state and therefore have the intended functionality.

\subsection{Correctness of COMP-N-SUB circuit I}
\label{app:correct-I}

 Consider the COMP-N-SUB circuit described in Section~\ref{subsec:cns-I} and illustrated in Figure~\ref{fig:compNsub}. We track the changes in the states of the input qubits after each step. The combined initial state of registers $B$ and $A$ and qubit $Z$ is
    \begin{eqnarray}
        \left(\bigotimes_{i=0}^{k-1}\ket{b_i}_{B_i}\ket{a_i}_{A_i}\right)\ket{0}_Z. \nonumber
    \end{eqnarray}

\begin{enumerate}
    \item  After applying the X gates, the state is
    \begin{eqnarray}
        \left(\bigotimes_{i=0}^{k-1}\ket{\conj{b_i}}_{B_i}\ket{a_i}_{A_i}\right)\ket{0}_Z.  \nonumber
        \label{eqn:csa_qrc_1}
    \end{eqnarray}

    \item  After applying the CNOT layer, the state is
    \begin{eqnarray}
     \ket{\conj{b_0}}_{B_0}\ket{a_0}_{A_0}\left( \bigotimes_{i=1}^{k-1}\ket{\conj{b_i}\oplus a_i}_{B_i}\ket{a_i}_{A_i} \right)\ket{0}_Z. \nonumber
    \label{eqn:csa_qrc_2}
\end{eqnarray}

    \item \begin{enumerate}
        \item  After applying $\CNOT_{(A_{k-1};Z)}$, the state is
        \begin{eqnarray}
     \ket{\conj{b_0}}_{B_0}\ket{a_0}_{A_0}\left( \bigotimes_{i=1}^{k-1}\ket{\conj{b_i}\oplus a_i}_{B_i}\ket{a_i}_{A_i} \right)\ket{a_{k-1}}_Z. \nonumber
    \label{eqn:csa_qrc_3a}
\end{eqnarray}

        \item  After applying the CNOT ladder, the state is
        \begin{eqnarray}
    \ket{\conj{b_0}}_{B_0}\ket{a_0}_{A_0}\ket{\conj{b_1}\oplus a_1}_{B_1}\ket{a_1}_{A_1}\left( \bigotimes_{i=2}^{k-1}\ket{\conj{b_i}\oplus a_i}_{B_i}\ket{a_i\oplus a_{i-1}}_{A_i} \right)\ket{a_{k-1}}_Z .  \nonumber
    \label{eqn:csa_qrc_3b} 
\end{eqnarray}
    \end{enumerate}

    \item \begin{enumerate}
        \item  After applying the Toffoli ladder, the state is
        \begin{eqnarray}
           && \ket{\conj{b_0}}_{B_0}\ket{a_0}_{A_0}\ket{\conj{b_1}\oplus a_1}_{B_1}\ket{a_0\conj{b_0}\oplus a_1}_{A_1}  \nonumber\\
           &&\qquad\cdot\left(\bigotimes_{i=2}^{k-1}\ket{\conj{b_i}\oplus a_i}_{B_i}\ket{ (\conj{b_{i-1}}\oplus a_{i-1})(c_{i-1}\oplus a_{i-1})\oplus a_i\oplus a_{i-1} }_{A_i}  \right)\ket{a_{k-1}}_Z   \nonumber\\
           &=& \ket{\conj{b_0}}_{B_0}\ket{a_0}_{A_0}\ket{\conj{b_1}\oplus a_1}_{B_1}\ket{c_1\oplus a_1}_{A_1}\left(\bigotimes_{i=2}^{k-1}\ket{\conj{b_i}\oplus a_i}_{B_i}\ket{ c_i\oplus a_i }_{A_i}  \right)\ket{a_{k-1}}_Z .  \nonumber
            \label{eqn:csa_qrc_4a}
        \end{eqnarray}

        \item After applying $\tof_{(B_{k-1},A_{k-1};Z)}$, the state is
        \begin{eqnarray}
          &&  \ket{\conj{b_0}}_{B_0}\ket{a_0}_{A_0}\left(\bigotimes_{i=1}^{k-1}\ket{\conj{b_i}\oplus a_i}_{B_i}\ket{ c_i\oplus a_i }_{A_i}  \right)\ket{ (\conj{b_{k-1}}\oplus a_{k-1})(c_{k-1}\oplus a_{k-1})\oplus a_{k-1}  }_Z     \nonumber \\
          &=& \ket{\conj{b_0}}_{B_0}\ket{a_0}_{A_0}\left(\bigotimes_{i=1}^{k-1}\ket{\conj{b_i}\oplus a_i}_{B_i}\ket{ c_i\oplus a_i }_{A_i}  \right)\ket{ c_k  }_Z.    \nonumber
          \label{eqn:csa_qrc_4b}
        \end{eqnarray}
    \end{enumerate}

    \item \begin{enumerate}
        \item  After applying $\tof_{(\conj{Z},A_{k-1};B_{k-1})}$, the state is as follows.
        \begin{eqnarray}
            && \ket{\conj{b_0}}_{B_0}\ket{a_0}_{A_0}\left(\bigotimes_{i=1}^{k-2}\ket{\conj{b_i}\oplus a_i}_{B_i}\ket{ c_i\oplus a_i }_{A_i} \right) \nonumber \\
            &&\cdot\ket{\conj{c_k}(c_{k-1}\oplus a_{k-1})\oplus \conj{b_{k-1}}\oplus a_{k-1}}_{B_{k-1}}\ket{c_{k-1}\oplus a_{k-1}}_{A_{k-1}} \ket{ c_k  }_Z \nonumber
        \end{eqnarray}

        \item  After applying the $i$th pair of Toffoli gates, the state changes are
        \begin{eqnarray}
        \ket{c_{k-i+1}\oplus a_{k-i+1}}&\mapsto&\ket{(\conj{b_{k-i}}\oplus a_{k-i})(c_{k-i}\oplus a_{k-i})\oplus c_{k-i+1} \oplus a_{k-i+1} }   \nonumber \\
    &=&\ket{ a_{k-i}\conj{b_{k-i }}\oplus \conj{b_{k-i }}c_{k-i }\oplus a_{k-i }c_{k-i }\oplus a_{k-i}\oplus c_{k-i+1 }\oplus a_{k-i+1} }    \nonumber \\
    &=& \ket{a_{k-i}\oplus a_{k-i+1}} \qquad [i = 2,\ldots,k-1]   \nonumber \\
    \ket{\conj{b_{k-i}}\oplus a_{k-i}}&\mapsto& \ket{\conj{c_k} (c_{k-i}\oplus a_{k-i})\oplus \conj{b_{k-i}} \oplus a_{k-i} } \qquad [i = 2,\ldots, k-1] .    \nonumber 
        \end{eqnarray}
     After applying the last pair of Toffoli gates, the state changes are
    \begin{eqnarray}
        \ket{a_1\oplus c_1}&\mapsto&\ket{a_0\conj{b_0}\oplus a_1\oplus c_1} = \ket{a_1}  \nonumber \\
        \ket{\conj{b_0}}&\mapsto&\ket{\conj{c_k}a_0\oplus\conj{b_0}} .   \nonumber
    \end{eqnarray}
     Thus, after applying the ladder of Toffoli pairs, the state is as follows.
    \begin{eqnarray}
        && \ket{\conj{c_k}a_0\oplus\conj{b_0} }_{B_0}\ket{a_0}_{A_0}\ket{\conj{c_k}(c_1\oplus a_1)\oplus \conj{b_1}\oplus a_1}_{B_1}\ket{a_1}_{A_1} \nonumber \\
        &&\cdot\left( \bigotimes_{i=2}^{k-1}\ket{ \conj{c_k}(c_i\oplus a_i)\oplus \conj{b_i}\oplus a_i }_{B_i} \ket{a_i\oplus a_{i-1}}_{A_i}  \right)\ket{c_k}_Z   \nonumber
    \end{eqnarray}
    \end{enumerate}

    \item  After applying the second CNOT ladder, the state is
    \begin{eqnarray}
         \ket{\conj{c_k}a_0\oplus\conj{b_0} }_{B_0}\ket{a_0}_{A_0}\ket{\conj{c_k}(c_1\oplus a_1)\oplus \conj{b_1}\oplus a_1}_{B_1}\ket{a_1}_{A_1} \left( \bigotimes_{i=2}^{k-1}\ket{ \conj{c_k}(c_i\oplus a_i)\oplus \conj{b_i}\oplus a_i }_{B_i} \ket{a_i}_{A_i}  \right)\ket{c_k}_Z.    \nonumber
    \end{eqnarray}

    \item  After applying the second CNOT layer, the state is
    \begin{eqnarray}
        \ket{\conj{c_k}a_0\oplus \conj{b_0} }_{B_0}\ket{a_0}_{A_0}\left( \bigotimes_{i=1}^{k-1}\ket{ \conj{c_k}(c_i\oplus a_i)\oplus \conj{b_i}  }_{B_i} \ket{a_i}_{A_i}  \right)\ket{c_k}_Z. \nonumber
    \end{eqnarray}
 Now, $\conj{c_k}(c_i\oplus a_i)\oplus\conj{b_i} = \conj{c_k}(c_i\oplus a_i\oplus\conj{b_i})\oplus \conj{b_i}(1\oplus\conj{c_k}) = \conj{c_k}s_i\oplus c_k\conj{b_i}$. Similarly, $\conj{c_k}a_0\oplus \conj{b_0} = \conj{c_k}s_0\oplus c_k\conj{b_0}$. Hence, we can rewrite the state above as
\begin{eqnarray}
        \ket{\conj{c_k}s_0\oplus c_k\conj{b_0} }_{B_0}\ket{a_0}_{A_0}\left( \bigotimes_{i=1}^{k-1}\ket{ \conj{c_k}s_i\oplus c_k\conj{b_i}  }_{B_i} \ket{a_i}_{A_i}  \right)\ket{c_k}_Z.   \nonumber
    \end{eqnarray}

    \item  We have $\conj{c_k}s_i\oplus c_k\conj{b_i} = \conj{c_k}(1\oplus\conj{s_i})\oplus c_k(1\oplus b_i) = 1\oplus \conj{c_k}\conj{s_i}\oplus c_kb_i$. Thus, the final state after applying the layer of X gates is as follows.
    \begin{eqnarray}
      &&  \ket{\conj{c_k}\conj{s_0}\oplus c_k b_0 }_{B_0}\ket{a_0}_{A_0}\left( \bigotimes_{i=1}^{k-1}\ket{ \conj{c_k}\conj{s_i}\oplus c_k b_i  }_{B_i} \ket{a_i}_{A_i}  \right)\ket{c_k}_Z    \nonumber \\
     &=& \ket{\conj{c_k}d_0\oplus c_k b_0 }_{B_0}\ket{a_0}_{A_0}\left( \bigotimes_{i=1}^{k-1}\ket{ \conj{c_k}d_i\oplus c_k b_i  }_{B_i} \ket{a_i}_{A_i}  \right)\ket{c_k}_Z \nonumber
        \label{cns_qrc_final}
    \end{eqnarray}
\end{enumerate}
 If $c_k=0$, implying $b\geq a$, we obtain $d=b-a$ in register $B$; otherwise, we obtain $b$ in register $B$. Thus, the circuit implements COMP-N-SUB as desired.

\subsection{Correctness of COMP-N-SUB circuit II}
\label{app:correct-II}

 Consider the COMP-N-SUB circuit described in Section~\ref{subsec:cns-II} and shown in Figure~\ref{fig:comPsub_qcla}. We track the changes in the states of the input qubits after each step. Only the states of the input qubits and ancilla qubits labeled $Z_i$ change during the application of unitaries $U_1$ and $U_2$. Hence, we write only the states of qubits $A_i$, $B_i$, and $Z_i$. Their combined initial state is
\begin{eqnarray}
    \bigotimes_{i=0}^{k-1}\ket{a_i}_{A_i}\ket{b_i}_{B_i}\ket{0}_{Z_i}.    \nonumber
\end{eqnarray}

\begin{enumerate}
    \item  After applying the X gates, the state is
    \begin{eqnarray}
    \bigotimes_{i=0}^{k-1}\ket{a_i}_{A_i}\ket{\conj{b_i}}_{B_i}\ket{0}_{Z_i}.    \nonumber
\end{eqnarray}

    \item  After applying the Toffoli layer, the state is
    \begin{eqnarray}
    \bigotimes_{i=0}^{k-1}\ket{a_i}_{A_i}\ket{\conj{b_i}}_{B_i}\ket{a_i\conj{b_i}}_{Z_i}.    \nonumber
\end{eqnarray}

    \item  After applying the CNOT layer, the state is
    \begin{eqnarray}
    \bigotimes_{i=0}^{k-1}\ket{a_i}_{A_i}\ket{a_i\oplus \conj{b_i}}_{B_i}\ket{a_i\conj{b_i}}_{Z_i}.    \nonumber
\end{eqnarray}

    \item  After applying $U_1$, the state is
    \begin{eqnarray}
    \bigotimes_{i=0}^{k-1}\ket{a_i}_{A_i}\ket{a_i\oplus \conj{b_i}}_{B_i}\ket{c_{i+1} }_{Z_i}.    \nonumber
\end{eqnarray}

    \item  After applying the CNOT layer, the state is
        \begin{eqnarray}
    \ket{a_0}_{A_0}\ket{a_0\oplus \conj{b_0}}_{B_0}\ket{c_1}_{Z_0}\left(\bigotimes_{i=1}^{k-1}\ket{a_i\oplus c_i}_{A_i}\ket{a_i\oplus \conj{b_i}}_{B_i}\ket{c_{i+1}}_{Z_i}\right).    \nonumber
\end{eqnarray}

    \item  After applying the Toffoli ladder, the state is
    \begin{eqnarray}
       && \ket{a_0}_{A_0}\ket{ \conj{c_k}(a_0\oplus a_0\oplus\conj{b_0} ) \oplus c_k (a_0\oplus\conj{b_0}) }_{B_0}\ket{c_1}_{Z_0} \nonumber  \\
       &&\quad\cdot\left(\bigotimes_{i=1}^{k-1}\ket{a_i\oplus c_i}_{A_i}\ket{ \conj{c_k} (a_i\oplus c_i\oplus a_i\oplus \conj{b_i} )\oplus c_k(a_i\oplus \conj{b_i}) }_{B_i} \ket{c_{i+1}}_{Z_i} \right)   \nonumber    \\
       &=& \ket{a_0}_{A_0}\ket{ \conj{c_k}(a_0\oplus s_0 ) \oplus c_k (a_0\oplus\conj{b_0}) }_{B_0}\ket{c_1}_{Z_0} \left(\bigotimes_{i=1}^{k-1}\ket{a_i\oplus c_i}_{A_i}\ket{ \conj{c_k} (a_i\oplus s_i )\oplus c_k(a_i\oplus \conj{b_i}) }_{B_i} \ket{c_{i+1}}_{Z_i} \right)   \nonumber 
       \end{eqnarray}

    \item  After applying the CNOT layer, the state is
    \begin{eqnarray}
        \bigotimes_{i=0}^{k-1}\ket{a_i}_{A_i}\ket{ \conj{c_k} (a_i\oplus s_i )\oplus c_k(a_i\oplus \conj{b_i}) }_{B_i} \ket{c_{i+1}}_{Z_i}.    \nonumber 
    \end{eqnarray}

    \item  After applying the multi-target CNOT, the state is
     \begin{eqnarray}
        \bigotimes_{i=0}^{k-1}\ket{a_i}_{A_i}\ket{ \conj{c_k} (a_i\oplus \conj{s_i} )\oplus c_k(a_i\oplus \conj{b_i}) }_{B_i} \ket{c_{i+1}}_{Z_i}.    \nonumber 
    \end{eqnarray}    

    \item  After applying $U_2$, the state is
     \begin{eqnarray}
       &&\left( \bigotimes_{i=0}^{k-2}\ket{a_i}_{A_i}\ket{ \conj{c_k} (a_i\oplus \conj{s_i} )\oplus c_k(a_i\oplus \conj{b_i}) }_{B_i} \ket{ \conj{c_k} a_i\conj{s_i} \oplus c_ka_i\conj{b_i} }_{Z_i} \right)  \nonumber \\
       &&\quad\cdot \ket{a_{k-1}}_{A_{k-1}}\ket{ \conj{c_k} (a_{k-1}\oplus \conj{s_{k-1}} )\oplus c_k(a_{k-1}\oplus \conj{b_{k-1} }) }_{B_{k-1}} \ket{c_k}_{Z_{k-1}} .    \nonumber 
    \end{eqnarray}

    \item  After applying the CNOT layer, the state is
     \begin{eqnarray}
        \left(\bigotimes_{i=0}^{k-2}\ket{a_i}_{A_i}\ket{\conj{c_k}\conj{s_i}\oplus c_k\conj{b_i}}_{B_i}\ket{\conj{c_k}a_i\conj{s_i}\oplus c_ka_i\conj{b_i}}_{Z_i}\right) \ket{a_{k-1}}_{A_{k-1}}\ket{\conj{c_k}\conj{s_{k-1}}\oplus c_k\conj{b_{k-1}}}_{B_{k-1}} \ket{c_k}_{Z_{k-1}}.    \nonumber
    \end{eqnarray}

    \item  After applying the Toffoli layer, the state is
    \begin{eqnarray}
       \left( \bigotimes_{i=0}^{k-2}\ket{a_i}_{A_i}\ket{\conj{c_k}\conj{s_i}\oplus c_k\conj{b_i}}_{B_i}\ket{0}_{Z_i} \right) \ket{a_{k-1}}_{A_{k-1}}\ket{\conj{c_k}\conj{s_{k-1}}\oplus c_k\conj{b_{k-1}}}_{B_{k-1}} \ket{c_k}_{Z_{k-1}}.    \nonumber
    \end{eqnarray}

    \item  After applying the final multi-target CNOT, the state is
    \begin{eqnarray}
     &&  \left( \bigotimes_{i=0}^{k-2}\ket{a_i}_{A_i}\ket{\conj{c_k}\conj{s_i}\oplus c_k b_i}_{B_i}\ket{0}_{Z_i}\right)\ket{a_{k-1}}_{A_{k-1}}\ket{\conj{c_k}\conj{s_{k-1}}\oplus c_k b_{k-1}}_{B_{k-1}} \ket{c_k}_{Z_{k-1}}   \nonumber \\
     &=& \left( \bigotimes_{i=0}^{k-2}\ket{a_i}_{A_i}\ket{\conj{c_k}d_i\oplus c_k b_i}_{B_i}\ket{0}_{Z_i}\right)\ket{a_{k-1}}_{A_{k-1}}\ket{\conj{c_k}d_{k-1}\oplus c_k b_{k-1}}_{B_{k-1}} \ket{c_k}_{Z_{k-1}}.    \nonumber
    \end{eqnarray}
    
\end{enumerate}
 Hence, if $c_k=0$, implying $b\geq a$, we obtain $d=b-a$ in register $B$; otherwise, we obtain $b$. This proves that the circuit performs as desired.

\begin{figure}[h]
    \centering
    \begin{subfigure}[b]{0.4\textwidth}
    \centering
    \makebox[\linewidth][c]{%
    \Qcircuit @C=0.5em @R=0.5em{
    \lstick{A_0} &\qw &\qw &\qw &\qw &\ctrl{1} &\qw    \\
    \lstick{B_0} &\qw &\qw &\qw &\qw &\targ &\qw    \\
    \lstick{A_1} &\qw &\qw &\qw &\ctrl{1} &\qw &\qw    \\
    \lstick{B_1} &\qw &\qw &\qw &\targ &\qw &\qw    \\
    \lstick{A_2} &\qw &\qw &\ctrl{1} &\qw &\qw &\qw    \\
    \lstick{B_2} &\qw &\qw &\targ &\qw &\qw &\qw    \\
    \lstick{A_3} &\qw &\ctrl{1} &\qw &\qw &\qw &\qw    \\
    \lstick{B_3} &\qw &\targ &\qw &\qw &\qw &\qw    \\
    \lstick{A_4} &\ctrl{1} &\qw &\qw &\qw &\qw &\qw    \\
    \lstick{B_4} &\targ &\qw &\qw &\qw &\qw &\qw    \\
    \lstick{Z_4} &\ctrlo{-1} &\ctrlo{-3} &\ctrlo{-5} &\ctrlo{-7} &\ctrlo{-9} &\qw
    }
    }
    \caption{}
    \label{fig:cns_qcla_box}
    \end{subfigure}
    \hfill
    \begin{subfigure}[b]{0.48\textwidth}
    \centering
    \makebox[\linewidth][c]{%
    \Qcircuit @C=0.5em @R=0.5em{
    \lstick{A_0} &\qw &\qw &\qw &\qw &\qw &\qw &\ctrl{1} &\qw &\qw &\qw    \\
    \lstick{B_0} &\qw &\qw &\qw &\qw &\qw &\qw &\targ &\qw &\qw &\qw    \\
    \lstick{A_1} &\qw &\qw &\qw &\qw &\qw &\ctrl{1} &\qw &\qw &\qw &\qw    \\
    \lstick{B_1} &\qw &\qw &\qw &\qw &\qw &\targ &\qw &\qw &\qw &\qw    \\
    \lstick{A_2} &\qw &\qw &\qw &\qw &\ctrl{1} &\qw &\qw &\qw &\qw &\qw    \\
    \lstick{B_2} &\qw &\qw &\qw &\qw &\targ &\qw &\qw &\qw &\qw &\qw    \\
    \lstick{A_3} &\qw &\qw &\qw &\ctrl{1} &\qw &\qw &\qw &\qw &\qw &\qw    \\
    \lstick{B_3} &\qw &\qw &\qw &\targ &\qw &\qw &\qw &\qw &\qw &\qw    \\
    \lstick{A_4} &\qw &\qw &\ctrl{1} &\qw &\qw &\qw &\qw &\qw &\qw &\qw    \\
    \lstick{B_4} &\qw &\qw &\targ &\qw &\qw &\qw &\qw &\qw &\qw &\qw    \\
    \lstick{Z_4} &\gate{X} &\ctrl{4} &\ctrl{-1} &\qw &\qw &\qw &\qw &\ctrl{4} &\gate{X} &\qw    \\
    \lstick{a_3} &\qw &\targ &\qw &\ctrl{-4} &\qw &\qw &\qw &\targ &\qw &\qw    \\
    \lstick{a_2} &\qw &\targ &\qw &\qw &\ctrl{-7} &\qw &\qw &\targ &\qw &\qw    \\
    \lstick{a_1} &\qw &\targ &\qw &\qw &\qw &\ctrl{-10} &\qw &\targ &\qw &\qw    \\
    \lstick{a_0} &\qw &\targ &\qw &\qw &\qw &\qw &\ctrl{-13} &\targ &\qw &\qw
    }
    }
    \caption{}
    \label{fig:cns_qcla_box_par}
    \end{subfigure}
    \caption{(a) The boxed portion of the circuit in Figure~\ref{fig:compnsub_qcla}. (b) Parallelization of the Toffoli gates that share the same control using additional ancilla qubits and CNOT gates. The resulting circuit has a Toffoli-depth of 1.}
    \label{fig:tof_par}
\end{figure}

 Figure~\ref{fig:tof_par} illustrates a technique for parallelizing Toffoli gates that share the same control. Such a pattern occurs in COMP-N-SUB circuit IIa; for example, see the boxed portion of the circuit in Figure~\ref{fig:compnsub_qcla}. Figure~\ref{fig:cns_qcla_box} reproduces this boxed portion, and Figure~\ref{fig:cns_qcla_box_par} shows an equivalent circuit with a Toffoli-depth of 1. For $k$ Toffoli gates sharing a control, we add $k-1$ ancilla qubits, all initialized to $\ket{0}$, and $k-1$ CNOT gates.

\subsection{Correctness of COMP-N-SUB circuit III}
\label{app:correct-III}

 Consider the COMP-N-SUB circuit described in Section~\ref{subsec:cns-III} and shown in Figure~\ref{fig:cns_and}. We track the changes in the states of the input qubits after each step. Their combined initial state is as follows.
\begin{eqnarray}
    \bigotimes_{i=0}^{k-1}\ket{b_i}_{B_i}\ket{a_i}_{A_i}\ket{0}_{Z_i}.  \nonumber
\end{eqnarray}
 The subsequent states of the qubits are as follows.
\begin{enumerate}
    \item  After applying X gates to register $B$, the state of the qubits is
    \begin{eqnarray}
        \bigotimes_{i=0}^{k-1}\ket{\conj{b_i}}_{B_i}\ket{a_i}_{A_i}\ket{0}_{Z_i}.  \nonumber
    \end{eqnarray}

    \item  The state of the qubits after applying the first logical-AND unitary in step 2 is
    \begin{eqnarray}
        \ket{\conj{b_0}}_{B_0}\ket{a_0}_{A_0}\ket{c_1}_{Z_0}\left(\bigotimes_{i=1}^{k-1}\ket{\conj{b_i}}_{B_i}\ket{a_i}_{A_i}\ket{0}_{Z_i}\right). \qquad [c_1=a_0\conj{b_0}]  \nonumber
    \end{eqnarray}

    \item  After step 3, the state of the qubits is
    \begin{eqnarray}
    && \ket{\conj{b_0}}_{B_0}\ket{a_0}_{A_0}\ket{c_1}_{Z_0}\left(\bigotimes_{i=1}^{k-1}\ket{\conj{b_i}\oplus c_i}_{B_i}\ket{a_i\oplus c_i}_{A_i}\ket{(\conj{b_i}\oplus c_i)(a_i\oplus c_i)\oplus c_i }_{Z_i}\right)    \nonumber \\
       &=& \ket{\conj{b_0}}_{B_0}\ket{a_0}_{A_0}\ket{c_1}_{Z_0}\left(\bigotimes_{i=1}^{k-1}\ket{\conj{b_i}\oplus c_i}_{B_i}\ket{a_i\oplus c_i}_{A_i}\ket{c_{i+1}}_{Z_i}\right) . \nonumber
    \end{eqnarray}

    \item  After step 4, the state of the qubits is as follows.
    \begin{eqnarray}
       && \ket{\conj{b_0}}_{B_0}\ket{a_0}_{A_0}\ket{c_1}_{Z_0}\left(\bigotimes_{i=1}^{k-2}\ket{\conj{b_i}\oplus c_i}_{B_i}\ket{a_i\oplus c_i}_{A_i}\ket{c_{i+1}}_{Z_i}\right) \nonumber \\
        &&\cdot  \ket{\conj{b_{k-1}}\oplus \conj{c_k}(a_{k-1}\oplus c_{k-1})}_{B_{k-1}}\ket{a_{k-1}}_{A_{k-1}}\ket{c_k}_{Z_{k-1}}  \nonumber
    \end{eqnarray}

    \item  After step 5, the state of the qubits is
    \begin{eqnarray}
        \ket{\conj{b_0}}_{B_0}\ket{a_0}_{A_0}\ket{c_1}_{Z_0}\left(\bigotimes_{i=1}^{k-1}\ket{\conj{b_i}\oplus \conj{c_k}(a_i\oplus c_i)}_{B_i}\ket{a_i }_{A_i}\ket{0 }_{Z_i}\right)  .  \nonumber
    \end{eqnarray}

    \item  After step 6, the state of the qubits is
    \begin{eqnarray}
        \ket{\conj{b_0}\oplus \conj{c_k}a_0  }_{B_0}\ket{a_0}_{A_0}\ket{0 }_{Z_0}\left(\bigotimes_{i=1}^{k-1}\ket{\conj{b_i}\oplus \conj{c_k}(a_i\oplus c_i)}_{B_i}\ket{a_i }_{A_i}\ket{0 }_{Z_i}\right)  .  \nonumber
    \end{eqnarray}

    \item  The final state after applying the X gates is as follows.
    \begin{eqnarray}
        \ket{\conj{c_k}\conj{s_0}\oplus c_kb_0}_{B_0}\ket{a_0}_{A_0}\ket{0}_{Z_0}\left(\bigotimes_{i=1}^{k-1}\ket{\conj{c_k}\conj{s_i}\oplus c_kb_i}_{B_i}\ket{a_i}_{A_i}\ket{0}_{Z_i}\right).  \nonumber
    \end{eqnarray}
\end{enumerate}
 Hence, if $b\geq a$, that is, if $c_k=0$, then we obtain $d=b-a$ in register $B$; otherwise, we obtain $b$, as desired.

\section{Resource estimates for RES-DIV circuits}
\label{app:resource_resDiv}

 In this section, we compute resource estimates for the RES-DIV circuits described in Section~\ref{subsec:restoreDiv-A}. Each circuit uses one $\cns^{(m)}$ unitary and $n-m'$ copies of the $\cns^{(m+1)}$ unitary, which act on $m$- and $(m+1)$-bit integers, respectively. Depending on the COMP-N-SUB circuit used, we refer to the resulting circuits as RES-DIV circuits I, IIa, IIb, and III. Table~\ref{tab:compNsub} gives the resource estimates for COMP-N-SUB.

\begin{itemize}
    \item \textbf{Toffoli-count:} The Toffoli-count upper bounds for RES-DIV circuit I, circuits IIa and IIb (which have the same bound), and circuit III are, respectively,
    \begin{eqnarray}
        && (3m-1)+(n-m')(3(m+1)-1) = 3m(n-m'+1)+2(n-m')-1;  \nonumber   
    \end{eqnarray} 
    \begin{eqnarray}
    (11m-4)+(n-m')(11(m+1)-4) = 11m(n-m'+1)+7(n-m')-4;    \nonumber
    \end{eqnarray}
     and
    \begin{eqnarray}
        m+(n-m')(m+1) = m(n-m'+1)+(n-m') .   \nonumber
    \end{eqnarray}

    \item \textbf{CNOT-count:} The CNOT counts of the Clifford+Toffoli implementations of the RES-DIV circuits are as follows.
    \begin{eqnarray}
        \text{Circuit-I:}&\quad& \n_{Tof}(Res_I) = (4m-5)+(n-m')(4(m+1)-5)  \nonumber \\
        &&=4m(n-m'+1)-(n-m')-5     \label{eqn:cnot_tof_ResI} \\
        \text{Circuit-IIa:}&\quad& \n_{Tof}(Res_{IIa}) = (6m-2)+(n-m')(6(m+1)-2) \nonumber \\
        &&=6m(n-m'+1)+4(n-m')-2     \label{eqn:cnot_tof_ResIIa} \\
        \text{Circuit-IIb:}&\quad& \n_{Tof}(Res_{IIb}) = (8m-4)+(n-m')(8(m+1)-4) \nonumber \\
        &&=8m(n-m'+1)+4(n-m')-4     \label{eqn:cnot_tof_ResIIb} 
    \end{eqnarray}
     The CNOT counts of the Clifford+T implementations of the RES-DIV circuits are as follows.
    \begin{eqnarray}
        \text{Circuit-I:}&\quad& \n_{T}(Res_I) = (25m-12)+(n-m')(25(m+1)-12) \nonumber \\
        &&=25m(n-m'+1)+13(n-m')-12     \label{eqn:cnot_t_ResI} \\
        \text{Circuit-IIa:}&\quad& \n_{T}(Res_{IIa}) \leq (83m-30)+(n-m')(83(m+1)-30) \nonumber\\
        &&= 83m(n-m'+1)+53(n-m')-30     \label{eqn:cnot_t_ResIIa} \\
        \text{Circuit-IIb:}&\quad& \n_{T}(Res_{IIb}) \leq (85m-32)+(n-m')(85(m+1)-32) \nonumber\\
        &&= 85m(n-m'+1)+53(n-m')-32    \label{eqn:cnot_t_ResIIb} \\
        \text{Circuit-III:}&\quad& \n_T(Res_{III})= (19m-7)+(n-m')(19(m+1)-7) \nonumber\\
        &&= 19m(n-m'+1)+12(n-m')-7 \label{eqn:cnot_t_ResIII}
    \end{eqnarray}

    \item \textbf{T-count:} The T-counts of the RES-DIV circuits are as follows.
    \begin{eqnarray}
        \text{Circuit-I:}&\quad& \tcount(Res_I) = (18m-4)+(n-m')(18(m+1)-4)  \nonumber \\
        &&=18m(n-m'+1)+14(n-m')-4    \label{eqn:t_ResI}  \\
        \text{Circuit-IIa:}&\quad& \tcount(Res_{IIa}) \leq (76m-21)+(n-m')(76(m+1)-21)   \nonumber \\
        &&=76m(n-m'+1)+55(n-m')-21    \label{eqn:t_ResIIa}    \\
        \text{Circuit-IIb:}&\quad& \tcount(Res_{IIb}) \leq (77m-21)+(n-m')(77(m+1)-21)   \nonumber \\
        && = 77m(n-m'+1)+56(n-m')-21    \label{eqn:t_ResIIb}    \\
        \text{Circuit-III:}&\quad& \tcount(Res_{III}) = 11m+(n-m')11(m+1)   \nonumber \\
        &&= 11m(n-m'+1)+11(n-m')  \label{eqn:t_ResIII}
    \end{eqnarray}

    \item \textbf{Qubit-count:} For all circuits, registers $A$ and $B$ require $m+1$ and $n+m-m'+2$ qubits, respectively. Although we append extra zeros, the quotient and remainder obtained by dividing an $n$-bit dividend by an $m$-bit divisor have lengths at most $n-m+1$ and $m$, respectively. The quotient and remainder are computed in register $B$. Thus, at the end of the computation, $(n+m-m'+2)-(n-m+1)-m=m-m'+1$ qubits of $B$ return to the $\ket{0}$ state. Register $A$ remains unchanged, and $A_m=\ket{0}$ at the end of the computation. Hence, all circuits have at least $m-m'+2$ ancilla qubits. In addition, $n$ and $m$ input qubits store the dividend and divisor, respectively.
    
    RES-DIV circuit I has no garbage qubits. Its total qubit count is $(n+m-m'+2)+(m+1)=n+2m-m'+3$, of which $m-m'+2$ are ancilla qubits, as explained above.
    
     For RES-DIV circuits IIa, IIb, and III, each iteration requires at most $2m-\omega(m)-\lfloor\log_2m\rfloor-1$, $3m-\omega(m)-\lfloor\log_2m\rfloor-2$, and $m-1$ additional ancilla qubits, respectively. These qubits are restored to their initial states at the end of each iteration and can therefore be reused in the next iteration. Thus, the total qubit counts of circuits IIa, IIb, and III are $n+4m-m'-\omega(m)-\lfloor\log_2m\rfloor+2$, $n+5m-m'-\omega(m)-\lfloor\log_2m\rfloor+1$, and $n+3m-m'+2$, respectively.

    \item \textbf{Toffoli-depth:} The Toffoli depths of the RES-DIV circuits are as follows.
    \begin{eqnarray}
        \text{Circuit-I:}&\quad&  (3m-1)+(n-m')(3(m+1)-1)   \nonumber \\
        &&=3m(n-m'+1)+2(n-m')-1 \nonumber   \\
        \text{Circuit-IIa:}&\quad&\leq (m+4\lfloor\log_2m\rfloor+8)+(n-m')(m+1+4\lfloor\log_2(m+1)\rfloor+8) \nonumber \\
        &&\leq (n-m'+1)\left(m+4\lfloor\log_2(m+1)\rfloor\right)+9(n-m')+8  \nonumber   \\
        \text{Circuit-IIb:}&\quad&  \leq (4\lfloor\log_2m\rfloor+9)+(n-m')(4\lfloor\log_2(m+1)\rfloor+9) \nonumber \\
        &&\leq (n-m'+1) \left(4\lfloor\log_2(m+1)\rfloor+9\right) \nonumber   \\
        \text{Circuit-III:}&\quad& m+(n-m')(m+1)=m(n-m'+1)+(n-m')  \nonumber  
    \end{eqnarray}
    
    \item \textbf{T-depth:} The T-depths of the RES-DIV circuits are as follows.
    \begin{eqnarray}
        \text{Circuit-I:}&\quad&\tcount_d(Res_I) = (9m-3)+(n-m')(9(m+1)-3)  \nonumber \\
        &&=9m(n-m'+1)+6(n-m')-3   \label{eqn:tdepth_ResI} \\
        \text{Circuit-IIa:}&\quad&\tcount_d(Res_{IIa}) \leq (3m+12\lfloor\log_2m\rfloor+24)+(n-m')(3m+3+12\lfloor\log_2(m+1)\rfloor+24) \nonumber \\
        &&\leq 3(n-m'+1)\left(m+4\lfloor\log_2(m+1)\rfloor\right)+27(n-m')+24      \label{eqn:tdepth_ResIIa} \\
        \text{Circuit-IIb:}&\quad&\tcount_d(Res_{IIb}) \leq (12\lfloor\log_2m\rfloor+27)+(n-m')(12\lfloor\log_2(m+1)\rfloor+27) \nonumber \\
        &&\leq 3(n-m'+1)\left(4\lfloor\log_2(m+1)\rfloor+9\right)  \label{eqn:tdepth_ResIIb} \\
        \text{Circuit-III:}&\quad&\tcount_d(Res_{III}) = 5m+(n-m')5(m+1) = 5m(n-m'+1)+5(n-m')  \label{eqn:tdepth_ResIII} 
    \end{eqnarray}

    \item \textbf{Measurements:} No measurements are required for RES-DIV circuits I, IIa, and IIb. The number of measurements required for RES-DIV circuit III is at most
    \begin{eqnarray}
        m+(n-m')(m+1) = nm+n+m-m'-mm'.  \nonumber
    \end{eqnarray}

\end{itemize}

\section{Clifford+T quantum circuits for pairs of Toffoli gates and pairs of CSWAP gates}
\label{app:tofPair}

 In this section, we show that if two Toffoli gates share a control or target qubit, then the T-count of the resulting unitary is at most 12. Figure~\ref{fig:tof_shared_ctrl} shows a quantum circuit for two Toffoli gates that share a control qubit. The resulting unitary is
\begin{eqnarray}
    \tof_{(q_1,q_2;q_3)}\cdot\tof_{(q_1,q_4;q_5)}.  \nonumber
\end{eqnarray}
 The circuit is obtained from the Clifford+T circuit for the Toffoli gate in Figure~\ref{fig:tof_ckt}. The input to each of the two T gates on qubit $q_1$ (outlined by dashed boxes in the figure) is the state $\ket{q_1}$. Hence, these T gates can be combined into a single $S$ (phase) gate, which is a Clifford gate. The T-depth and CNOT-count of this circuit are at most 5 and 14, respectively.

 Similarly, in the circuit for two Toffoli gates that share a target qubit, shown in Figure~\ref{fig:tof_shared_tgt}, two T gates can be combined into a single $S$ gate. Thus, in both the shared-control and shared-target cases, the T-count of the overall unitary is at most 12. For the shared-target circuit, the T-depth and CNOT-count are at most 5 and 12, respectively.

 A similar argument applies to two CSWAP (Fredkin) gates that share a control, using the identity
$$ \cswap_{(q_1;q_2,q_3)} = \CNOT_{(q_3;q_2)}\cdot\tof_{(q_1,q_2;q_3)}\cdot\CNOT_{(q_3;q_2)}. $$
 The unitary $\cswap_{(q_1;q_2,q_3)}\cdot\cswap_{(q_1;q_4,q_5)}$ has T-count, T-depth, and CNOT-count at most 12, 5, and 18, respectively. If two CSWAP gates share both the control and one target (for example, $\cswap_{(q_1;q_2,q_3)}\cdot\cswap_{(q_1;q_3,q_4)}$), then the T-count of the combined unitary is 8, as shown in \cite{2025_KJM}.

\begin{figure}[h!]
    \centering
    
    \begin{subfigure}[b]{\textwidth}
        \hspace*{\fill}
        \scalebox{0.8}{
        \Qcircuit @C=1em @R=1em{
        \lstick{q_1} &\qw &\gate{T} &\targ &\qw &\ctrl{2} &\qw &\ctrl{1} &\gate{T^{\dagger}} &\qw &\ctrl{2} &\targ &\gate{T} &\targ &\qw &\ctrl{4} &\qw &\ctrl{3} &\gate{T^{\dagger}} &\qw &\ctrl{4} &\targ &\qw     \\
        \lstick{q_2} &\qw &\gate{T} &\ctrl{-1} &\targ &\qw &\gate{T^{\dagger}} &\targ &\gate{T^{\dagger}} &\targ &\qw &\ctrl{-1} &\qw &\qw &\qw &\qw &\qw &\qw &\qw &\qw &\qw &\qw &\qw    \\
        \lstick{q_3} &\gate{H} &\gate{T} &\qw &\ctrl{-1} &\targ &\qw &\qw &\gate{T^{\dagger}} &\ctrl{-1} &\targ &\gate{H} &\qw &\qw &\qw &\qw &\qw &\qw &\qw &\qw &\qw &\qw &\qw    \\
        \lstick{q_4} &\qw &\gate{T} &\qw &\qw &\qw &\qw &\qw &\qw &\qw &\qw &\qw &\qw &\ctrl{-3} &\targ &\qw &\gate{T^{\dagger}} &\targ &\gate{T^{\dagger}} &\targ &\qw &\ctrl{-3} &\qw    \\
        \lstick{q_5} &\gate{H} &\gate{T} &\qw &\qw &\qw &\qw &\qw &\qw &\qw &\qw &\qw &\qw &\qw &\ctrl{-1} &\targ &\qw &\gate{T^{\dagger}} &\qw &\ctrl{-1} &\targ &\gate{H} &\qw \gategroup{1}{3}{1}{3}{1em}{--} \gategroup{1}{13}{1}{13}{1em}{--}
        }
        }
        \hspace*{\fill}
        \caption{}
        \label{fig:tof_shared_ctrl}
    \end{subfigure}

    \vspace{1.5em}

    \begin{subfigure}[b]{\textwidth}
        \hspace*{\fill}
        \scalebox{0.8}{
        \Qcircuit @C=1em @R=1em{
        \lstick{q_1} &\qw &\gate{T} &\targ &\qw &\ctrl{4} &\qw &\ctrl{1} &\gate{T^{\dagger}} &\qw &\ctrl{4} &\targ &\qw &\qw &\qw &\qw &\qw &\qw &\qw &\qw &\qw &\qw    \\
        \lstick{q_2} &\qw &\gate{T} &\ctrl{-1} &\targ &\qw &\gate{T^{\dagger}} &\targ &\gate{T^{\dagger}} &\targ &\qw &\ctrl{-1} &\qw &\qw &\qw &\qw &\qw &\qw &\qw &\qw &\qw &\qw    \\
        \lstick{q_3} &\qw &\qw &\qw &\qw &\qw &\qw &\qw &\qw &\qw &\qw &\gate{T} &\targ &\qw &\ctrl{2} &\qw &\ctrl{1} &\gate{T^{\dagger}} &\qw &\ctrl{2} &\targ &\qw    \\
        \lstick{q_4} &\qw &\qw &\qw &\qw &\qw &\qw &\qw &\qw &\qw &\qw &\gate{T} &\ctrl{-1} &\targ &\qw &\gate{T^{\dagger}} &\targ &\gate{T^{\dagger}} &\targ &\qw &\ctrl{-1} &\qw    \\
        \lstick{q_5} &\gate{H} &\gate{T} &\qw &\ctrl{-3} &\targ &\gate{T^{\dagger}} &\qw &\qw &\ctrl{-3} &\targ &\gate{T} &\qw &\ctrl{-1} &\targ &\qw &\qw &\gate{T^{\dagger}} &\ctrl{-1} &\targ &\gate{H} &\qw \gategroup{5}{3}{5}{3}{1em}{--} \gategroup{5}{12}{5}{12}{1em}{--}
        }
        }
        \hspace*{\fill}
        \caption{}
        \label{fig:tof_shared_tgt}
    \end{subfigure}
    \caption{(a) Circuit for $\tof_{(q_1,q_2;q_3)}\cdot\tof_{(q_1,q_4;q_5)}$, in which the two Toffoli gates share a control qubit. (b) Circuit for $\tof_{(q_1,q_2;q_5)}\cdot\tof_{(q_3,q_4;q_5)}$, in which the two Toffoli gates share a target qubit. In both circuits, the boxed T gates can be combined and replaced by the Clifford phase gate $\phase$.}
    \label{fig:tofPair}
\end{figure}

\section{Removing trailing zeros from binary integers}
\label{app:even2odd}

 We first describe a quantum circuit that removes the trailing zeros from a binary integer and outputs the resulting odd integer. We then describe a quantum circuit that performs a similar operation on a pair of integers.

\subsection{Quantum circuit for removing trailing zeros from a single integer}
\label{app:Uodd}

 Suppose that an integer $v$ has $t$ trailing zero bits, so that $v=2^t\cdot v'$ for some odd integer $v'$ and nonnegative integer $t$. We want to compute the binary representation of $v'$ by removing the $t$ trailing zeros from $v$. A quantum circuit for this task is described in \cite{2018_HSRS} and reproduced in Figure~\ref{fig:Uodd}. The circuit consists of two parts. The first part (the unboxed portion of Figure~\ref{fig:Uodd}) counts the trailing zeros, and the second part (the boxed portion of Figure~\ref{fig:Uodd}) computes $v'$ by cyclically shifting the bits of $v$. We refer to the unitary implemented by this circuit as $U_{odd}$.

 If $v=(v_{n-1},\ldots,v_0)$ is an $n$-bit integer, we allocate $n$ qubits to store its binary representation. We allocate $\lceil\log n\rceil$ qubits, $(t_{\lceil\log n\rceil-1},\ldots,t_0)$, all initialized to $\ket{0}$, to compute the binary representation of $t$, the number of trailing zeros of $v$. An additional qubit, $a$, is initialized to $\ket{1}$ by applying $\X$ to the $\ket{0}$ state. If $n$ is not a power of two, we allocate $n'-n$ additional ancilla qubits, where $n'=2^{\lceil\log n\rceil}$. These qubits are initialized to $\ket{0}$ and remain in that state throughout the computation. They are required for the second part of the circuit and are labeled $v_{n'-1},\ldots,v_n$.

 We first describe the part of the circuit that computes $t$ (the unboxed portion of Figure~\ref{fig:Uodd}). We write $C^{|\mathcal{S}|}X_{(\mathcal{S};x)}$ for a multi-controlled X gate whose controls are the $|\mathcal{S}|$ qubits in the set $\mathcal{S}$ and whose target is qubit $x$. We refer to this part of the circuit as $\mathcal{C}_{t}$.
\begin{enumerate}
    \item Apply $\CNOT_{(v_0;a)}$.

    \item  For $i=1,\ldots,n-1$:
    \begin{enumerate}
        \item  Write the binary decomposition of $i$ as $(i_{\lceil\log n\rceil-1},\ldots,i_0)$.

        \item  Set $\mathcal{S}_i=\{j : i_j=1\}$.
        
        \item  Apply $\prod_{j\in\mathcal{S}_i}\tof_{(a,v_i;t_j)}$.

        \item  Apply $C^{|\mathcal{S}_i|}X_{(\mathcal{S}_i;a)}$.
    \end{enumerate}
     
\end{enumerate}
 Given $t$, we can compute $v'=v/2^t$ by cyclically shifting the binary representation of $v$ by $t$ places so that the least significant bit becomes 1 and all trailing zeros are moved to the most significant positions. This can be done by applying groups of CSWAP gates, as shown in the boxed portion of Figure~\ref{fig:Uodd}. Recall that $n'=2^{\lceil\log n\rceil}$. We refer to this part of the circuit as $\mathcal{C}_{shift}$.
\begin{enumerate}
    \item  For $j=0,\ldots,\log n'-1$:
    \begin{enumerate}
        \item  For $i=0,\ldots,n'-2^j-1$, apply $\cswap_{(t_j;v_i,v_{i+2^j})}$.
    \end{enumerate}
\end{enumerate}

\begin{figure}[h]
        \centering
        \makebox[\textwidth][c]{%
        \Qcircuit @C=0.5em @R=0.75em{
        \lstick{\ket{1}_a} &\targ &\ctrl{1} &\targ &\ctrl{2} &\targ &\ctrl{1} &\ctrl{2} &\targ &\ctrl{3} &\targ &\ctrl{1} &\ctrl{3} &\targ &\ctrl{2} &\ctrl{3} &\targ &\ctrl{1} &\ctrl{2} &\ctrl{3} &\targ &\qw &\qw &\qw &\qw &\qw &\qw &\qw &\qw &\qw &\qw &\qw &\qw &\qw &\qw &\qw &\qw &\qw &\qw &\qw &\qw   \\
        \lstick{t_0} &\qw &\targ &\ctrl{-1} &\qw &\qw &\targ &\qw &\ctrl{-1} &\qw &\qw &\targ &\qw &\ctrl{-1} &\qw &\qw &\qw &\targ &\qw &\qw &\ctrl{-1} &\qw &\qw &\ctrl{3} &\ctrl{4} &\ctrl{5} &\ctrl{6} &\ctrl{7} &\ctrl{8} &\ctrl{9} &\qw &\qw &\qw &\qw &\qw &\qw &\qw &\qw &\qw &\qw &\qw    \\
        \lstick{t_1} &\qw &\qw &\qw &\targ &\ctrl{-2} &\qw &\targ &\ctrl{-1} &\qw &\qw &\qw &\qw &\qw &\targ &\qw &\ctrl{-2} &\qw &\targ &\qw &\ctrl{-1} &\qw &\qw &\qw &\qw &\qw &\qw &\qw &\qw &\qw &\ctrl{3} &\ctrl{4} &\ctrl{5} &\ctrl{6} &\ctrl{7} &\ctrl{8} &\qw &\qw &\qw &\qw &\qw     \\
        \lstick{t_2} &\qw &\qw &\qw &\qw &\qw &\qw &\qw &\qw &\targ &\ctrl{-3} &\qw &\targ &\ctrl{-2} &\qw &\targ &\ctrl{-1} &\qw &\qw &\targ &\ctrl{-1} &\qw &\qw &\qw &\qw &\qw &\qw &\qw &\qw &\qw &\qw &\qw &\qw &\qw &\qw &\qw &\ctrl{4} &\ctrl{5} &\ctrl{6} &\ctrl{7} &\qw    \\
        \lstick{v_0} &\ctrl{-4} &\qw &\qw &\qw &\qw &\qw &\qw &\qw &\qw &\qw &\qw &\qw &\qw &\qw &\qw &\qw &\qw &\qw &\qw &\qw &\qw &\qw &\qswap &\qw &\qw &\qw &\qw &\qw &\qw &\qswap &\qw &\qw &\qw &\qw &\qw &\qswap &\qw &\qw &\qw &\qw    \\
        \lstick{v_1} &\qw & \ctrl{-4} &\qw &\qw &\qw &\qw &\qw &\qw &\qw &\qw &\qw &\qw &\qw &\qw &\qw &\qw &\qw &\qw &\qw &\qw &\qw &\qw &\qswap\qwx &\qswap &\qw &\qw &\qw &\qw &\qw &\qw &\qswap &\qw &\qw &\qw &\qw &\qw &\qswap &\qw &\qw &\qw    \\
        \lstick{v_2} &\qw &\qw &\qw &\ctrl{-4} &\qw &\qw &\qw &\qw &\qw &\qw &\qw &\qw &\qw &\qw &\qw &\qw &\qw &\qw &\qw &\qw &\qw &\qw &\qw &\qswap\qwx &\qswap &\qw &\qw &\qw &\qw &\qswap\qwx &\qw &\qswap &\qw &\qw &\qw &\qw &\qw &\qswap &\qw &\qw    \\
        \lstick{v_3} &\qw &\qw &\qw &\qw &\qw &\ctrl{-6} &\ctrl{-5} &\qw &\qw &\qw &\qw &\qw &\qw &\qw &\qw &\qw &\qw &\qw &\qw &\qw &\qw &\qw &\qw &\qw &\qswap\qwx &\qswap &\qw &\qw &\qw &\qw &\qswap\qwx &\qw &\qswap &\qw &\qw &\qw &\qw &\qw &\qswap &\qw    \\
        \lstick{v_4} &\qw &\qw &\qw &\qw &\qw &\qw &\qw &\qw &\ctrl{-5} &\qw &\qw &\qw &\qw &\qw &\qw &\qw &\qw &\qw &\qw &\qw &\qw &\qw &\qw &\qw &\qw &\qswap\qwx &\qswap &\qw &\qw &\qw &\qw &\qswap\qwx &\qw &\qswap &\qw &\qswap\qwx &\qw &\qw &\qw &\qw   \\
        \lstick{v_5} &\qw &\qw &\qw &\qw &\qw &\qw &\qw &\qw &\qw &\qw &\ctrl{-8} &\ctrl{-6} &\qw &\qw &\qw &\qw &\qw &\qw &\qw &\qw &\qw &\qw &\qw &\qw &\qw &\qw &\qswap\qwx &\qswap &\qw &\qw &\qw &\qw &\qswap\qwx &\qw &\qswap &\qw &\qswap\qwx &\qw &\qw &\qw    \\
        \lstick{v_6} &\qw &\qw &\qw &\qw &\qw &\qw &\qw &\qw &\qw &\qw &\qw &\qw &\qw &\ctrl{-8} &\ctrl{-7} &\qw &\qw &\qw &\qw &\qw &\qw &\qw &\qw &\qw &\qw &\qw &\qw &\qswap\qwx &\qswap &\qw &\qw &\qw &\qw &\qswap\qwx &\qw &\qw &\qw &\qswap\qwx &\qw &\qw    \\
        \lstick{v_7} &\qw &\qw &\qw &\qw &\qw &\qw &\qw &\qw &\qw &\qw &\qw &\qw &\qw &\qw &\qw &\qw &\ctrl{-10} &\ctrl{-9} &\ctrl{-8} &\qw &\qw &\qw &\qw &\qw &\qw &\qw &\qw &\qw &\qswap\qwx &\qw &\qw &\qw &\qw &\qw &\qswap\qwx &\qw &\qw &\qw &\qswap\qwx &\qw  \gategroup{1}{23}{12}{41}{0.9em}{--}  
        }
        }
        \caption{ A quantum circuit that counts the trailing zeros in the binary representation of an 8-bit integer $v$ and then computes $v'=v/2^t$. The tuples $(v_7,\ldots,v_0)$ and $(t_2,t_1,t_0)$ encode the integers $v$ and $t$, respectively. The boxed portion of the circuit shifts the bits of $v$ by $t$ positions so that the least significant bit of the output integer $v'$ is $1$. This circuit was described in \cite{2018_HSRS}.}
        \label{fig:Uodd}
\end{figure}

 Before deriving the resource estimates, we describe a useful circuit optimization.

\subsubsection*{Optimization of the circuit in Figure~\ref{fig:tofGroup}(a)}

 Consider a unitary implemented by a set of $k$ Toffoli gates that share both control qubits, as shown in Figure~\ref{fig:tofGroup}(a). The corresponding unitary is
\begin{eqnarray}
    \prod_{j=1}^k\tof_{(c_1,c_2;q_j)}.  \nonumber
\end{eqnarray}
 Figure~\ref{fig:tofGroup}(b) shows an equivalent circuit, implemented as follows.
\begin{enumerate}
    \item  For $j=2,\ldots,k$, apply $\CNOT_{(q_1;q_j)}$.

    \item Apply $\tof_{(c_1,c_2;q_1)}$.

    \item  For $j=2,\ldots,k$, apply $\CNOT_{(q_1;q_j)}$.
\end{enumerate}
 We can verify the equivalence by tracing the qubit states after each step. The output state of the circuit in Figure~\ref{fig:tofGroup}(a) is
\begin{eqnarray}
    \ket{c_1}\left(\bigotimes_{j=1}^k\ket{c_1c_2\oplus q_j}  \right)\ket{c_2}.  \label{eqn:state_pattern1}
\end{eqnarray}

 Suppose that the input state of the circuit in Figure~\ref{fig:tofGroup}(b) is
\begin{eqnarray}
    \ket{c_1}\left(\bigotimes_{j=1}^k\ket{q_j}  \right)\ket{c_2}.  \nonumber
\end{eqnarray}
 After step 1, the state is
\begin{eqnarray}
    \ket{c_1}\ket{q_1}\left(\bigotimes_{j=2}^k\ket{q_1\oplus q_j}  \right)\ket{c_2}.  \nonumber
\end{eqnarray}
 After step 2, the state is
\begin{eqnarray}
    \ket{c_1}\ket{c_1c_2\oplus q_1}\left(\bigotimes_{j=2}^k\ket{q_1\oplus q_j}  \right)\ket{c_2}.  \nonumber
\end{eqnarray}
 After step 3, the state is
\begin{eqnarray}
   && \ket{c_1}\ket{c_1c_2\oplus q_1}\left(\bigotimes_{j=2}^k\ket{c_1c_2\oplus q_1\oplus q_1\oplus q_j}  \right)\ket{c_2} \nonumber   \\
   &=&\ket{c_1}\left(\bigotimes_{j=1}^k\ket{c_1c_2\oplus q_j}  \right)\ket{c_2};  \nonumber
\end{eqnarray}
 This is the same as the state in Equation~\ref{eqn:state_pattern1}, proving that the circuits in Figures~\ref{fig:tofGroup}(a) and~\ref{fig:tofGroup}(b) are equivalent.

\begin{figure}[h]
    \centering
    \begin{subfigure}[b]{0.15\textwidth}
    \centering
    \makebox[\linewidth][c]{%
    \Qcircuit @C=0.3em @R=0.3em{
    \lstick{c_1} &\ctrl{1} &\ctrl{2} &\ctrl{3} &\ctrl{4} &\qw    \\
    \lstick{q_1} &\targ &\qw &\qw &\qw &\qw    \\
    \lstick{q_2} &\qw &\targ &\qw &\qw &\qw    \\
    \lstick{q_3} &\qw &\qw &\targ &\qw &\qw    \\
    \lstick{q_4} &\qw &\qw &\qw &\targ &\qw    \\
    \lstick{c_2} &\ctrl{-4} &\ctrl{-3} &\ctrl{-2} &\ctrl{-1} &\qw    \\
    }
    }
    \caption{}
    \end{subfigure}
    \hfill
    \begin{subfigure}[b]{0.15\textwidth}
    \centering
    \makebox[\linewidth][c]{%
    \Qcircuit @C=0.3em @R=0.3em{
    \lstick{c_1} &\qw &\ctrl{1} &\qw &\qw    \\
    \lstick{q_1} &\ctrl{3} &\targ &\ctrl{3} &\qw    \\
    \lstick{q_2} &\targ &\qw &\targ &\qw    \\
    \lstick{q_3} &\targ &\qw &\targ &\qw    \\
    \lstick{q_4} &\targ &\qw &\targ &\qw    \\
    \lstick{c_2} &\qw &\ctrl{-4} &\qw &\qw    \\
    }
    }
    \caption{}
    \end{subfigure}
    \hfill
    \begin{subfigure}[b]{0.25\textwidth}
        \centering
        \makebox[\linewidth][c]{%
        \Qcircuit @C=0.3em @R=0.3em{
        \lstick{c_1} &\ctrl{1} &\ctrl{2} &\ctrl{3} &\ctrl{4} &\ctrl{1} &\ctrl{2} &\ctrl{3} &\ctrl{4} &\qw    \\
        \lstick{q_1} &\targ &\qw &\qw &\qw &\targ &\qw &\qw &\qw &\qw    \\
        \lstick{q_2} &\qw &\targ &\qw &\qw &\qw &\targ &\qw &\qw &\qw    \\
        \lstick{q_3} &\qw &\qw &\targ &\qw &\qw &\qw &\targ &\qw &\qw    \\
        \lstick{q_4} &\qw &\qw &\qw &\targ &\qw &\qw &\qw &\targ &\qw    \\
        \lstick{c_2} &\ctrl{-4} &\ctrl{-3} &\ctrl{-2} &\ctrl{-1} &\ctrlo{-4} &\ctrlo{-3} &\ctrlo{-2} &\ctrlo{-1} &\qw    \\
        \lstick{c_3} &\qw &\qw &\qw &\qw &\ctrl{-1} &\ctrl{-1} &\ctrl{-1} &\ctrl{-1} &\qw
        }
        }
        \caption{}
    \end{subfigure}
    \hfill
    \begin{subfigure}[b]{0.25\textwidth}
        \centering
        \makebox[\linewidth][c]{%
        \Qcircuit @C=0.2em @R=0.2em{
        \lstick{c_1} &\qw &\ctrl{1} &\qw &\qw &\ctrl{1} &\qw &\qw &\qw    \\
        \lstick{q_1} &\ctrl{3} &\targ &\qw &\qw &\targ &\ctrl{3} &\qw &\qw    \\
        \lstick{q_2} &\targ &\qw &\qw &\qw &\qw &\targ &\qw &\qw    \\
        \lstick{q_3} &\targ &\qw &\qw &\qw &\qw &\targ &\qw &\qw    \\
        \lstick{q_4} &\targ &\qw &\qw &\qw &\qw &\targ &\qw &\qw    \\
        \lstick{\ket{0}_c} &\qw &\qw &\qw &\targ &\ctrl{-4} &\targ &\qw &\qw    \\
        \lstick{c_2} &\qw &\ctrl{-5} &\gate{X} &\ctrl{-1} &\qw &\ctrl{-1} &\gate{X} &\qw     \\
        \lstick{c_3} &\qw &\qw &\qw &\ctrl{-1} &\qw &\ctrl{-1} &\qw &\qw
        }
        }
        \caption{  }
    \end{subfigure}
    \caption{ (a) A quantum circuit consisting of a group of Toffoli gates that share both control qubits. (b) A circuit equivalent to (a). (c) A quantum circuit consisting of a group of Toffoli gates that share both control qubits and a group of $C^3X$ gates that share all three control qubits. (d) A circuit equivalent to (c).}
    \label{fig:tofGroup}
\end{figure}

\subsubsection*{Resource estimates}

We now count the gates and qubits required to implement $U_{odd}$. We will use the following identity repeatedly.
\begin{eqnarray}
     && \sum_{j=2}^{\lceil\log n\rceil }\binom{\lceil \log n\rceil }{j}(j-1)   \nonumber \\
   &=& \left(-\binom{\lceil\log n\rceil }{1}+\sum_{j=1}^{\lceil\log n\rceil }\binom{\lceil\log n\rceil }{j} \right)-\left(-1-\lceil\log n\rceil+\sum_{j=0}^{\lceil\log n\rceil }\binom{\lceil\log n\rceil }{j} \right)     \nonumber \\
   &=& \left(\lceil\log n\rceil 2^{\lceil\log n\rceil-1}-\lceil\log n\rceil \right)- \left(2^{\lceil\log n\rceil }-\lceil\log n\rceil-1  \right)  \nonumber\\
   &=& \lceil\log n\rceil 2^{\lceil\log n\rceil-1} -2^{\lceil\log n\rceil} +1   \label{eqn:Uodd_identity}
\end{eqnarray}

\begin{itemize}
    \item \textbf{Toffoli-count:} Consider $\mathcal{C}_t$, the unboxed part of the circuit in Figure \ref{fig:Uodd}, which computes $t$. In step 2(c), we apply a product of Toffoli gates similar to that in Figure \ref{fig:tofGroup}(a). As discussed above, a product of $k$ Toffoli gates can be replaced by an equivalent circuit containing one Toffoli gate and $k-1$ CNOT gates. Thus, the $n-1$ iterations require $n-1$ Toffoli gates.

    In step 2(d), we require multi-controlled X gates whenever the binary representation of $i$ contains at least two 1s. For any $k\geq 2$, a $C^kX$ gate can be implemented with $k-1$ Toffoli gates and $k-1$ additional ancilla qubits. To reuse these ancillas, we add Toffoli gates to uncompute them.

Thus, the number of Toffoli gates required to implement $\mathcal{C}_t$ is at most
\begin{eqnarray}
   && (n-1)+2\sum_{j=2}^{\lceil\log n\rceil }\binom{\lceil \log n\rceil }{j}(j-1)   \nonumber \\
   &=&  \lceil\log n\rceil2^{\lceil\log n\rceil} +n -2^{\lceil\log n\rceil+1} +1    \qquad[\text{Using Equation \ref{eqn:Uodd_identity}}]   \nonumber \\
   &\leq& \lceil\log n\rceil 2^{\lceil\log n\rceil } -n+1. \qquad [\because 2^{\lceil\log n\rceil}\geq n ]   \nonumber
\end{eqnarray}

Next, consider $\mathcal{C}_{shift}$, the boxed part of the circuit in Figure \ref{fig:Uodd}, which applies a cyclic shift. In iteration $j$, we require $2^{\lceil\log n\rceil}-2^j$ CSWAP gates. Each CSWAP gate can be implemented with one Toffoli gate. Thus, the number of Toffoli gates required is at most
\begin{eqnarray}
    \sum_{j=0}^{\lceil\log n\rceil-1}\left(2^{\lceil\log n\rceil}-2^j\right) = \lceil\log n\rceil 2^{\lceil\log n\rceil}-2^{\lceil\log n\rceil}+1 \leq \lceil\log n\rceil 2^{\lceil\log n\rceil}+1.  \label{eqn:Uodd_tof_shift}
\end{eqnarray}

Hence, a Clifford+Toffoli implementation of $U_{odd}$ requires at most the following number of Toffoli gates:
\begin{eqnarray}
    \lceil\log n\rceil 2^{\lceil\log n\rceil+1} +n -3\cdot 2^{\lceil\log n\rceil}+2 \leq \lceil\log n\rceil 2^{\lceil\log n\rceil+1}-2n+2.  \label{eqn:Uodd_tof}
\end{eqnarray}

    \item \textbf{T-count:} The compute--uncompute pairs of multi-controlled X gates in step 2(d) of $\mathcal{C}_t$ can be implemented with logical-AND gadgets \cite{2018_G}. The remaining Toffoli gates are implemented using the construction in \cite{2010_NC}, which has a T-count of 7. Thus, the number of T gates required to implement $\mathcal{C}_t$ is at most
    \begin{eqnarray}
   && 7(n-1)+4\sum_{j=2}^{\lceil\log n\rceil }\binom{\lceil \log n\rceil }{j}(j-1)   \nonumber \\
   &=&  \lceil\log n\rceil2^{\lceil\log n\rceil+1} +7n -4\cdot2^{\lceil\log n\rceil} -3 \qquad [\text{Using Equation \ref{eqn:Uodd_identity}}]  \nonumber   \\
   &\leq& \lceil\log n\rceil 2^{\lceil\log n\rceil+1 } +3n-3.    \nonumber 
\end{eqnarray}

    Next, consider $\mathcal{C}_{shift}$. The CSWAP gates in the first group share a control and a target, so their T-count has the following upper bound \cite{2025_KJM}:
\begin{eqnarray}
    4\cdot 2^{\lceil\log n\rceil}-1, &\qquad& \text{since } 2^{\lceil\log n\rceil} \text{ is even}.    \nonumber
\end{eqnarray}
 The CSWAP gates in the second group share a control, so we can apply the construction discussed in Section \ref{app:tofPair}. Because $2^{\lceil\log n\rceil}-2^j$ is even, the T-count of group $j$ is at most
\begin{eqnarray}
    6(2^{\lceil\log n\rceil}-2^j ).   \nonumber
\end{eqnarray}
Thus, the T-count of $\mathcal{C}_{shift}$ is at most
\begin{eqnarray}
    && 4\cdot 2^{\lceil\log n\rceil}-1+\sum_{j=1}^{\lceil\log n\rceil-1} 6(2^{\lceil\log n\rceil}-2^j )      \nonumber \\
    &=& 4\cdot 2^{\lceil\log n\rceil}-1+6\sum_{j=1}^{\lceil\log n\rceil-1} 2^{\lceil\log n\rceil} -12\sum_{j=0}^{\lceil\log n\rceil-2}2^j  \nonumber \\
    &=& 6\lceil\log n\rceil 2^{\lceil\log n\rceil}-8\cdot 2^{\lceil\log n\rceil}+11 .   \label{eqn:Uodd_t_shift}
\end{eqnarray}
Hence, a Clifford+T implementation of $U_{odd}$ requires at most the following number of T gates:
\begin{eqnarray}
    8\lceil\log n\rceil 2^{\lceil\log n\rceil} +7n -12\cdot 2^{\lceil\log n\rceil}+8\leq 8\lceil\log n\rceil 2^{\lceil\log n\rceil}-5n+8.   \label{eqn:Uodd_t}
\end{eqnarray}

     \item \textbf{CNOT-count:} First, consider $\mathcal{C}_t$. We apply a CNOT gate in step 1. In step 2(c), using the optimization shown in Figures \ref{fig:tofGroup}(a) and (b), we apply $2(k-1)$ CNOT gates for each group of $k$ Toffoli gates, where $k\geq 2$. Thus, CNOT gates are required whenever the binary representation of $i$ contains at least two 1s. In step 2(d), we apply a CNOT gate when $|\mathcal{S}_i|=1$. Hence, the number of CNOT gates required is at most
    \begin{eqnarray}
       && 1 + \left( 2\sum_{j=2}^{\lceil\log n\rceil} \binom{\lceil\log n\rceil}{j} (j-1) \right) + \left( \lceil\log n\rceil -1 \right)    \nonumber \\
       &=& (\lceil\log n\rceil-2) 2^{\lceil\log n\rceil} + \lceil\log n\rceil +2.   \qquad [\text{Using Equation \ref{eqn:Uodd_identity}}]  \nonumber
    \end{eqnarray}

    Next, consider $\mathcal{C}_{shift}$. Each CSWAP gate can be implemented by a Toffoli gate conjugated by two CNOT gates. Thus, the number of CNOT gates required is at most twice the Toffoli count in Equation \ref{eqn:Uodd_tof_shift}. Hence, a Clifford+Toffoli implementation of $U_{odd}$ requires at most the following number of CNOT gates:
    \begin{eqnarray}
        && (\lceil\log n\rceil-2) 2^{\lceil\log n\rceil} + \lceil\log n\rceil +2+ 2\left( \lceil\log n\rceil 2^{\lceil\log n\rceil}-2^{\lceil\log n\rceil}+1 \right)    \nonumber   \\
        &=&(3\lceil\log n\rceil-4) 2^{\lceil\log n\rceil} + \lceil\log n\rceil +4. \label{eqn:Uodd_cnot_tof}
    \end{eqnarray}

    For a Clifford+T implementation, we account for the 7, 6, and 9 CNOT gates required by each Toffoli, logical-AND, and CSWAP implementation, respectively. Hence, the number of CNOT gates required is at most
    \begin{eqnarray}
        && 7(n-1) + 6\left( \sum_{j=2}^{\lceil\log n\rceil} \binom{\lceil\log n\rceil}{j} (j-1) \right) + 9\left( \lceil\log n\rceil 2^{\lceil\log n\rceil} - \lceil\log n\rceil +1 \right) \nonumber \\
        &&+ \left( (3\lceil\log n\rceil-4) 2^{\lceil\log n\rceil} + \lceil\log n\rceil +4 \right)   \nonumber \\
        &=& 15\lceil\log n\rceil 2^{\lceil\log n\rceil} +\lceil\log n\rceil + 7n -19\cdot 2^{\lceil\log n\rceil}+12 \qquad [\text{Using Equation \ref{eqn:Uodd_identity}}]\nonumber \\
        &\leq& 15\lceil\log n\rceil 2^{\lceil\log n\rceil} +\lceil\log n\rceil -12n+12 . \label{eqn:Uodd_cnot_t}
    \end{eqnarray}

    \item \textbf{Qubit-count:} The circuit uses $n$ input qubits to store the integer $v$. We prepend $2^{\lceil\log n\rceil}-n$ additional ancilla qubits, which return to their initial $\ket{0}$ states and can therefore be reused in later computations. We also require $\lceil\log n\rceil-1$ ancilla qubits to apply the MCX gates. These ancillas are restored to their initial states after each compute--uncompute MCX pair. The circuit leaves $\lceil\log n\rceil+1$ qubits in non-restored states: qubit $a$ and the $\lceil\log n\rceil$ qubits that store $t$. At the component level, the latter are output qubits, while $a$ is a garbage qubit. When $U_{odd}$ is used within the GCD circuit, $t$ is not part of the final GCD output, so all $\lceil\log n\rceil+1$ non-restored qubits count as garbage. In summary, in addition to the $n$ input qubits, we require $2^{\lceil\log n\rceil}+\lceil\log n\rceil-n-1$ ancilla qubits, $\lceil\log n\rceil$ output qubits, and one garbage qubit; within the GCD circuit, the output and garbage qubits together are counted as $\lceil\log n\rceil+1$ garbage qubits.

    \item \textbf{Toffoli-depth:} For a Clifford+Toffoli implementation, the Toffoli depth does not exceed the Toffoli count:
    \begin{eqnarray}
        \lceil\log n\rceil 2^{\lceil\log n\rceil+1} +n -3\cdot 2^{\lceil\log n\rceil}+2 \leq \lceil\log n\rceil 2^{\lceil\log n\rceil+1}-2n+2.  \label{eqn:Uodd_tofDepth}
    \end{eqnarray}

    \item \textbf{T-depth:} The optimal T-depth of the Toffoli construction in \cite{2010_NC} is 3 \cite{2022_GMM}, while that of a logical-AND gadget is 2. Hence, the T-depth of $\mathcal{C}_t$ is at most
   \begin{eqnarray}
   && 3(n-1)+2\sum_{j=2}^{\lceil\log n\rceil }\binom{\lceil \log n\rceil }{j}(j-1)   \nonumber \\
   &=&  \lceil\log n\rceil2^{\lceil\log n\rceil} +3n -2\cdot2^{\lceil\log n\rceil} -1\qquad [\text{Using Equation \ref{eqn:Uodd_identity}}] \nonumber   \\
   &\leq& \lceil\log n\rceil 2^{\lceil\log n\rceil } +n-1.    \nonumber 
\end{eqnarray}
The optimal T-depth of a CSWAP gate is 3 \cite{2022_GMM}. Thus, the T-depth of $\mathcal{C}_{shift}$ is at most
\begin{eqnarray}
    3\sum_{j=0}^{\lceil\log n\rceil-1} \left( 2^{\lceil\log n\rceil}-2^j\right) = 3\lceil\log n\rceil 2^{\lceil\log n\rceil}-3\cdot 2^{\lceil\log n\rceil}+3.    \label{eqn:Uodd_Tdepth_shift}
\end{eqnarray}
Hence, the T-depth of a Clifford+T implementation of $U_{odd}$ is at most
\begin{eqnarray}
    4\lceil\log n\rceil 2^{\lceil\log n\rceil} + 3n - 5\cdot 2^{\lceil\log n\rceil}+2 \leq 4\lceil\log n\rceil 2^{\lceil\log n\rceil} - 2n+2.   \label{eqn:Uodd_Tdepth}
\end{eqnarray}

\end{itemize}

\subsection{Removing trailing zeros from a pair of integers}
\label{app:Upair}

\begin{figure}[h]
    \centering
    \Qcircuit @C=0.5em @R=0.5em{
    \lstick{\ket{1}_a} &\targ &\targ &\ctrl{1} &\ctrl{1} &\targ &\ctrl{2} &\ctrl{2} &\targ &\ctrl{1} &\ctrl{2} &\ctrl{1} &\ctrl{2} &\targ &\ctrl{3} &\ctrl{3} &\targ &\ctrl{1} &\ctrl{3} &\ctrl{1} &\ctrl{3} &\targ &\ctrl{3} &\ctrl{2} &\ctrl{3} &\ctrl{2} &\targ &\ctrl{1} &\ctrl{2} &\ctrl{3} &\ctrl{1} &\ctrl{2} &\ctrl{3} &\targ &\qw    \\
    \lstick{t_0} &\qw &\qw &\targ &\targ &\ctrl{-1} &\qw &\qw &\qw &\targ &\qw &\targ &\qw &\ctrl{-1} &\qw &\qw &\qw &\targ &\qw &\targ &\qw &\ctrl{-1} &\qw &\qw &\qw &\qw &\qw &\targ &\qw &\qw &\targ &\qw &\qw &\ctrl{-1} &\qw    \\
    \lstick{t_1} &\qw &\qw &\qw &\qw &\qw &\targ &\targ &\ctrl{-2} &\qw &\targ &\qw &\targ &\ctrl{-1} &\qw &\qw &\qw &\qw &\qw &\qw &\qw &\qw &\qw &\targ &\qw &\targ &\ctrl{-2} &\qw &\targ &\qw &\qw &\targ &\qw &\ctrl{-1} &\qw    \\
    \lstick{t_2} &\qw &\qw &\qw &\qw &\qw &\qw &\qw &\qw &\qw &\qw &\qw &\qw &\qw &\targ &\targ &\ctrl{-3} &\qw &\targ &\qw &\targ &\ctrl{-2} &\targ &\qw &\targ &\qw &\ctrl{-1} &\qw &\qw &\targ &\qw &\qw &\targ &\ctrl{-1} &\qw    \\
    \lstick{v_0} &\ctrl{-4} &\ctrlo{-4} &\qw &\qw &\qw &\qw &\qw &\qw &\qw &\qw &\qw &\qw &\qw &\qw &\qw &\qw &\qw &\qw &\qw &\qw &\qw &\qw &\qw &\qw &\qw &\qw &\qw &\qw &\qw &\qw &\qw &\qw &\qw &\qw    \\
    \lstick{v_1} &\qw &\qw &\ctrl{-4} &\ctrlo{-4} &\qw &\qw &\qw &\qw &\qw &\qw &\qw &\qw &\qw &\qw &\qw &\qw &\qw &\qw &\qw &\qw &\qw &\qw &\qw &\qw &\qw &\qw &\qw &\qw &\qw &\qw &\qw &\qw &\qw &\qw    \\
    \lstick{v_2} &\qw &\qw &\qw &\qw &\qw &\ctrl{-4} &\ctrlo{-4} &\qw &\qw &\qw &\qw &\qw &\qw &\qw &\qw &\qw &\qw &\qw &\qw &\qw &\qw &\qw &\qw &\qw &\qw &\qw &\qw &\qw &\qw &\qw &\qw &\qw &\qw &\qw    \\
    \lstick{v_3} &\qw &\qw &\qw &\qw &\qw &\qw &\qw &\qw &\ctrl{-6} &\ctrl{-5} &\ctrlo{-6} &\ctrlo{-5} &\qw &\qw &\qw &\qw &\qw &\qw &\qw &\qw &\qw &\qw &\qw &\qw &\qw &\qw &\qw &\qw &\qw &\qw &\qw &\qw &\qw &\qw    \\
    \lstick{v_4} &\qw &\qw &\qw &\qw &\qw &\qw &\qw &\qw &\qw &\qw &\qw &\qw &\qw &\ctrl{-5} &\ctrlo{-5} &\qw &\qw &\qw &\qw &\qw &\qw &\qw &\qw &\qw &\qw &\qw &\qw &\qw &\qw &\qw &\qw &\qw &\qw &\qw    \\
    \lstick{v_5} &\qw &\qw &\qw &\qw &\qw &\qw &\qw &\qw &\qw &\qw &\qw &\qw &\qw &\qw &\qw &\qw &\ctrl{-8} &\ctrl{-6} &\ctrlo{-8} &\ctrlo{-6} &\qw &\qw &\qw &\qw &\qw &\qw &\qw &\qw &\qw &\qw &\qw &\qw &\qw &\qw    \\
    \lstick{v_6} &\qw &\qw &\qw &\qw &\qw &\qw &\qw &\qw &\qw &\qw &\qw &\qw &\qw &\qw &\qw &\qw &\qw &\qw &\qw &\qw &\qw &\ctrl{-7} &\ctrl{-8} &\ctrlo{-7} &\ctrlo{-8} &\qw &\qw &\qw &\qw &\qw &\qw &\qw &\qw &\qw    \\
    \lstick{v_7} &\qw &\qw &\qw &\qw &\qw &\qw &\qw &\qw &\qw &\qw &\qw &\qw &\qw &\qw &\qw &\qw &\qw &\qw &\qw &\qw &\qw &\qw &\qw &\qw &\qw &\qw &\ctrl{-10} &\ctrl{-9} &\ctrl{-8} &\ctrlo{-10} &\ctrlo{-9} &\ctrlo{-8} &\qw &\qw    \\
    \lstick{u_0} &\qw &\ctrl{-8} &\qw &\qw &\qw &\qw &\qw &\qw &\qw &\qw &\qw &\qw &\qw &\qw &\qw &\qw &\qw &\qw &\qw &\qw &\qw &\qw &\qw &\qw &\qw &\qw &\qw &\qw &\qw &\qw &\qw &\qw &\qw &\qw    \\
    \lstick{u_1} &\qw &\qw &\qw &\ctrl{-8} &\qw &\qw &\qw &\qw &\qw &\qw &\qw &\qw &\qw &\qw &\qw &\qw &\qw &\qw &\qw &\qw &\qw &\qw &\qw &\qw &\qw &\qw &\qw &\qw &\qw &\qw &\qw &\qw &\qw &\qw    \\
    \lstick{u_2} &\qw &\qw &\qw &\qw &\qw &\qw &\ctrl{-8} &\qw &\qw &\qw &\qw &\qw &\qw &\qw &\qw &\qw &\qw &\qw &\qw &\qw &\qw &\qw &\qw &\qw &\qw &\qw &\qw &\qw &\qw &\qw &\qw &\qw &\qw &\qw    \\
    \lstick{u_3} &\qw &\qw &\qw &\qw &\qw &\qw &\qw &\qw &\qw &\qw &\ctrl{-8} &\ctrl{-8} &\qw &\qw &\qw &\qw &\qw &\qw &\qw &\qw &\qw &\qw &\qw &\qw &\qw &\qw &\qw &\qw &\qw &\qw &\qw &\qw &\qw &\qw    \\
    \lstick{u_4} &\qw &\qw &\qw &\qw &\qw &\qw &\qw &\qw &\qw &\qw &\qw &\qw &\qw &\qw &\ctrl{-8} &\qw &\qw &\qw &\qw &\qw &\qw &\qw &\qw &\qw &\qw &\qw &\qw &\qw &\qw &\qw &\qw &\qw &\qw &\qw    \\
    \lstick{u_5} &\qw &\qw &\qw &\qw &\qw &\qw &\qw &\qw &\qw &\qw &\qw &\qw &\qw &\qw &\qw &\qw &\qw &\qw &\ctrl{-8} &\ctrl{-8} &\qw &\qw &\qw &\qw &\qw &\qw &\qw &\qw &\qw &\qw &\qw &\qw &\qw &\qw    \\
    \lstick{u_6} &\qw &\qw &\qw &\qw &\qw &\qw &\qw &\qw &\qw &\qw &\qw &\qw &\qw &\qw &\qw &\qw &\qw &\qw &\qw &\qw &\qw &\qw &\qw &\ctrl{-8} &\ctrl{-8} &\qw &\qw &\qw &\qw &\qw &\qw &\qw &\qw &\qw    \\
    \lstick{u_7} &\qw &\qw &\qw &\qw &\qw &\qw &\qw &\qw &\qw &\qw &\qw &\qw &\qw &\qw &\qw &\qw &\qw &\qw &\qw &\qw &\qw &\qw &\qw &\qw &\qw &\qw &\qw &\qw &\qw &\ctrl{-10} &\ctrl{-9} &\ctrl{-8} &\qw &\qw
    }
    \caption{A quantum circuit that takes two 8-bit integers, $u$ and $v$, with binary representations $(u_7,\ldots,u_0)$ and $(v_7,\ldots,v_0)$, respectively. It computes the largest integer $t$ such that the binary representations of both $u$ and $v$ end in at least $t$ consecutive zeros. The tuple $(t_2,t_1,t_0)$ is the binary representation of $t$.}
    \label{fig:C_t_2}
\end{figure}

In this section, we consider a pair of $n$-bit integers, $u$ and $v$, such that $u=2^t\cdot u'$ and $v=2^t\cdot v'$ for integers $u'$ and $v'$, at least one of which is odd. Thus, the $t$ least significant bits in the binary representations of both $u$ and $v$ are 0, and the next bit is 1 in at least one representation. Our goal is to compute $u'$ and $v'$. We follow a procedure similar to that in Appendix \ref{app:Uodd}: first, we compute $t$, and then we cyclically shift the bits of $u$ and $v$ by $t$ places. We denote this unitary by $U_{odd}^{(2)}$.

As in the circuit for $U_{odd}$ (Figure \ref{fig:Uodd}), we allocate $\lceil\log n\rceil$ qubits to store the binary representation of $t$. We also allocate $n$ qubits each to store the binary representations $u=(u_{n-1},\ldots,u_0)$ and $v=(v_{n-1},\ldots,v_0)$. We call these the $u$ and $v$ registers, respectively. Qubit $a$ is initialized to the $\ket{1}$ state. If $n$ is not a power of two, we prepend $2^{\lceil\log n\rceil}-n$ ancilla qubits to each of the $u$ and $v$ registers. These qubits remain in the $\ket{0}$ state throughout the computation.

First, we describe the circuit that computes $t$. It is obtained by modifying $\mathcal{C}_t$, the unboxed part of the circuit in Figure \ref{fig:Uodd}, to operate on a pair of integers. We refer to the resulting circuit as $\mathcal{C}_t^{(2)}$; Figure \ref{fig:C_t_2} illustrates it.
\begin{enumerate}
    \item Apply $\CNOT_{(v_0;a)} \cdot\tof_{(u_0,\conj{v_0});a}$.

    \item For $i = 1, \ldots, n-1$:
    \begin{enumerate}
        \item Set $(i_{\lceil\log n\rceil-1},\ldots,i_0)\leftarrow$ the binary representation of $i$.

        \item Set $\mathcal{S}_i = \{j : i_j = 1\}$.
        
        \item Apply $\left( \prod_{j\in\mathcal{S}_i} \tof_{(a,v_i; t_j) } \right) \cdot \left( \prod_{j\in\mathcal{S}_i} C^3X_{(a,\conj{v_i},u_i;t_j)} \right)$.

        \item Apply $C^{|\mathcal{S}_i|} X_{(\mathcal{S}_i;a)}$.
    \end{enumerate}     
\end{enumerate}

The circuit that cyclically shifts the bits of $u$ and $v$ by $t$ places is the same as $\mathcal{C}_{shift}$, the boxed part of the circuit in Figure \ref{fig:Uodd}. Here, however, we apply it twice: once to register $u$ and once to register $v$. Both applications are controlled by qubits $t_{\lceil\log n\rceil-1},\ldots,t_0$.

\subsubsection*{Resource estimates}

We now count the gates and qubits required to implement $U_{odd}^{(2)}$.

\begin{itemize}
    \item \textbf{Toffoli-count:} We require one Toffoli gate in step 1. Another significant difference between $\mathcal{C}_{t}$ and $\mathcal{C}_t^{(2)}$ occurs in step 2(c), where we apply an additional group of $C^3X$ gates. As shown in Figures \ref{fig:tofGroup}(a) and (b), the first group of Toffoli gates can be replaced by one Toffoli gate and a few CNOT gates. We can apply the same optimization to the second group of $C^3X$ gates, as illustrated in Figures \ref{fig:tofGroup}(c) and (d). We first decompose each $C^3X$ gate into Toffoli gates using an additional ancilla qubit $c$, initialized to $\ket{0}$. This qubit can be restored to $\ket{0}$ later. Hence, we obtain the following identity.
    \begin{eqnarray}
    \prod_{j\in\mathcal{S}_i} C^3X_{(a,\conj{v_i},u_i;t_j)} = \tof_{(\conj{v_i},u_i;c)} \cdot \left( \prod_{j\in\mathcal{S}_i} \tof_{(a,c;t_j)} \right)\cdot\tof_{(\conj{v_i},u_i;c)}   \nonumber
    \end{eqnarray}
    We then apply the circuit equivalence shown in Figures \ref{fig:tofGroup}(a) and (b) and cancel the intermediate CNOT gates.

    The remaining steps are identical to those of $\mathcal{C}_t$. Hence, the number of Toffoli gates required for $\mathcal{C}_t^{(2)}$ is at most
    \begin{eqnarray}
        && 1 + 4(n-1) + 2\sum_{j=2}^{\lceil\log n\rceil }\binom{\lceil \log n\rceil }{j}(j-1)   \nonumber \\
        &=& \lceil\log n\rceil 2^{\lceil\log n\rceil} +4n-2\cdot 2^{\lceil\log n\rceil}-1 \qquad[\text{Using Equation~\ref{eqn:Uodd_identity}}]   \nonumber  \\
        &\leq& \lceil\log n\rceil 2^{\lceil\log n\rceil} +2n-1.  \nonumber
    \end{eqnarray}

    We then apply two copies of $\mathcal{C}_{shift}$. Using Equation \ref{eqn:Uodd_tof_shift}, the number of Toffoli gates required for a Clifford+Toffoli implementation of $U_{odd}^{(2)}$ is therefore at most
    \begin{eqnarray}
        &&  \lceil\log n\rceil 2^{\lceil\log n\rceil} +4n-2\cdot 2^{\lceil\log n\rceil}-1 + 2\left( \lceil\log n\rceil 2^{\lceil\log n\rceil}-2^{\lceil\log n\rceil}+1 \right)  \nonumber \\
        &=& 3\lceil\log n\rceil 2^{\lceil\log n\rceil} + 4n -4\cdot 2^{\lceil\log n\rceil} +1 \leq 3\lceil\log n\rceil 2^{\lceil\log n\rceil} +1.    \label{eqn:Uodd2_tof}
    \end{eqnarray}

    \item \textbf{T-count:} The multi-controlled X gates in step 2(d) can be implemented with logical-AND gadgets, as described for $\mathcal{C}_t$. Step 2(c) requires four Toffoli gates, two of which form a compute--uncompute pair and can therefore be implemented using a logical-AND gadget. Thus, the number of T gates required to implement $\mathcal{C}_t^{(2)}$ is at most
    \begin{eqnarray}
   && 7+7\cdot 2(n-1)+4(n-1)+4\sum_{j=2}^{\lceil\log n\rceil }\binom{\lceil \log n\rceil }{j}(j-1)   \nonumber \\
   &=&  2\lceil\log n\rceil 2^{\lceil\log n\rceil} +18n -4\cdot 2^{\lceil\log n\rceil}-7 \qquad [\text{Using Equation \ref{eqn:Uodd_identity}}]  \nonumber   \\
   &\leq& 2\lceil\log n\rceil 2^{\lceil\log n\rceil } +14n-7.    \nonumber 
\end{eqnarray}
    Because we require two copies of $\mathcal{C}_{shift}$, Equation \ref{eqn:Uodd_t_shift} implies that the number of T gates required for a Clifford+T implementation of $U_{odd}^{(2)}$ is at most
    \begin{eqnarray}
        &&  2\lceil\log n\rceil 2^{\lceil\log n\rceil} +18n -4\cdot 2^{\lceil\log n\rceil}-7 + 2\left( 6\lceil\log n\rceil 2^{\lceil\log n\rceil}-8\cdot 2^{\lceil\log n\rceil}+11 \right)  \nonumber\\
        &=& 14\lceil\log n\rceil 2^{\lceil\log n\rceil} +18n -20\cdot 2^{\lceil\log n\rceil} +15 \leq 14\lceil\log n\rceil 2^{\lceil\log n\rceil} -2n +15.   \label{eqn:Uodd2_t}
    \end{eqnarray}

    \item \textbf{CNOT-count:} The number of CNOT gates required for a Clifford+Toffoli implementation of $U_{odd}^{(2)}$ is the same as that for $U_{odd}$ and is at most
    \begin{eqnarray}
        (3\lceil\log n\rceil-4) 2^{\lceil\log n\rceil} + \lceil\log n\rceil +4. \label{eqn:Uodd2_cnot_tof}
    \end{eqnarray}

    For a Clifford+T implementation of $U_{odd}^{(2)}$, we account for the CNOT counts of the Toffoli, CSWAP, and logical-AND implementations. The number of CNOT gates required is therefore at most
    \begin{eqnarray}
        && 7+7\cdot 2(n-1)+6(n-1)+6\sum_{j=2}^{\lceil\log n\rceil }\binom{\lceil \log n\rceil }{j}(j-1)     \nonumber   \\
        &&+9\cdot 2\left( \lceil\log n\rceil 2^{\lceil\log n\rceil} - \lceil\log n\rceil +1 \right)     \nonumber   \\
        &=& 21\lceil\log n\rceil 2^{\lceil\log n\rceil} + 20n -6\cdot 2^{\lceil\log n\rceil} -18\lceil\log n\rceil + 11\qquad [\text{Using Equation \ref{eqn:Uodd_identity}}]    \nonumber \\
        &\leq& 21\lceil\log n\rceil 2^{\lceil\log n\rceil} + 14n -18\lceil\log n\rceil + 11
        \label{eqn:Uodd2_cnot_t}
    \end{eqnarray}

    \item \textbf{Qubit-count:} We require $2n$ input qubits to store $u$ and $v$. We prepend $2\left(2^{\lceil\log n\rceil}-n\right)$ additional ancilla qubits, which return to their initial $\ket{0}$ states. We also require $\left(\lceil\log n\rceil-1\right)+1$ ancilla qubits to apply the MCX gates in step 2(d) and the $C^3X$ gates in step 2(c). These ancillas are also restored to their initial states. The circuit leaves $\lceil\log n\rceil+1$ qubits in non-restored states: qubit $a$ and the $\lceil\log n\rceil$ qubits that store $t$. At the component level, the latter are output qubits, while $a$ is a garbage qubit. When $U_{odd}^{(2)}$ is used within the GCD circuit, $t$ is not part of the final GCD output, so all $\lceil\log n\rceil+1$ non-restored qubits count as garbage. In summary, in addition to the $2n$ input qubits, we require $2^{\lceil\log n\rceil+1}+\lceil\log n\rceil-2n$ ancilla qubits, $\lceil\log n\rceil$ output qubits, and one garbage qubit; within the GCD circuit, the output and garbage qubits together are counted as $\lceil\log n\rceil+1$ garbage qubits.

    \item \textbf{Toffoli-depth:} For a Clifford+Toffoli implementation, the Toffoli depth does not exceed the Toffoli count:
    \begin{eqnarray}
       3\lceil\log n\rceil 2^{\lceil\log n\rceil} + 4n -4\cdot 2^{\lceil\log n\rceil} +1 \leq 3\lceil\log n\rceil 2^{\lceil\log n\rceil} +1. \label{eqn:Uodd2_tofDepth}
    \end{eqnarray}

    \item \textbf{T-depth:} Using the optimal T-depths of the Toffoli and logical-AND implementations, the T-depth of $\mathcal{C}_t^{(2)}$ is at most
\begin{eqnarray}
    && 3+3\cdot 2(n-1)+2(n-1)+2\sum_{j=2}^{\lceil\log n\rceil }\binom{\lceil \log n\rceil }{j}(j-1)   \nonumber \\
   &=&  \lceil\log n\rceil 2^{\lceil\log n\rceil} +8n -2\cdot 2^{\lceil\log n\rceil}-3 \qquad [\text{Using Equation \ref{eqn:Uodd_identity}}]  \nonumber   \\
   &\leq& \lceil\log n\rceil 2^{\lceil\log n\rceil } +6n-3.    \nonumber 
\end{eqnarray}
Because two copies of $\mathcal{C}_{shift}$ are required, Equation \ref{eqn:Uodd_Tdepth_shift} gives the following upper bound on the T-depth of $U_{odd}^{(2)}$:
\begin{eqnarray}
    && \lceil\log n\rceil 2^{\lceil\log n\rceil} +8n -2\cdot 2^{\lceil\log n\rceil}-3 + 2\left( 3\lceil\log n\rceil 2^{\lceil\log n\rceil}-3\cdot 2^{\lceil\log n\rceil}+3 \right)  \nonumber \\
    &=& 7\lceil\log n\rceil 2^{\lceil\log n\rceil} +8n -8\cdot 2^{\lceil\log n\rceil} +3\leq 7\lceil\log n\rceil 2^{\lceil\log n\rceil} +3  \label{eqn:Uodd2_Tdepth}
\end{eqnarray}
     
\end{itemize}

\section{Resource estimates for a quantum circuit computing the GCD}
\label{app:gcd}

In this section, we compute the numbers of gates and qubits required to implement the quantum circuit in Figure \ref{fig:gcd} (Section \ref{subsec:gcd}), which computes the GCD of two $n$-bit integers. The pseudocode for the binary GCD algorithm used here is given in Algorithm \ref{alg:gcd}.

A more detailed description of the quantum circuit and its constituent unitaries is given in Section \ref{subsec:gcd}. The variants of the GCD circuit obtained by using COMP-N-SUB circuits I, IIa, IIb, and III are called GCD-I, GCD-IIa, GCD-IIb, and GCD-III, respectively. Here, $N_{iter}$ denotes the number of iterations of the loop in steps 5--17 and is at most $2n$.

\begin{itemize}
    \item \textbf{Toffoli-count:} Let $\tcount^{of}(.)$ denote the Toffoli count of a unitary. The number of Toffoli gates required is at most
    \begin{eqnarray}
        &&\tcount^{of}\left( U_{odd}^{(2)} \right) + \tcount^{of}\left(\mathcal{C}_{shift}\right) + N_{iter} \left( 2\tcount^{of}(U_{odd}) + 2\tcount^{of}(\cns) \right)    \nonumber \\
        &\leq& \left(3\lceil\log n\rceil 2^{\lceil\log n\rceil}+1\right)+\left(\lceil\log n\rceil 2^{\lceil\log n\rceil} +1 \right) +2N_{iter} \left( 2\lceil\log n\rceil 2^{\lceil\log n\rceil}-2n+2  \right) \nonumber \\
        && + 2N_{iter}\tcount^{of}(\cns) \qquad [\text{Using the estimates in Appendix \ref{app:even2odd}}]    \nonumber    \\
        &=& 4(N_{iter}+1)\lceil\log n\rceil 2^{\lceil\log n\rceil} - 4N_{iter}(n-1) + 2 + 2N_{iter}\tcount^{of}(\cns).   \nonumber
    \end{eqnarray}
    Substituting the costs of the various COMP-N-SUB circuits from Section \ref{sec:cns} gives the following upper bounds on the Toffoli counts of Clifford+Toffoli implementations of the GCD circuits.
    \begin{eqnarray}
      \tcount^{of}(GCD_I)&\leq&  4(N_{iter}+1)\lceil\log n\rceil 2^{\lceil\log n\rceil} - 4N_{iter}(n-1) + 2 + 2N_{iter}(3n-1) \nonumber    \\
       &=& 4(N_{iter}+1)\lceil\log n\rceil 2^{\lceil\log n\rceil} +2N_{iter}(n+1)+2  \label{eqn:gcdI_tof_count}  \\
       \tcount^{of}(GCD_{IIa}) &=& \tcount^{of}(GCD_{IIb})  \nonumber \\
       &\leq& 4(N_{iter}+1)\lceil\log n\rceil 2^{\lceil\log n\rceil} - 4N_{iter}(n-1) + 2 + 2N_{iter}(11n-4)   \nonumber \\
       &=& 4(N_{iter}+1)\lceil\log n\rceil 2^{\lceil\log n\rceil} +2N_{iter}(9n-2)+2  \label{eqn:gcdII_tof_count}  \\
       \tcount^{of}(GCD_{III}) &\leq& 4(N_{iter}+1)\lceil\log n\rceil 2^{\lceil\log n\rceil} - 4N_{iter}(n-1) + 2 + 2N_{iter}n   \nonumber \\
       &=& 4(N_{iter}+1)\lceil\log n\rceil 2^{\lceil\log n\rceil} -2N_{iter}(n-2)+2  \label{eqn:gcdIII_tof_count}
    \end{eqnarray}

    \item \textbf{T-count:} Let $\tcount(.)$ denote the T-count of a unitary. The number of T gates required is at most
    \begin{eqnarray}
        &&\tcount\left( U_{odd}^{(2)} \right) + \tcount\left(\mathcal{C}_{shift}\right) + N_{iter} \left( 2\tcount(U_{odd}) + 2\tcount(\cns) \right)    \nonumber \\
        &\leq& \left(14\lceil\log n\rceil 2^{\lceil\log n\rceil}+15\right)+\left(6\lceil\log n\rceil 2^{\lceil\log n\rceil} +11 \right) +2N_{iter} \left( 8\lceil\log n\rceil 2^{\lceil\log n\rceil}-5n+8  \right) \nonumber \\
        && + 2N_{iter}\tcount(\cns) \qquad [\text{Using the estimates in Appendix \ref{app:even2odd}}]    \nonumber    \\
        &=& 4(4N_{iter}+5)\lceil\log n\rceil 2^{\lceil\log n\rceil} - 2N_{iter}(5n-8) + 26 + 2N_{iter}\tcount(\cns).   \nonumber
    \end{eqnarray}
    Substituting the costs of the various COMP-N-SUB circuits from Section \ref{sec:cns} gives the following upper bounds on the numbers of T gates required by Clifford+T implementations of the GCD circuits.
    \begin{eqnarray}
      \tcount(GCD_I)&\leq& 4(4N_{iter}+5)\lceil\log n\rceil 2^{\lceil\log n\rceil} - 2N_{iter}(5n-8) + 26 + 2N_{iter}(18n-4) \nonumber    \\
       &=& 4(4N_{iter}+5)\lceil\log n\rceil 2^{\lceil\log n\rceil} + 2N_{iter}(13n+4) + 26  \label{eqn:gcdI_t_count}  \\
       \tcount(GCD_{IIa}) &\leq& 4(4N_{iter}+5)\lceil\log n\rceil 2^{\lceil\log n\rceil} - 2N_{iter}(5n-8) + 26 + 2N_{iter}(76n-21) \nonumber \\
       &\leq& 4(4N_{iter}+5)\lceil\log n\rceil 2^{\lceil\log n\rceil} +142N_{iter}n  \label{eqn:gcdIIa_t_count}  \\
       \tcount(GCD_{IIb}) &\leq& 4(4N_{iter}+5)\lceil\log n\rceil 2^{\lceil\log n\rceil} - 2N_{iter}(5n-8) + 26 + 2N_{iter}(77n-21) \nonumber \\
       &\leq& 4(4N_{iter}+5)\lceil\log n\rceil 2^{\lceil\log n\rceil} +144N_{iter}n  \label{eqn:gcdIIb_t_count} \\
       \tcount(GCD_{III}) &\leq& 4(4N_{iter}+5)\lceil\log n\rceil 2^{\lceil\log n\rceil} - 2N_{iter}(5n-8) + 26 + 2N_{iter}\cdot 11n \nonumber    \\
       &=& 4(4N_{iter}+5)\lceil\log n\rceil 2^{\lceil\log n\rceil} + 2N_{iter}(6n+8) + 26  \label{eqn:gcdIII_t_count}
    \end{eqnarray}

    \item \textbf{CNOT-count:} Let $\n_{Tof}(.)$ and $\n_T(.)$ denote the CNOT counts of a unitary implemented using the Clifford+Toffoli and Clifford+T gate sets, respectively. For a Clifford+Toffoli implementation, the number of CNOT gates required is at most
    \begin{eqnarray}
        &&\n_{Tof}\left( U_{odd}^{(2)} \right) + \n_{Tof}\left(\mathcal{C}_{shift}\right) + N_{iter} \left( 2\n_{Tof}(U_{odd}) + 2\n_{Tof}(\cns) \right)    \nonumber \\
        &\leq& \left( 3\lceil\log n\rceil 2^{\lceil\log n\rceil} +\lceil\log n\rceil +4 \right) +\left( 2\lceil\log n\rceil 2^{\lceil\log n\rceil} +2 \right) \nonumber \\
        &&+2N_{iter}\left( 3\lceil\log n\rceil 2^{\lceil\log n\rceil}+\lceil\log n\rceil+4 \right) +2N_{iter}\n_{Tof}(\cns) \nonumber \\
        &=& (6N_{iter}+5)\lceil\log n\rceil 2^{\lceil\log n\rceil} + (2N_{iter}+1)\lceil\log n\rceil + 6 + 8N_{iter} +2N_{iter}\n_{Tof}(\cns).  \nonumber
    \end{eqnarray}
    Similarly, the number of CNOT gates required for a Clifford+T implementation is at most
    \begin{eqnarray}
        &&\n_{T}\left( U_{odd}^{(2)} \right) + \n_{T}\left(\mathcal{C}_{shift}\right) + N_{iter} \left( 2\n_{T}(U_{odd}) + 2\n_{T}(\cns) \right)    \nonumber \\
        &\leq& \left( 21\lceil\log n\rceil 2^{\lceil\log n\rceil} +14n+11 \right) + \left( 9\lceil\log n\rceil 2^{\lceil\log n\rceil} +9 \right) + 2N_{iter} \left( 15\lceil\log n\rceil 2^{\lceil\log n\rceil} +12 \right) \nonumber \\
        &&+2N_{iter}\n_T(\cns)  \nonumber \\
        &=& 30(N_{iter}+1)\lceil\log n\rceil 2^{\lceil\log n\rceil} + 14n+24N_{iter}+20+2N_{iter}\n_T(\cns).    \nonumber
    \end{eqnarray}
    Substituting the costs of the different COMP-N-SUB variants gives the following upper bounds on the CNOT counts of Clifford+Toffoli implementations of the GCD circuits.
    \begin{eqnarray}
        \n_{Tof}(GCD_I)&\leq& (6N_{iter}+5)\lceil\log n\rceil 2^{\lceil\log n\rceil} + (2N_{iter}+1)\lceil\log n\rceil + 6 + 8N_{iter} +2N_{iter}\left(4n-5\right) \nonumber    \\
       &\leq& (6N_{iter}+5)\lceil\log n\rceil 2^{\lceil\log n\rceil} + (2N_{iter}+1)\lceil\log n\rceil + 8N_{iter}n + 6  \label{eqn:gcdI_cnot_tof}  \\
       \n_{Tof}(GCD_{IIa}) &\leq& (6N_{iter}+5)\lceil\log n\rceil 2^{\lceil\log n\rceil} + (2N_{iter}+1)\lceil\log n\rceil + 6 + 8N_{iter} +2N_{iter}\left(6n-2\right) \nonumber \\
       &=& (6N_{iter}+5)\lceil\log n\rceil 2^{\lceil\log n\rceil} + (2N_{iter}+1)\lceil\log n\rceil + 4N_{iter}(3n+1) + 6  \label{eqn:gcdIIa_cnot_tof}  \\
       \n_{Tof}(GCD_{IIb}) &\leq& (6N_{iter}+5)\lceil\log n\rceil 2^{\lceil\log n\rceil} + (2N_{iter}+1)\lceil\log n\rceil + 6 + 8N_{iter} +2N_{iter}\left(8n-4\right) \nonumber \\
       &=& (6N_{iter}+5)\lceil\log n\rceil 2^{\lceil\log n\rceil} + (2N_{iter}+1)\lceil\log n\rceil + 16N_{iter}n+ 6   \label{eqn:gcdIIb_cnot_tof} 
    \end{eqnarray}
    
    Similarly, the following are upper bounds on the CNOT counts of Clifford+T implementations of the GCD circuits.
    \begin{eqnarray}
        \n_{T}(GCD_I)&\leq& 30(N_{iter}+1)\lceil\log n\rceil 2^{\lceil\log n\rceil} + 14n+24N_{iter}+20 +2N_{iter}\left( 25n-12\right) \nonumber    \\
       &=& 30(N_{iter}+1)\lceil\log n\rceil 2^{\lceil\log n\rceil} + 14n+50N_{iter}n+20  \label{eqn:gcdI_cnot_t}  \\
       \n_{T}(GCD_{IIa}) &\leq& 30(N_{iter}+1)\lceil\log n\rceil 2^{\lceil\log n\rceil} + 14n+24N_{iter}+20 +2N_{iter}\left(83n-30\right) \nonumber \\
       &\leq& 30(N_{iter}+1)\lceil\log n\rceil 2^{\lceil\log n\rceil} + 14n+166N_{iter}n   \label{eqn:gcdIIa_cnot_t}  \\
       \n_{T}(GCD_{IIb}) &\leq& 30(N_{iter}+1)\lceil\log n\rceil 2^{\lceil\log n\rceil} + 14n+24N_{iter}+20 +2N_{iter}\left(85n-32\right) \nonumber \\
       &\leq& 30(N_{iter}+1)\lceil\log n\rceil 2^{\lceil\log n\rceil} + 14n+ 170N_{iter}n \label{eqn:gcdIIb_cnot_t} \\
       \n_{T}(GCD_{III}) &\leq& 30(N_{iter}+1)\lceil\log n\rceil 2^{\lceil\log n\rceil} + 14n+24N_{iter}+20 +2N_{iter}\left(19n-7\right) \nonumber    \\
       &=& 30(N_{iter}+1)\lceil\log n\rceil 2^{\lceil\log n\rceil} + 14n+38N_{iter}n+20+10N_{iter} \label{eqn:gcdIII_cnot_t}
    \end{eqnarray}

    \item \textbf{Qubit-count:} We require $2n$ input qubits to store the two input integers.

    Let $\q_g(.)$ denote the number of garbage qubits required to implement a unitary. The number of garbage qubits required to implement the GCD circuits is
    \begin{eqnarray}
       && \q_g\left( U_{odd}^{(2)} \right) + \q_g\left(\mathcal{C}_{shift}\right) + N_{iter} \left( 2\q_g(U_{odd}) + 2\q_g(\cns) \right)    \nonumber \\
       &=& \left( 1+\lceil\log n\rceil \right) +0 + N_{iter} \left( 2(1+\lceil\log n\rceil)+1 \right)   \nonumber   \\
       &=& 2\lceil\log n\rceil N_{iter}+ 3N_{iter} + \lceil\log n\rceil +1. \nonumber
    \end{eqnarray}
     We require one garbage qubit to implement any variant of COMP-N-SUB. Next, let $\q_a(.)$ denote the number of ancilla qubits required to implement a unitary. These qubits are returned to their initial states and can therefore be reused in later computations. Thus, the number of ancilla qubits required to implement the GCD circuits is
    \begin{eqnarray}
        && \max\left\{  \q_a\left( U_{odd}^{(2)} \right) , \q_a\left(\mathcal{C}_{shift}\right) ,  2\q_a\left(U_{odd}\right)  ,  \q_a\left(\cns\right)  \right\}    \nonumber \\
        &=& \max\left\{ \left(2\cdot 2^{\lceil\log n\rceil} +\lceil\log n\rceil -2n\right), 0, \left( 2\cdot 2^{\lceil\log n\rceil} +2\lceil\log n\rceil-2n-2  \right), \q_a\left( \cns\right) \right\}  \nonumber   \\
       &=& \max\left\{  \left( 2\cdot 2^{\lceil\log n\rceil} +2\lceil\log n\rceil-2n-2  \right), \q_a\left( \cns\right) \right\} . \nonumber
    \end{eqnarray}
    Substituting the ancilla-qubit requirements of the various COMP-N-SUB unitaries gives the following requirements for the GCD circuits.
    \begin{eqnarray}
        \q_a(GCD_I) &=& 2\cdot 2^{\lceil\log n\rceil}+2\lceil\log n\rceil-2n-2   \nonumber \\
        \q_a(GCD_{IIa}) &=& \max\left\{2\cdot 2^{\lceil\log n\rceil}+2\lceil\log n\rceil-2n-2,\,2n-\omega(n)-\lfloor\log n\rfloor-1\right\} \nonumber \\
        \q_a(GCD_{IIb}) &=& \max\left\{2\cdot 2^{\lceil\log n\rceil}+2\lceil\log n\rceil-2n-2,\,3n-\omega(n)-\lfloor\log n\rfloor-2\right\} \nonumber \\
        \q_a(GCD_{III}) &=& \max\left\{2\cdot 2^{\lceil\log n\rceil}+2\lceil\log n\rceil-2n-2,\,n-1\right\}. \nonumber
    \end{eqnarray}
    Let $\q_T(.)$ denote the total qubit count. The total qubit counts of the GCD circuits are then as follows.
    \begin{eqnarray}
        \q_T(GCD_I) &=& 2\cdot 2^{\lceil\log n\rceil}+2\lceil\log n\rceil N_{iter}+3N_{iter}+3\lceil\log n\rceil-1 \label{eqn:gcdI_qubit} \\
        \q_T(GCD_{IIa}) &=& (2N_{iter}+1)\lceil\log n\rceil+2n+3N_{iter}+1 \nonumber \\
        &&+\max\left\{2\cdot 2^{\lceil\log n\rceil}+2\lceil\log n\rceil-2n-2,\,2n-\omega(n)-\lfloor\log n\rfloor-1\right\} \label{eqn:gcdIIa_qubit} \\
        \q_T(GCD_{IIb}) &=& (2N_{iter}+1)\lceil\log n\rceil+2n+3N_{iter}+1 \nonumber \\
        &&+\max\left\{2\cdot 2^{\lceil\log n\rceil}+2\lceil\log n\rceil-2n-2,\,3n-\omega(n)-\lfloor\log n\rfloor-2\right\} \label{eqn:gcdIIb_qubit} \\
        \q_T(GCD_{III}) &=& (2N_{iter}+1)\lceil\log n\rceil+2n+3N_{iter}+1 \nonumber \\
        &&+\max\left\{2\cdot 2^{\lceil\log n\rceil}+2\lceil\log n\rceil-2n-2,\,n-1\right\}. \label{eqn:gcdIII_qubit}
    \end{eqnarray}

    \item \textbf{Toffoli-depth:} Let $\tcount_d^{of}(.)$ denote the Toffoli-depth of a unitary. The Toffoli-depth of the GCD circuits is at most
    \begin{eqnarray}
        &&\tcount_d^{of}\left( U_{odd}^{(2)} \right) + \tcount_d^{of}\left(\mathcal{C}_{shift}\right) + N_{iter} \left( \tcount_d^{of}(U_{odd}) + 2\tcount_d^{of}(\cns) \right)    \nonumber \\
        &\leq& \left(3\lceil\log n\rceil 2^{\lceil\log n\rceil}+1 \right)+\left(\lceil\log n\rceil 2^{\lceil\log n\rceil}+1\right)+N_{iter}\left(2\lceil\log n\rceil 2^{\lceil\log n\rceil} -2n+2  \right) \nonumber \\
        &&+2N_{iter}\tcount_d^{of}(\cns)   \nonumber   \\
        &\leq& 2(N_{iter}+2)\lceil\log n\rceil 2^{\lceil\log n\rceil} -2N_{iter}(n-1)+2 + 2N_{iter}\tcount_d^{of}(\cns).    \nonumber
    \end{eqnarray}
    For a Clifford+Toffoli implementation using the COMP-N-SUB circuits in Section~\ref{sec:cns}, the Toffoli-depths of the GCD circuits are bounded as follows.
    \begin{eqnarray}
        \tcount_d^{of}(GCD_I) &\leq& 2(N_{iter}+2)\lceil\log n\rceil 2^{\lceil\log n\rceil} -2N_{iter}(n-1)+2 + 2N_{iter}\left(3n-1\right)  \nonumber   \\
        &=& 2(N_{iter}+2)\lceil\log n\rceil 2^{\lceil\log n\rceil} +4N_{iter}n+2 \label{eqn:gcdI_tof_depth}  \\
        \tcount_d^{of}(GCD_{IIa}) &\leq& 2(N_{iter}+2)\lceil\log n\rceil 2^{\lceil\log n\rceil} -2N_{iter}(n-1)+2 + 2N_{iter}\left(n+4\lfloor\log n\rfloor+8\right) \nonumber \\
        &=& 2(N_{iter}+2)\lceil\log n\rceil 2^{\lceil\log n\rceil} + 2N_{iter}\left( 4\lfloor\log n\rfloor+9 \right)+2 \label{eqn:gcdIIa_tof_depth}    \\
        \tcount_d^{of}(GCD_{IIb}) &\leq& 2(N_{iter}+2)\lceil\log n\rceil 2^{\lceil\log n\rceil} -2N_{iter}(n-1)+2 + 2N_{iter}\left(4\lfloor\log n\rfloor +9\right) \nonumber   \\
        &=& 2(N_{iter}+2)\lceil\log n\rceil 2^{\lceil\log n\rceil} + 2N_{iter}\left(4\lfloor\log n\rfloor-n+10 \right)+2 \label{eqn:gcdIIb_tof_depth}    \\
        \tcount_d^{of}(GCD_{III}) &\leq& 2(N_{iter}+2)\lceil\log n\rceil 2^{\lceil\log n\rceil} -2N_{iter}(n-1)+2 + 2N_{iter}n  \nonumber   \\
        &=& 2(N_{iter}+2)\lceil\log n\rceil 2^{\lceil\log n\rceil} + 2(N_{iter}+1) \label{eqn:gcdIII_tof_depth}
    \end{eqnarray}

    \item \textbf{T-depth:} Let $\tcount_d(.)$ denote the T-depth of a unitary. The T-depth of the GCD circuits is at most
    \begin{eqnarray}
        &&\tcount_d\left( U_{odd}^{(2)} \right) + \tcount_d\left(\mathcal{C}_{shift}\right) + N_{iter} \left( \tcount_d(U_{odd}) + 2\tcount_d(\cns) \right)    \nonumber \\
        &\leq& \left( 7\lceil\log n\rceil 2^{\lceil\log n\rceil}+3 \right) + \left( 3\lceil\log n\rceil 2^{\lceil\log n\rceil}+3 \right) + 4N_{iter}\lceil\log n\rceil 2^{\lceil\log n\rceil} + 2N_{iter}\tcount_d(\cns)    \nonumber   \\
        &=& 2\left( 5+ 2N_{iter} \right)\lceil\log n\rceil 2^{\lceil\log n\rceil} +6 +2N_{iter}\tcount_d(\cns). \nonumber
    \end{eqnarray}
    For a Clifford+T implementation using the various COMP-N-SUB circuits from Section~\ref{sec:cns}, the T-depths of the GCD circuits are bounded as follows.
    \begin{eqnarray}
        \tcount_d(GCD_I) &\leq& 2\left( 5+ 2N_{iter} \right)\lceil\log n\rceil 2^{\lceil\log n\rceil} +6 +2N_{iter}\left( 9n-3\right)   \nonumber \\
        &\leq& 2\left( 5+ 2N_{iter} \right)\lceil\log n\rceil 2^{\lceil\log n\rceil} + 18N_{iter}n   \label{eqn:gcdI_tdepth} \\
        \tcount_d(GCD_{IIa}) &\leq& 2\left( 5+ 2N_{iter} \right)\lceil\log n\rceil 2^{\lceil\log n\rceil} +6 +2N_{iter}\left(3n+12\lfloor\log n\rfloor+24 \right)   \nonumber \\
        &=& 2\left( 5+ 2N_{iter} \right)\lceil\log n\rceil 2^{\lceil\log n\rceil} +6N_{iter} \left( n+4\lfloor\log n\rfloor+8  \right)+6 \label{eqn:gcdIIa_tdepth}   \\
        \tcount_d(GCD_{IIb}) &\leq& 2\left( 5+ 2N_{iter} \right)\lceil\log n\rceil 2^{\lceil\log n\rceil} +6 +2N_{iter}\left( 12\lfloor\log n\rfloor+27 \right)   \nonumber \\
        &=& 2\left( 5+ 2N_{iter} \right)\lceil\log n\rceil 2^{\lceil\log n\rceil} + 6N_{iter}\left( 4\lfloor\log n\rfloor +9 \right) +6   \label{eqn:gcdIIb_tdepth}   \\
        \tcount_d(GCD_{III}) &\leq& 2\left( 5+ 2N_{iter} \right)\lceil\log n\rceil 2^{\lceil\log n\rceil} +6 +2N_{iter}\left( 5n \right)   \nonumber   \\
        &=& 2\left( 5+ 2N_{iter} \right)\lceil\log n\rceil 2^{\lceil\log n\rceil} + 10N_{iter}n+6     \label{eqn:gcdIII_tdepth}
    \end{eqnarray}
    
\end{itemize}

 \end{document}